\documentclass[reprint,aps,pre,superscriptaddress,longbibliography]{revtex4-2}
\usepackage{amsmath}
\usepackage{amssymb}
\usepackage{bm}
\usepackage{color}
\usepackage{commath}
\usepackage{dcolumn}
\usepackage{graphicx}
\usepackage{hyperref}
\usepackage{multirow}
\usepackage{subfigure}

\begin{document}

\preprint{ITP307/2025-01}

\title{Discontinuous phase transitions of feature detection in lateral predictive coding}

\author{Zhen-Ye Huang}
\email{Current address: School of Science, Westlake University, Hangzhou 310030, China}
\affiliation{
  Institute of Theoretical Physics, Chinese Academy of Sciences, Beijing 100190, China
}
\affiliation{
  School of Physical Sciences, University of Chinese Academy of Sciences, Beijing 100049, China
}

\author{Weikang Wang}
\email{wangwk@itp.ac.cn}
\affiliation{
  Institute of Theoretical Physics, Chinese Academy of Sciences, Beijing 100190, China
}

\author{Hai-Jun Zhou}
\email{zhouhj@itp.ac.cn}
\affiliation{
  Institute of Theoretical Physics, Chinese Academy of Sciences, Beijing 100190, China
}
\affiliation{
  School of Physical Sciences, University of Chinese Academy of Sciences, Beijing 100049, China
}
\affiliation{
  MinJiang Collaborative Center for Theoretical Physics, MinJiang University, Fuzhou 350108, China}

\date{\today}

\begin{abstract}
The brain may adopt the strategy of lateral predictive coding (LPC) to construct optimal internal representations for salient features in input sensory signals, reducing the energetic cost of information transmission. Here we first consider the task of detecting one non-Gaussian signal by LPC from Gaussian background signals of the same magnitude, which is intractable by principal component decomposition. We study the emergence of feature detection function from the perspective of tradeoff between energetic cost $E$ and information robustness, and implement this tradeoff by a thermodynamic free energy. We define $E$ as the mean $L_1$-norm of the internal state vectors, and quantify the level of information robustness by an entropy measure $S$. There are at least three types of optimal LPC matrices, one type with very weak synaptic weights and $S \approx 0$, and two functional types either with low energy $E$ or with high entropy $S$ in which one single unit selectively responds to the non-Gaussian input feature.  Energy--information tradeoff induce two discontinuous phase transitions between these three types of optimal LPC networks. We then extend the discussion to detecting and distinguishing between two non-Gaussian input features and observe similar discontinuous phase transitions.
\end{abstract}

\maketitle


\section{Introduction}

Predictive coding is a basic strategy adopted by the brain to reduce energetic cost of signal transmission~\cite{Srinivasan-etal-1982,Rao-Ballard-1999,Huang-Rao-2011,Ali-etal-2022}. Between different hierarchical layers of the brain feedforward and feedback signals are constantly exchanged, and at each hierarchical layer the bottom-up signals are partially canceled by top-down signals to produce residual prediction-error output messages back to higher and lower layers~\cite{VanZwol-etal-2024,Millidge-etal-2022}. Besides these between-layer interactions, lateral predictive coding (LPC) interactions  within individual layers are also extremely important for efficient and robust neural signal processing. There are statistical correlations between the input signals of different neurons. Through lateral interactions with appropriate synaptic weights $w_{i j}$, the response of one neuron $j$ can help to predict and cancel the input to another neuron $i$~\cite{Srinivasan-etal-1982,Huang-etal-2022}. The competition caused by such lateral interactions may be a major microscopic mechanism underlying the selective response and sparse coding of biological neurons~\cite{Rozell-etal-2008,Yu-etal-2018,Yang-Zhou-etal-2017}. Lateral predictive coding may also support associative memory in the hippocampus of the brain~\cite{Tang-etal-2023b}. 

Lateral interactions greatly reduce the output pair correlations such that the outputs from different neurons are representing different collective features (patterns) of the input data, offering biologically plausible implementations of principal component analysis and independent component analysis~\cite{Hyvarinen-etal-2009}. As an acquired internal model encoding the statistical regularity of input signals, the LPC weight matrix $\bm{W}$ is highly nonrandom and non-symmetric ($w_{i j} \neq w_{j i}$). Exploring non-symmetric LPC interactions and the emergence of structural patterns and collective behaviors in optimal LPC networks become an interesting subject of statistical physics, with implications for the design of artificial neural networks.

Recently we performed a theoretical study of phase transitions in the optimal LPC network driven by energy--information tradeoff~\cite{Huang-etal-2023}. In line with the efficient-coding principle~\cite{Barlow-1972,Bialek-2024}, we posited that the optimal LPC matrix $\bm{W}$ is the outcome of balance between two conflicting demands: reducing the energetic cost of transmitting the output signal on the one hand and retaining information robustness against noise on the other hand. We found that, as the tradeoff control parameter (the temperature $T$) decreases, the optimal weight matrix changes qualitatively at several critical points, and rich internal structures such as cyclic dominance and excitation--inhibition balance emerge without the need of imposing any additional assumptions and regularization terms.

The optimal LPC network identifies a set of (not necessarily orthogonal) independent components of the input signal vectors after a continuous phase transition, and it is located at the edge of chaos at still lower temperatures in the sense that the minimum real part $\lambda_0$ of the complex eigenvalues of $(\bm{I}+\bm{W})$ becomes close to zero. However, because the mean energetic cost of the model only depends on the correlation matrix of the input data but not on any of the higher-order moments, the optimal network is not capable of distinguishing between non-Gaussian and Gaussian distributed signals.

Non-Gaussian signals are ubiquitous in natural environments~\cite{Hyvarinen-etal-2009,Jutten-Herault-1991}. In the present work, we study theoretically and by computer simulations the conditions for the emergence of feature detection function in a linear LPC model system using the same energy--information tradeoff framework. Different from our earlier model~\cite{Huang-etal-2023}, here we assume that the energetic cost is the $L_1$-norm (absolute value) of the output signal. We regard this energy form as a better approximation to biological reality, as the energetic cost of information transmission may be roughly a linear function of the firing rate of action potentials in a real nervous system. We demonstrate that discontinuous phase transitions can occur in the optimal LPC matrix, and a single hidden non-Gaussian feature of the input data can be captured and represented by a single unit at both high and low temperatures. The corresponding optimal LPC networks either have relatively low mean energy or have relatively high information robustness. Our numerical algorithm also reports LPC matrices whose energy values are local minima but not the global minimum, indicating that the energy landscape of the LPC system is complex.

We also consider the more difficult case of two non-Gaussian features being hidden in the input signal vectors. The tradeoff between energetic cost and information robustness can again induce discontinuous phase transitions in this task. After either a discontinuous drop in the mean $L_1$-norm energy, or a discontinuous jump in information robustness, the resulting optimal LPC networks attain spontaneously the ability of capturing both features. These two (not necessarily orthogonal) non-Gaussian features will be represented separately by two different single units.

Our work brings new theoretical insights into the spontaneous emergence of structural patterns and functions in lateral predictive coding networks. It may also encourage future exploration on artificial neural networks with lateral interactions. The synaptic weights of the LPC system are not symmetric and they are not independent random variables. As a clear demonstration of non-randomness and non-reciprocity, we show that the complex eigenvalues of the optimal LPC system are pushed towards lying on a semicircle by the stress of energetic cost minimization. 

It may be interesting to point out that, our theoretical prediction of discontinuous phase transitions is consistent with some empirical observations in the literature, which reported that learning to recognize complex patterns or rules is a slow process with sudden transitions (see, e.g., Refs.~\cite{Hosenfeld-etal-1997,Boshuizen-2004,Collins-etal-2019,Rosenberg-etal-2021}). On the much longer time scale of evolution the human brain has been the result of several major phase transitions~\cite{Bretas-etal-2020,Ginsburg-Jablonka-2021}, and our work may also be relevant for appreciating the evolution of brain structure and functions in terms of energy--information tradeoff.

Lateral predictive coding is still a rarely touched topic in the statistical physics community. Our work is an attempt to determine all the synaptic weights of LPC completely through the tradeoff between energetic cost and information robustness, but we have only considered some simplest random problem instances. We have not yet applied the LPC model to real-world feature detection problems and have only started to address the problem of separating multiple non-Gaussian features. We hope the present work can stimulate more deeper and broader investigations in the near future.

\section{Theoretical framework}

Linear LPC is a simplified model for energy-efficient information processing in the nervous system. The system is formed by $N$ units and the synaptic interactions between them. Each unit with index $i \in \{1, \ldots, N\}$ may represent a single neuron or a collection of neurons. The unit $i$ has a real-valued internal (and output) state  $x_i$ and it receives real-valued input signals $s_i$. An internal state of the whole system is denoted by a column vector $\vec{\bm{x}} = (x_1, \ldots, x_N)^\top$ with the superscript $\top$ indicating transpose, and an input vector is denoted by $\vec{\bm{s}} = (s_1, \ldots, s_N)^\top$. The instantaneous response of the system to an input $\vec{\bm{s}}$ is described by the following linear recursive dynamics
\begin{equation}
\tau_0 \frac{\textrm{d} \vec{\bm{x}}}{\textrm{d} t}
  \, = \,  \vec{\bm{s}} - \vec{\bm{x}} - \bm{W} \vec{\bm{x}} \; .
  \label{eq:lpc}
\end{equation}
Here $\bm{W}$ is the synaptic weight matrix with elements $w_{i j}$. (We use bold upper-case Roman symbols to denote matrices and lower-case roman symbols with subscripts to denote matrix elements.)  The lateral influence $\sum_{j\neq i} w_{i j} x_j$ of all the other units $j$ on unit $i$ is interpreted as a prediction about the input $s_i$. We only consider predictive interactions between different units, so all the diagonal elements are set to zero ($w_{i i} = 0$). The matrix $\bm{W}$ is not necessarily symmetric and so $w_{i j} \neq w_{j i}$~\cite{Huang-etal-2022} in general.

The parameter $\tau_0$ is the time scale of the instantaneous dynamics. (We can simply set $\tau_0 = 1$ after rescaling time $t$ by $\tau_0$.) We can add an unbiased noise term to the right-hand side of Eq.~(\ref{eq:lpc}) to account for fluctuating environmental effects. It is anticipated that the environmental noise is changing on a time scale much faster than $\tau_0$. If the input signal vector $\vec{\bm{s}}$ changes much slower than $\tau_0$, the steady-state mean response of Eq.~(\ref{eq:lpc}) will be
\begin{equation}
  \vec{\bm{x}} \, = \, \frac{\bm{I}}{\bm{I} + \bm{W}}\, \vec{\bm{s}} \; ,
  \label{eq:xss}
\end{equation}
where $\bm{I}$ is the identity matrix. This steady-state output $\vec{\bm{x}}$ is equal to $\vec{\bm{s}} - \bm{W} \vec{\bm{x}}$, so it also serves as the prediction-error vector~\cite{Srinivasan-etal-1982}. Notice that the real parts of all the eigenvalues of the matrix $(\bm{I}+\bm{W})$ must be positive to ensure the convergence of $\vec{\bm{x}}$~\cite{Huang-etal-2023}. The determinant of the matrix $(\bm{I}+\bm{W})$ is guaranteed to be positive when the real parts of all its eigenvalues are positive. 

The major energetic costs in the mammalian cortex are associated with action potential generation and synaptic transmission~\cite{Niven-2016,Howarth-etal-2012}. In our present work the energetic cost $E$ is defined as the summed mean absolute value of the internal states (prediction errors) $x_i$:
\begin{equation}
  E \, \equiv \,  \sum_{i=1}^{N} \Bigl\langle \bigl| x_i \bigr| \Bigr\rangle
  \, = \,  \sum_{i=1}^{N} \Bigl\langle \Bigl| \sum_{j=1}^{N}
  \bigl(\frac{\bm{I}}{\bm{I}+\bm{W}}\bigr)_{i j} s_j \Bigr| \Bigr\rangle
  \; ,
  \label{eqL1E}
\end{equation}
where $\langle A \rangle \equiv \int \mathrm{d} \vec{\bm{s}} A(\vec{\bm{s}}) p_{\textrm{in}}(\vec{\bm{s}})$ denotes the mean value of variable $A(\vec{\bm{s}})$ over the probability distribution $p_{\text{in}}(\vec{\bm{s}})$ of inputs. We assume that the LPC system will try to minimize the energy $E$ by adapting the weight matrix $\bm{W}$ to the distribution $p_{\textrm{in}}(\vec{\bm{s}})$ of input signals.

Because of the linear mapping between $\vec{\bm{s}}$ and $\vec{\bm{x}}$, we can derive (see Appendix~\ref{app:entropy} for details) that the entropy difference $S$ between the probability distribution of the output signal $\vec{\bm{x}}$ and that of the input signal $\vec{\bm{s}}$ is
\begin{equation}
  S = -\ln \Bigl[ \textrm{det} \bigl(\bm{I} + \bm{W} \bigr) \Bigr] \; ,
  \label{entropy}
\end{equation}
where $\textrm{det}(\bm{\cdot})$ reports the determinant of a matrix. Since the entropy of the input vectors $\vec{\bm{s}}$ is independent of the weight matrix, in the following discussions we simply refer to the entropy difference $S$ as the entropy of the output vectors $\vec{\bm{x}}$.

The geometric picture underlying the expression (\ref{entropy}) is that a volume of the input $\vec{\bm{s}}$-space is mapped to a volume of the output $\vec{\bm{x}}$-space with a rescaling  (Jacobian) factor $1/\det( \bm{I}+\bm{W})$. It is obviously desirable for this volume ratio to be as large as possible, so that the outputs $\vec{\bm{x}}^{(1)}$ and $\vec{\bm{x}}^{(2)}$ of two input signals $\vec{\bm{s}}^{(1)}$ and $\vec{\bm{s}}^{(2)}$ might still be well separated after they are corrupted by the inevitable transmission noise~\cite{Huang-etal-2023}. Indeed, the entropy $S$ is a quantitative measure of the mutual information between input $\vec{\bm{s}}$ and output $\vec{\bm{x}}$ under transmission noise (see Appendix~\ref{app:IR}). We assume that the functional benefit of information robustness is another intrinsic force which drives the evolution of $\bm{W}$ towards entropy $S$ maximization~\cite{Barlow-1972,Jutten-Herault-1991,Bell-Sejnowski-1995,Bialek-2024}. 

But entropy maximization and energy minimization are conflicting objectives. Following the earlier work \cite{Huang-etal-2023}, we introduce a tradeoff parameter $T$ to balance energy efficiency and information robustness, and define a free energy quantity $F$ as
\begin{equation}
  F \, =  \, E - T\, S \; .
  \label{freeenergy}
\end{equation}

At each fixed value of $T$ the global minimum of $F$ determines the optimal weight matrix $\bm{W}$. The parameter $T$ represents the fitness pressure which forces the system to reduce energy consumption when $T$ is small and encourages it to increase the output entropy when $T$ is large.  We call $T$ the temperature of the LPC system. The free energy minimization problem (\ref{freeenergy}) is equivalent to the problem of Pareto optimal front with the minimization goal being $E/(1+T) - T S /(1+T)$~\cite{Seoane-Sole-2015}. When the number $\mathcal{M}$ of input samples $\vec{\bm{s}}$ approaches infinity, the accumulated total free energy is $\mathcal{M} F$. In this sense of statistical counting~\cite{Qian-2022}, generic phase transitions will occur even for finite system sizes $N$ if the minimum $F$ is singular at certain critical values of temperature $T$. We emphasize that the thermodynamic limit for the LPC system is taking to be $\mathcal{M}\rightarrow \infty$ but with $N$ being finite~\cite{Huang-etal-2023} (see also Refs.~\cite{Seoane-Sole-2015,Kocillari-etal-2018} for related discussions). 

A very interesting observation of our earlier work \cite{Huang-etal-2023} was that, within certain temperature range, the optimal LPC system achieves a decomposition of the input signal vector $\vec{\bm{s}}$ as 
\begin{equation}
    \vec{\bm{s}} \, = \, x_1 \vec{\bm{u}}_1  + x_2 \vec{\bm{u}}_2  + \ldots +  x_N \vec{\bm{u}}_N \; ,
    \label{eqa290425}
\end{equation}
such that the pair correlation $\langle x_i x_j \rangle = 0$ for any $i \neq j$ and the second moments $\langle x_k^2 \rangle = T /2$ for all the $x_k$ coefficients. The $i$-th entry of the vector $\vec{\bm{u}}_i$ is identical to unity and its $j$-th entry with $j\neq i$ is the synaptic weight $w_{j i}$. Let us emphasize that these vectors $\vec{\bm{u}}_i$ are \emph{not} the principal components of the ensemble of input vectors $\vec{\bm{s}}$. First, $\vec{\bm{u}}_i$ are not vectors of unit length, and second and more significantly, they are in general not mutually orthogonal to each other. Therefore Eq.~(\ref{eqa290425}) is a non-orthogonal decomposition with ``independent" coefficients $x_i$.

We present in Appendix~\ref{app:CGsystem} some analytical results on this interesting decomposition for correlated Gaussian input signals, under our modified assumption of $L_1$-norm mean energy (\ref{eqL1E}). These results indicate that, for correlated Gaussian input signals, the $L_1$-norm energy form (\ref{eqL1E}) does not bring qualitative difference. The next two sections focus on input signals which contain non-Gaussian components, and we will demonstrate that optimal LPC systems with $L_1$-norm mean energy are capable of separating non-Gaussian features from the Gaussian background signals.

\section{Detection of a single non-Gaussian feature}
\label{sec:1freature}

Natural signals contain both background noises and nonrandom features~\cite{Hyvarinen-etal-2009}. In our present theoretical work, we first consider the problem of a single feature $\vec{\bm{\phi}}_1$ hidden in Gaussian random backgrounds~\cite{Wang-Lu-2019},
\begin{equation}
  \vec{\bm{s}} \, = \, a_1 \vec{\bm{\phi}}_1 \, + \,
  b_2 \vec{\bm{\phi}}_2 + \ldots + b_N \vec{\bm{\phi}}_N \; .
  \label{eqsform}
\end{equation}
Here $\vec{\bm{\phi}}_i = (\phi_{1,i}, \ldots, \phi_{N,i})^\top$ are  $N$-dimensional real vector of unit length ($\sum_j \phi_{j, i}^2 = 1$) and they are orthogonal to each other ($\sum_j \phi_{j, i} \phi_{j, k} = 0$ for $i\neq k$), and $\{b_i\}_{i=2}^N$ are independent Gaussian random coefficients with zero mean and unit variance. The coefficient $a_1$ also has zero mean and unit variance, but it is sampled from a non-Gaussian probability distribution $q(a_1)$. The task for the LPC network is to distinguish $\vec{\bm{\phi}}_1$ from all the other directions $\vec{\bm{\phi}}_j$.

In our present problem setting, the correlation matrix of the input signal vectors $\vec{\bm{s}}$ is the $N$-dimensional identity matrix, 
\begin{equation}
  \bigl\langle \vec{\bm{s}} \, {\vec{\bm{s}}}^\top \bigr\rangle \,  = \,
  \bm{I}
  \; ,
\end{equation}
which contains no information about the feature direction $\vec{\bm{\phi}}_1$. It is therefore impossible to infer the direction $\vec{\bm{\phi}}_1$ by performing principal-component analysis on this correlation matrix. As the mean energy of the LPC model of Ref.~\cite{Huang-etal-2023} only depends on this correlation matrix and not on the higher moments of $p_{\textrm{in}}(\vec{\bm{s}})$, the resulting optimal LPC system is naturally incapable of accomplishing feature detection for the challenging task (\ref{eqsform}). This motivates us to adopt the $L_1$-norm mean energy (\ref{eqL1E}).

\subsection{Energy and order parameter}
\label{sec:setting}

At a fixed value of the non-Gaussian coefficient $a_1$, the conditional probability distribution $p_{\textrm{out}}(x_i | a_1)$ of the output state $x_i$ of the $i$-th unit is Gaussian,
\begin{equation}
  p_{\textrm{out}}\bigl(x_i | a_1 \bigr) \, = \, \frac{1}{\sqrt{2 \pi \sigma_i^2}}
  \exp\Bigl( - \frac{(x_i - a_1 \mu_i)^2}{2 \sigma_i^2} \Bigr) \; .
  \label{eq241217a}
\end{equation}
The expectation value of $x_i$ is proportional to $a_1$ with coefficient $\mu_i$ and its variance is $\sigma_i^2$. The analytical expressions for $\mu_i$ and $\sigma_i^2$ are easy to derive (see Appendix~\ref{app:pouti}):
\begin{eqnarray}
  \mu_i  \, & \equiv & \,
    \Bigl[ \frac{\bm{I}}{\bm{I} + \bm{W}} \vec{\bm{\phi}}_1 \Bigr]_i
  =  \sum_j \Bigl[\frac{\bm{I}}{\bm{I}+\bm{W}}\Bigr]_{i j} \phi_{j, 1} \; ,
  \label{eq:muiexp} \\
  \sigma_i^2 \, & \equiv & \, 
  \Bigl[ \frac{\bm{I}}{(\bm{I}+\bm{W})^\top (\bm{I}+\bm{W})} \Bigr]_{i i}
  -  \mu_i^2 \; .
  \label{eq:sigmai2}
\end{eqnarray}

From the expression (\ref{eq:muiexp}) we understand that $\mu_i$ is the representation of $\vec{\bm{\phi}}_1$ by the $i$-th unit of the network. In other words, $a_1 \mu_i$ is the mean response of unit $i$ to the non-Gaussian feature $a_1 \vec{\bm{\phi}}_1$. We can define the relative responsiveness parameter $Q_i$ as
\begin{equation}
  Q_i \equiv  \sqrt{ \frac{\mu_i^2}{ \sum_{j=1}^N  \mu_j^2} } \; ,
  \label{eq:Qi}
\end{equation}
such that $\sum_{i=1}^{N} Q_i^2 = 1$. The unit $i$ whose $Q_i$ is the maximum among all the $N$ units is referred to as the most responsive unit. We define an order parameter (the overlap $Q$) as 
\begin{equation}
  Q \, = \, \max\limits_{i} Q_i \; ,
  \label{eq:Qoverlap}
\end{equation}
which is the maximum of $Q_i$ over all the $N$ units. If $Q$ approaches the lower-bound value $1/\sqrt{N}$, all the units are responding equally and weakly to the feature $\vec{\bm{\phi}}_1$. In the opposite situation of $Q \approx 1$, a single unit is responding to $\vec{\bm{\phi}}_1$ very strongly and all the other units are indifferent to this feature, and it means that feature detection has been accomplished.

For the non-Gaussian probability distribution $q(a_1)$, a discrete form convenient for analytical analysis is
\begin{equation}
  q( a_1 ) \, = \,
  \left\{
  \begin{array}{ll}
    (1-p_0)/2 \; ,
    & \quad \quad a_1 = 1/ \sqrt{1-p_0} \; , \\
    p_0 \; , & \quad \quad a_1 = 0 \; , \\  
    (1-p_0) / 2 \; , & \quad\quad a_1 = -1 / \sqrt{1-p_0} \; .
  \end{array}
  \right.
  \label{eqPa1}
\end{equation}
The mean of $a_1$ is zero and its variance is unity, for any value of the adjustable parameter $p_0 \in [0, 1)$. It is then easy to derive an analytical expression for the mean $L_1$-norm energy (\ref{eqL1E}) as
\begin{equation}
  E  = \sum\limits_{i=1}^{N} \Bigl[ \sqrt{ \frac{2 \sigma_i^2}{\pi} }
    \bigl(  (1-p_0) e^{-\zeta_i^2} + p_0 \bigr) +
    \sqrt{ (1-p_0) \mu_i^2} \textrm{erf}( \zeta_i )  \Bigr] \; ,
    \label{eqL1Edisc}
\end{equation}
where $\zeta_i \equiv \sqrt{\mu_i^2/2 (1-p_0) \sigma_i^2}$ and $\textrm{erf}(\zeta_i)$ is the standard error function (see Appendix~\ref{app:mE} for technical details).

Besides this simple discrete prior distribution, we also consider other forms of $q(a_1)$, including the continuous Laplace distribution $q(a_1) = e^{-\sqrt{2} |a_1|}/\sqrt{2}$ and the long-tailed power-law distribution $q(a_1) \sim |a_1|^{-\gamma}$ with exponent $\gamma$. We list in Appendix~\ref{app:mE} the mean energy expressions corresponding to these two prior distributions.

A more general situation is to assume that the coefficient $a_1$ is Gaussian with probability $p_g$ and is non-Gaussian with the remaining probability $(1-p_g)$. We expect that as $p_g$ increases from zero, the feature detection problem will become more and more difficult. There may exist a critical point along this $p_g$ axis. For simplicity, we do not explore this interesting issue here and will restrict $p_g=0$ in the present work.

\subsection{Energy minimization at fixed entropy}
\label{subsec:algorithm}

We carry out extensive numerical computations on many problem ensembles, which differ in the number $N$ of units, the feature direction $\vec{\bm{\phi}}_1$, and the coefficient distribution $q(a_1)$.  The numerical results obtained by our algorithm on these different ensembles turn out to be qualitatively similar. To be concrete, here we mainly discuss results obtained on the representative ensemble of size $N=36$, uniform  $\vec{\bm{\phi}}_1 \propto \bigl(1, 1, \ldots, 1 \bigr)^{\top}$ and the discrete distribution (\ref{eqPa1}) with $p_0 = 0.7$. We also report some results obtained on a system of larger size $N=100$. 

\begin{figure*}
  \centering
  \subfigure[$\ S\, =\, 0$]{
    \includegraphics[angle=270,width=0.3\linewidth]{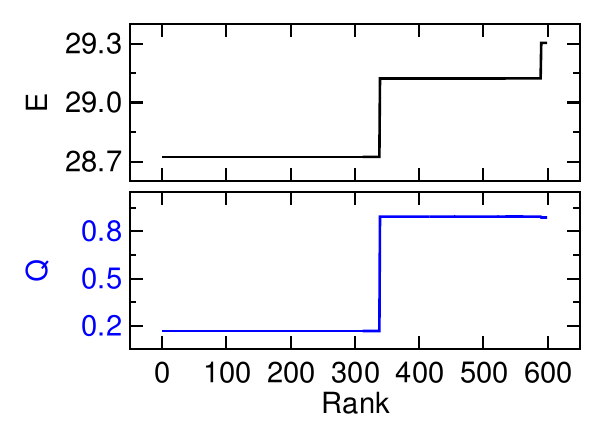}
    \label{fig:E_Psi_lD0_N36}
  } 
  \subfigure[$\ S\, =\, 0$]{
    \includegraphics[angle=270,width=0.3\linewidth]{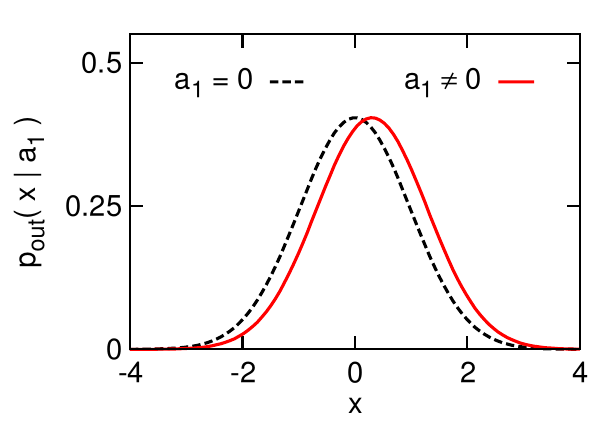}
    \label{fig:PxN36lD0}
  }
  \\
  \subfigure[$\ S\, =\, -1.5$]{
    \includegraphics[angle=270,width=0.3\linewidth]{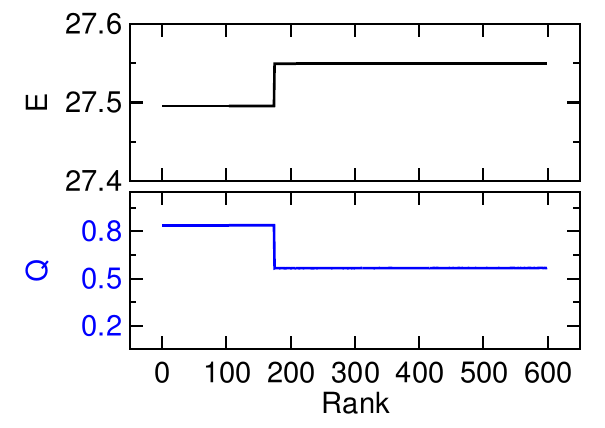}
    \label{fig:E_Psi_lD1p5_N36}
  } 
  \subfigure[$\ S\, =\,  -1.5$]{
    \includegraphics[angle=270,width=0.3\linewidth]{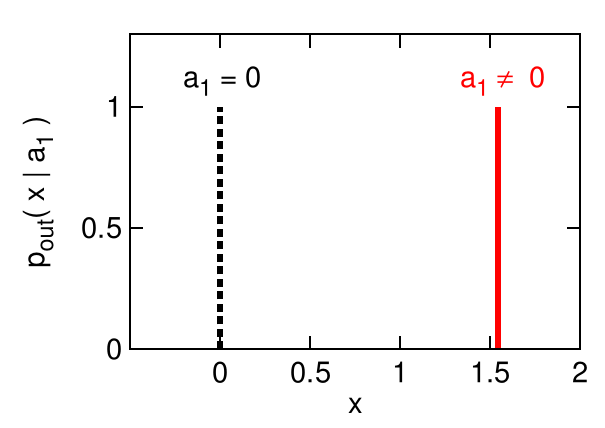}
    \label{fig:PxN36lD1p5}
  }
  \caption{
    (left) Minimal energies $E$ (sorted in ascending order) and the corresponding overlap values $Q$ obtained through $600$ independent runs of the stochastic search dynamics at fixed value of $S=0$ (a) and $S=-1.5$ (c). (right) Probability distribution of the internal state $x$ of the most responsive unit conditional on the coefficient $a_1$, for the optimal weight matrix with $S=0$ (b) and $S=-1.5$ (d). System size $N=36$ and $p_0=0.7$.
    }
  \label{fig:E_Psi_N36}
\end{figure*}
\begin{figure*}
  \centering
  \subfigure[$\ S = -1.5, \ E\, \approx\, 27.50$]{
    \includegraphics[angle=270,width=0.3\linewidth]{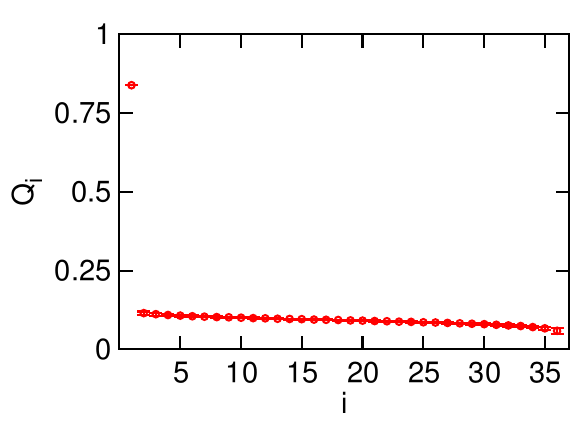}
    \label{fig:Qi_lD1p5_N36A}
  }
  \subfigure[$\ S = -1.5, \ E\, \approx\, 27.55$]{
    \includegraphics[angle=270,width=0.3\linewidth]{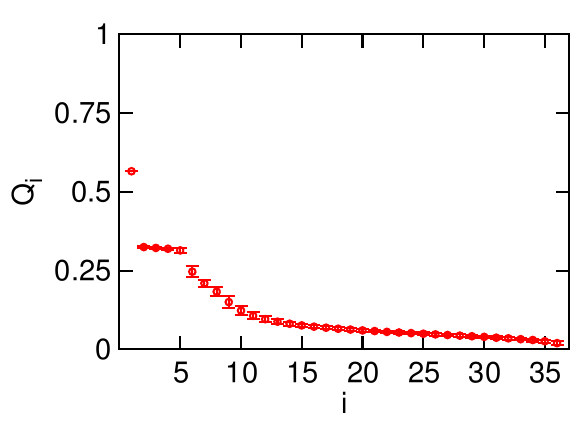}
    \label{fig:Qi_lD1p5_N36B}
  }
  \caption{
    Rank plots of the $N=36$ responsiveness quantities $Q_i$ (Eq.~(\ref{eq:Qi})), computed from the $600$ independently sampled matrices of Fig.~\ref{fig:E_Psi_lD1p5_N36} with fixed entropy $S = -1.5$. Each data point (mean and standard deviation) is the average over the $175$ weight matrices with energy values $E\approx 27.50$ (a) and over the $425$ weight matrices with energy values $E\approx 27.55$ (b).
  }
  \label{fig:Qiprofile}
\end{figure*}

We adopt a microcanonical (entropy-clamped) annealing approach to solve the optimal LPC problem (details of this algorithm have been described in Ref.~\cite{Huang-etal-2023}). The entropy range $S \in [-6, 9]$ around $S = 0$ is examined, and at each value of $S$ the hard constraint $\textrm{det}\bigl(\bm{I}+\bm{W}\bigr) = e^{-S}$ is imposed on the weight matrix $\bm{W}$. At each elementary step of the stochastic search dynamics, we perturb a randomly chosen row or column of the current matrix under the constraints of fixed $S$ and zero diagonal elements, and compute the associated energy change $\delta E$. We accept the perturbed matrix with certainty if $\delta E \leq 0$ or with probability $e^{-\kappa \delta E}$ if $\delta E > 0$. After a large number of such trials (typically $10^6$) the annealing parameter $\kappa$ is then increased by a factor $1+\varepsilon$ (typically $\varepsilon = 0.02$). The initial value of $\kappa$ is set to be $100$. When $\kappa$ reaches a final threshold value (typically $10^8$) we terminate the annealing process and output the minimum energy value $E$ reached during the whole evolution trajectory and the corresponding matrix $\bm{W}$. We always verify that the minimum eigenvalue of $(\bm{I}+\bm{W})$ has positive real part. 

We now demonstrate by two concrete examples that, the optimal weight matrices reached at different values of entropy $S$ may have qualitative differences in their feature detection property. 

Figure~\ref{fig:E_Psi_lD0_N36} plots in ascending order the obtained minimal energies $E$ and the corresponding overlaps $Q$ from $600$ independent runs of the matrix annealing algorithm at fixed $S = 0$, all starting from the same initial weight matrix. The minimal energies form several bands, indicating the existence of many local minimal energies. The global minimum energy is $E=28.7235$, and the corresponding overlap $Q=0.1667$ is equal to the theoretical lower-bound, meaning that the optimal LPC system at $S=0$ is not capable of detecting the hidden feature direction $\vec{\bm{\phi}}_1$. This conclusion also holds when the entropy is positive but relatively small (e.g., $S=1$). The conditional probabilities $p_{\textrm{out}}(x | a_1)$ of the internal state $x$ of the most responsive unit are largely indistinguishable at $a_1=0$ and $a_1 = 1/\sqrt{1-p_0}=1.8257$, see Fig.~\ref{fig:PxN36lD0}. 

We should emphasize that actually some of the sampled weight matrices with fixed entropy $S = 0$ are capable of detecting the feature direction $\vec{\phi}_1$. Indeed about $40\%$ of the sampled minimal-energy matrices have very high overlap values $Q \approx 0.89$ (Fig.~\ref{fig:E_Psi_lD0_N36}). But the minimal energies $E \approx 29.15$ of these matrices are remarkably higher than the global minimum value and therefore they can not win the competition in energetic cost.

\begin{figure*}
  \centering
  \subfigure[]{
    \includegraphics[angle=270,width=0.3\linewidth]{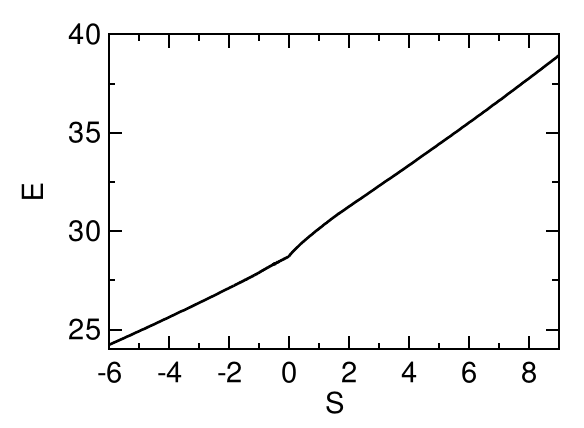}
    \label{fig:S_E_N36}
  }
  \subfigure[]{
    \includegraphics[angle=270,width=0.3\linewidth]{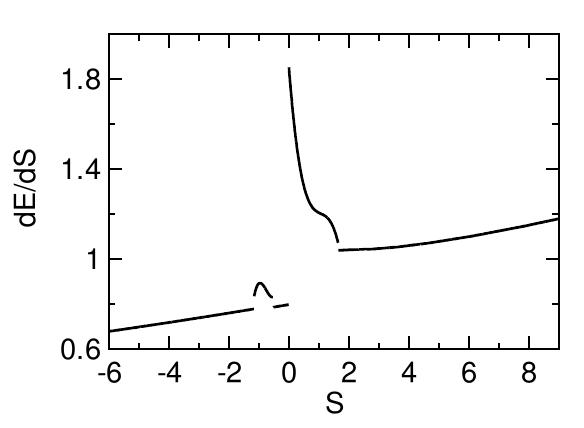}
    \label{fig:S_T_N36}
  }
  \\
  \subfigure[]{
    \includegraphics[angle=270,width=0.3\linewidth]{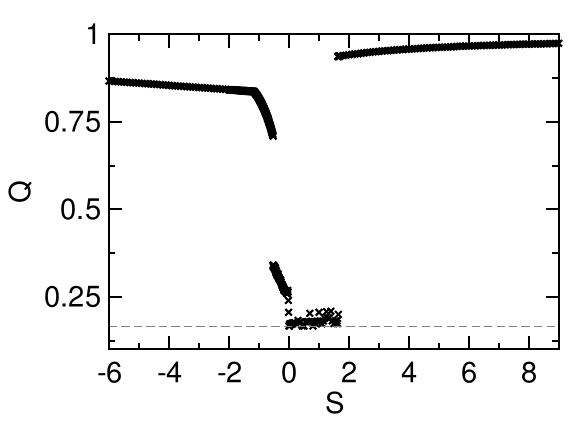}
    \label{fig:S_Q_N36}
  } 
  \subfigure[]{
    \includegraphics[angle=270,width=0.3\linewidth]{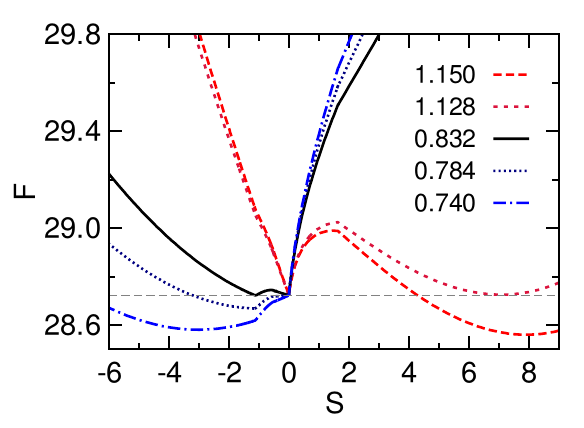}
    \label{fig:S_F_N36}
  }
  \caption{
  Thermodynamic quantities versus entropy $S$ for the system of size $N=36$ and $p_0 = 0.7$. (a) Minimum energy $E$. (b) Energy slope $ \textrm{d} E /\textrm{d} S$. (c) Overlap $Q$. (d) Free energy $F$ at five different temperatures $T$ ranging from $0.740$ to $1.150$.
  }
  \label{fig:ESN36}
\end{figure*}

Feature detection becomes achievable if the entropy is sufficiently positive ($S > 1.63$) or is sufficiently negative ($S < -1.16$). As an example, we list $600$ independently sampled minimal energy values and the corresponding overlaps at $S = -1.5$, all starting from a single initial matrix (Fig.~\ref{fig:E_Psi_lD1p5_N36}). The optimal weight matrix with the global minimum energy $E = 27.4955$ has high overlap $Q=0.8387$. As shown in Fig.~\ref{fig:PxN36lD1p5}, the most responsive unit is strongly active (with output $|x| \approx 1.52$) when the feature is present ($a_1 \neq 0$) and it is completely silent ($x \approx 0$) when the feature is absent ($a_1 = 0$). All the other $(N-1)$ units are mainly responding to the Gaussian background signals and their responses in the presence and absence of $\vec{\bm{\phi}}_1$ are indistinguishable (similar to Fig.~\ref{fig:PxN36lD0}). To further demonstrate this fact, we plot in Fig.~\ref{fig:Qi_lD1p5_N36A} the response quantities $Q_i$ (Eq.~(\ref{eq:Qi})) of all the $N$ units in descending order. We clearly see that all the $Q_i$ values are less than $0.116$ except for the largest one, which is $0.8387$.   

Similar to the case of $S = 0$, Fig.~\ref{fig:E_Psi_lD1p5_N36} reveals that about $71\%$ of the reported matrices at $S=-1.5$ by our optimization algorithm are local optimal solutions with higher energy values $E \approx 27.55$ and moderate overlap values $Q \approx 0.565$. Looking into these locally optimal matrices, we find that multiple units are selectively responding to the feature direction $\vec{\bm{\phi}}_1$. Indeed the rank plot of the relative responsiveness quantities $Q_i$ in Fig.~\ref{fig:Qi_lD1p5_N36B} reveals that, besides the most responsive unit, there are four units $i$ with $Q_i > 0.31$ and another five units $j$ with $Q_j > 0.12$. Representing a single non-Gaussian feature by multiple units at $S = -1.5$ appears to be a non-optimal strategy in terms of energetic cost.

\subsection{Discontinuous phase transitions}

We determine the global minimum energy values $E$ at various fixed entropy values $S$ to get an energy curve $E(S)$. Figure~\ref{fig:S_E_N36} reveals that the minimum energy $E$ is a continuous and monotonic function of entropy in the examined range of $S \in [-6, 9]$. However, the function $E(S)$ is convex only for $S < -1.16$ and $S > 1.63$. In the intermediate range of $S \in (-1.16, 1.63)$, including the point $S=0$, the energy slope $\textrm{d}E/\textrm{d}S$ is discontinuous and nonmonotonic (Fig.~\ref{fig:S_T_N36}) and the overlap $Q(S)$ is discontinuous (Fig.~\ref{fig:S_Q_N36}). The non-convexity of $E(S)$ and the discontinuity of $Q(S)$ indicate qualitative changes of the optimal weight matrix $\bm{W}$ and the occurrence of discontinuous phase transitions.

\begin{figure*}
\centering
   \subfigure[$\ S\, = \, 0$]{
    \includegraphics[angle=0,width=0.325\linewidth]{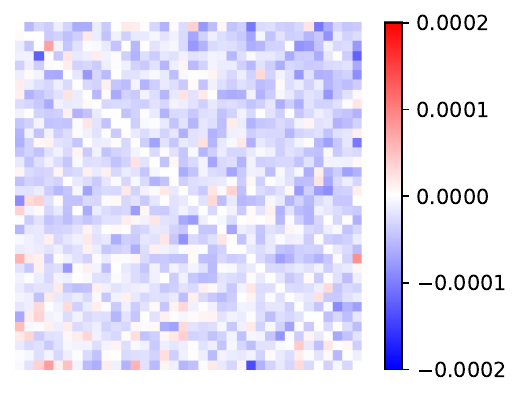}
    \label{fig:WN36_lD0}
  } 
  \subfigure[$\ S\, =\, -2$]{
    \includegraphics[angle=0,width=0.3\linewidth]{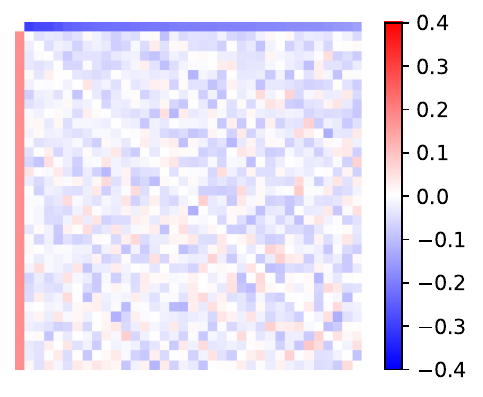}
    \label{fig:WN36_lD2}
  }
  \subfigure[$\ S\, =\, 8$]{
    \includegraphics[angle=0,width=0.3\linewidth]{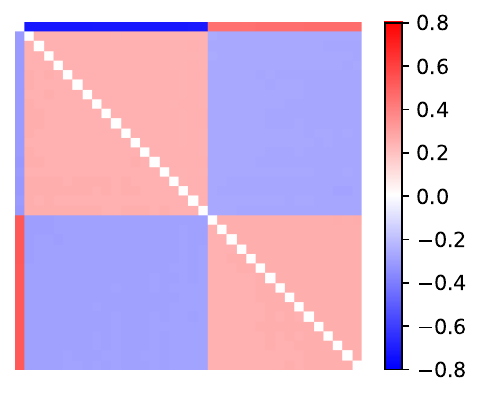}
    \label{fig:WN36_lDm8}
  }
  \caption{
  Example optimal weight matrices for the system of $N=36$ and $p_0=0.7$ with entropy  $S=0$ (a), $S= -2$ (b) and $S=8$ (c), corresponding to the three different phases of Fig.~\ref{fig:ESN36}. 
  }
  \label{fig:WmtxN36}
\end{figure*}

To explicitly visualize energy--information tradeoff, we plot the free energy $F = E - T S$ as a function of $S$ at several fixed temperature values $T$ (Fig.~\ref{fig:S_F_N36}). We find that if $T$ is higher than $1.1283$ the minimum value of $F$ is achieved at a large positive value of $S > 7$ with high overlap $Q > 0.9$. At $T = 1.1283$ two degenerate free energy minima are present, one at $S = 7.10$ with  $Q=0.97$ and energy $E=36.73$ and the other at $S=0$ with $Q=0.1667$ and $E=28.72$, leading to a discontinuous phase transition.  When $T \in (0.8320, 1.1283)$ there is only one minimum $F$ and it is located exactly at  $S=0$. Then at $T = 0.8320$ another global minimum $F$ appears at $S=-1.12$ with $Q=0.83$ and energy $E=27.79$, indicating another discontinuous phase transition. As $T$ further decreases, the minimum free energy is achieved at $S \leq -1.12$ and the overlap $Q$ is high and is slowly increasing as $T$ decreases.

Our results therefore establish that feature detection is feasible (for $p_0=0.7$) at both high and low temperatures but impossible at intermediate temperatures. We draw in Fig.~\ref{fig:ESN36} three optimal weight matrices as representative examples, with different entropy values $S=0$, $-2$ and $8$.

For $T \in (0.8320, 1.1283)$, the optimal matrix with  $S = 0$ is rather weak and homogeneous (all the synaptic weights $w_{i j}$ are of order $10^{-4}$), and the different rows and columns can not be distinguished (Fig.~\ref{fig:WN36_lD0}). There are no significant lateral interactions in this system and naturally it can not perform feature detection.

The optimal matrix at $S = -2$ contains a single unit (index $i_0=1$) which most strongly inhibits all the other units $j$ (with positive weights $w_{j i_0} \approx 0.176$) and is most strongly excited by these units (with negative weights $w_{i_0 j}$ dispersed from $-0.148$ to $-0.282$). The subsystem formed by the other units are itself homogeneous with the weights $w_{i j}$ being much weaker (of order $10^{-2}$) (Fig.~\ref{fig:WN36_lD2}). The mean energy of this system is $E = 27.112$. This example demonstrates that the non-reciprocal excitation--inhibition between a single unit $i_0$ and the remaining homogeneous subsystem helps to reduce the energetic cost of lateral predictive coding. Feature detection at $T < 0.8320$ is a byproduct of this structural organization.

The optimal matrix at $S=8$ is quite different (Fig.~\ref{fig:WN36_lDm8}). Here the input feature $\vec{\bm{\phi}}_1$ is detected by a single unit $i_0=1$, and this unit is strongly excited by a group (say $A$) of $19$ units and is strongly inhibited by the other group (say $B$) of $16$ units. Unit $i_0$ excites group $A$ and inhibits group $B$ in return, demonstrating reciprocal interactions. There are also relatively strong inhibitory (positive) interactions within both groups $A$ and $B$, while these two groups mutually excite each other with relatively strong negative weights. Overall, the interactions of this three-component optimal network are reciprocal, with the weights $w_{i j}$ and $w_{j i}$ between two units $i$ and $j$ being of the same sign. The mean energy of this system is $E = 37.752$. This example demonstrates that the optimal LPC system may form multiple components with both reciprocal excitatory and reciprocal inhibitory interactions to improve information robustness. This structural organization leads to feature detection at $T > 1.1283$.

We have checked that the discontinuous emergence of feature detection function will also be observed if the feature direction  $\vec{\bm{\phi}}_1$ is a random unit vector~\cite{SI2025a}. When the $p_0$ value of Eq.~(\ref{eqPa1}) decreases, $q(a_1)$ becomes less deviated from Gaussian. As a result, we find that the entropy value $S$ needs to be more negatively or more positively deviated from zero to achieve the feature detection function. We have constructed a phase diagram with $p_0$ and temperature $T$ as control parameters for systems with smaller size $N=10$ (see Fig.~S1 of Ref.~\cite{SI2025a}). When we change the feature distribution $q(a_1)$ to be the continuous Laplace distribution or the discretized power-law distribution as mentioned in the end of Sec.~\ref{sec:setting}, the numerical results are qualitatively similar to the results reported here (see sections S5 and S6 of Ref.~\cite{SI2025a}). These additional simulation results confirm that the discontinuous emergence of feature detection capability is a general property of our LPC model.

\subsection{Spectrum analysis}
\label{subsec:spectrum}

To gain further insight into the optimal LPC matrices $\bm{W}$, we now study the spectrum property of the sampled matrices $(\bm{I}+\bm{W})$. We increase the system size to $N=100$ to have more eigenvalues, fixing $\vec{\bm{\phi}}_1 \propto (1, \ldots, 1)^\top$ and $p_0=0.7$ as before. Another motivation for considering a much larger system size $N$ is to see its effect on the feature detection function.

\begin{figure*}
  \centering
  \subfigure[ $\ S\, =\, -10$]{
    \includegraphics[angle=270,width=0.3\linewidth]{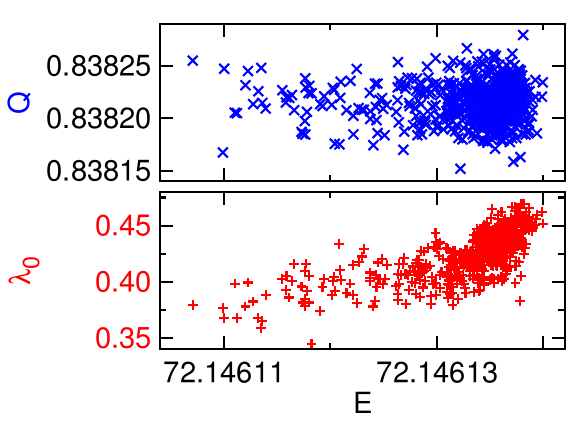}
    \label{fig:EL0QlD10N100}
  }
  \subfigure[ $\ S\, =\, -15$]{
    \includegraphics[angle=270,width=0.3\linewidth]{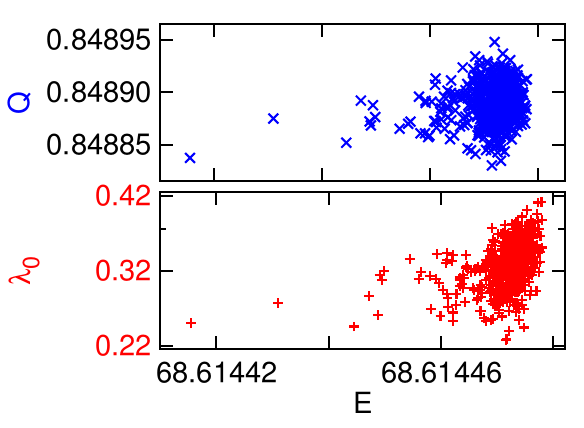}
    \label{fig:EL0QlD15N100}
  }
  \\
  \subfigure[]{
    \includegraphics[angle=270,width=0.3\linewidth]{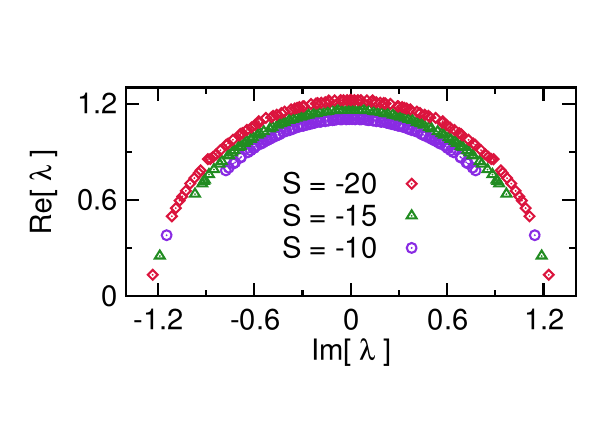}
    \label{fig:LambdaRI}
  }
  \subfigure[]{
    \includegraphics[angle=270,width=0.3\linewidth]{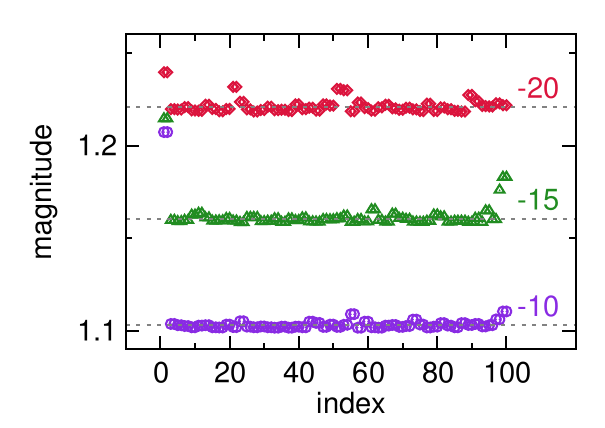}
    \label{fig:LambdaRadius}
  }
  \caption{
    Spectrum analysis for system size $N=100$. (a-b) The overlap $Q$ and minimum eigenvalue real part $\lambda_0$ of $600$ independently sampled minimal-energy matrices $\bm{I}+\bm{W}$ versus the corresponding minimal energy $E$, at fixed entropy $S=-10$ (a) and $-15$ (b). (c) The real and imaginary parts of all the complex eigenvalues for three single example matrices with fixed entropy $S = -10$ (circles), $-15$ (triangles), and $-20$ (diamonds). All the eigenvalues are located approximately on a semicircle at each fixed $S$. (d) The magnitudes $\sqrt{|\lambda|^2}$ of all the eigenvalues. The horizontal dotted lines mark the mean magnitudes averaged over all the eigenvalues except for the first two with minimum real part $\lambda_0$.
  }
  \label{fig:N100LambdaProfile}
\end{figure*}

As there is only one non-Gaussian feature direction and all the other $(N-1)$ dimensions are Gaussian inputs, the input signal $s_i$ to each unit $i$ becomes more and more closer to Gaussian as $N$ increases. Consistent with this fact, we find that the onset of feature detection for the system of size $N=100$ is shifted to entropy values $S$ being even further deviated away from $S=0$. At $S = -5$, for example, all the $600$ sampled LPC matrices by our annealing algorithm have modest overlap values $Q \approx 0.39$ (feature detection is largely failed); at $S=-9$, among the $600$ independently sampled LPC matrices, we find that only three of them have the global minimum energy $E \approx 72.8740$ and high overlap $Q \approx 0.8360$, while all the other $597$ matrices are local optimal ones with energy $E \approx 72.9183$ and $Q\approx 0.58$. On the other hand, when $S \leq -10$, we find that all the $600$ sampled LPC matrices have very similar energy values and very high overlap values $Q \geq 0.838$. For example, at $S=-10$ the energy values are $E\approx 72.1461$ and $Q\approx 0.8382$ (Fig.~\ref{fig:EL0QlD10N100}, and at $S=-15$ the energy values are close to $68.6144$ and $Q\approx 0.8449$ (Fig.~\ref{fig:EL0QlD15N100}). 

The real parts of all the eigenvalues of the matrix $(\bm{I}+\bm{W})$ need to be positive to guarantee the convergence of Eq.~(\ref{eq:lpc}). We denote by $\lambda_0$ the minimum real part of all the eigenvalues of $(\bm{I}+\bm{W})$. We find that, when the entropy $S$ is not too much deviated from zero, the condition $\lambda_0 > 0$ is automatically satisfied without the need of explicitly imposing this constraint in our annealing algorithm.  As two concrete examples, we show the two sets of $600$ $\lambda_0$ values obtained at $S=-10$ and $S=-15$ in Fig.~\ref{fig:EL0QlD10N100} and \ref{fig:EL0QlD15N100}, respectively. At each value of $S$, there is a weak trend of $\lambda_0$ increasing with mean $L_1$-norm energy $E$. The mean value of $\lambda_0$ decreases as $S$ becomes more negative. For example, $\lambda_0 = 0.42 \pm 0.02$ (mean and standard deviation) at $S=-10$ and $\lambda_0 = 0.33 \pm 0.03$ at $S=-15$.  The minimum value $\lambda_0$ approaches zero at $S \approx 24$.  This means that, when the entropy is fixed to a value more negative than $-24$, we will have to impose the constraint of $\lambda_0 > 0$ explicitly in our matrix annealing algorithm. In other words, at sufficiently negative values of $S$, the optimal LPC matrices are located at the edge of chaos with $\lambda_0 \approx 0^+$ (slightly above zero), which has also been observed in our earlier work \cite{Huang-etal-2023}. Notice that when $\lambda_0$ becomes smaller, the response dynamics (\ref{eq:lpc}) will take more time to converge and therefore the system will be more slower in catching the input features. This speed of response is functionally relevant~\cite{Lan-etal-2012,Nicoletti-Busiello-2024,Olsen-etal-2024}, and an extension of the present work is to consider the tradeoff between response speed (measured by $\lambda_0$) and the free energy $F$ at fixed temperature $T$.  We will study these interesting issues of criticality and speed tradeoff~\cite{Sompolinsky-etal-1988,Qiu-Huang-2024,Calvo-etal-2024,Safavi-etal-2024,Barzon-etal-2025} in more detail in a separate paper.

As the weight matrix is not symmetric, the eigenvalues of $(\bm{I}+\bm{W})$ are complex. To illustrate the distribution of these complex eigenvalues, we plot all the eigenvalues for three optimal systems with different entropy values $S = -10$, $-15$ and $-20$ in Fig.~\ref{fig:LambdaRI}. We observe that, as the magnitude of the imaginary part $\textrm{Im}[ \lambda]$ of an eigenvalue $\lambda$ increases, its real part $\textrm{Re}[ \lambda]$ decreases. Most of the eigenvalues appear to be sitting on a semicircle. This semicircle property is demonstrated more clearly in Fig.~\ref{fig:LambdaRadius}, which reveals that the magnitudes $\sqrt{ |\lambda^2|}$ of all the $N$ eigenvalues are approximately equal, except for the pair of eigenvalues with the minimum real part $\lambda_0$. It is well known that the complex eigenvalues of a purely random matrix are distributed uniformly within the whole area of a circle. These optimal LPC matrices are therefore qualitatively distinct from purely random matrices. They are the results of optimization: when the entropy $S$ is fixed, energy minimization will push the complex eigenvalues of the optimal LPC system onto a semicircle as much as possible (see Appendix~\ref{app:semicircle} for an analytical explanation).

\section{Detection and separation of two non-Gaussian features}

The visual and auditory sensory perception systems of the biological brain are capable of detecting multiple features and distinguishing between them~\cite{Jutten-Herault-1991,Hyvarinen-Oja-2000,Wang-Lu-2019}. To help appreciate these important information processing functions, in this section we investigate whether the feature detection and separation function can emerge spontaneously in our simple linear LPC model. For simplicity of theoretical analysis and numerical computations, here we assume that there are only two non-Gaussian features. The input signal vector $\vec{\bm{s}}$ are generated according to
\begin{equation}
\begin{aligned}
     \vec{\bm{s}} \ = &  \ a_1 \, \bigl(\cos(\theta /2)\vec{\bm{\phi}}_1  + \sin(\theta /2) \vec{\bm{\phi}}_2\bigr) \, + \\
     & \  a_2\,  \bigl(\cos(\theta /2)\vec{\bm{\phi}}_1  - \sin(\theta /2) \vec{\bm{\phi}}_2\bigr) 
      \ + \ 
     \sum\limits_{k=3}^N b_k \vec{\bm{\phi}}_k \; ,
     \end{aligned}
     \label{eq2finput}
\end{equation}
where $\vec{\bm{\phi}}_i$ are again $N$ orthogonal unit vectors as in Eq.~(\ref{eqsform}), and $b_k$ are $(N-2)$ independent Gaussian random variables with zero mean and unit variance. The coefficients $a_1$ and $a_2$ are two non-Gaussian random variables with zero mean and unit variance, and the associated two non-Gaussian feature directions are denoted as
\begin{equation}
    \begin{aligned}
      \hat{\bm{\phi}}_1 \ & \equiv \ 
      \cos(\theta /2 ) \vec{\bm{\phi}}_1 + 
      \sin(\theta /2 ) \vec{\bm{\phi}}_2 \; , \\
      \hat{\bm{\phi}}_2 \ & \equiv \ 
      \cos(\theta /2 ) \vec{\bm{\phi}}_1 - 
      \sin(\theta /2 ) \vec{\bm{\phi}}_2 \; ,
      \end{aligned}
      \label{eq:hatphi12}
\end{equation}
with $\theta \in (0, \pi /2]$ being the angle between $\hat{\bm{\phi}}_1$ and $\hat{\bm{\phi}}_2$. If $\theta = \pi / 2$ then $\hat{\bm{\phi}}_1$ and $\hat{\bm{\phi}}_2$ are orthogonal, and otherwise they are partially aligned with each other.

At fixed values of $a_1$ and $a_2$, the steady-state output vector $\vec{\bm{x}}$ of the LPC system follows a Gaussian distribution. The variance $\sigma_i^2$ of each individual output signal $x_i$  is
\begin{equation}
    \sigma_i^2 \, = \,  \sum\limits_{j=1}^N \Bigl[ \frac{\bm{I}}{\bm{I}+\bm{W}}\Bigr]_{i j}^2 - \Bigl[ \frac{\bm{I}}{\bm{I}+\bm{W}} \vec{\bm{\phi}}_1 \Bigr]_i^2
    - \Bigl[ \frac{\bm{I}}{\bm{I}+\bm{W}} \vec{\bm{\phi}}_2 \Bigr]_i^2 \; ,
\end{equation}
which is actually independent of $a_1$ and $a_2$. On the other hand, the mean value of $x_i$ is linear in $a_1$ and $a_2$:
\begin{equation}
\bigl\langle x_i\bigr\rangle \, = \, 
a_1 \hat{\mu}_i^{(1)} + a_2 \hat{\mu}_i^{(2)} \; .
\label{eqc030525}
\end{equation}
Here $\hat{\mu}_i^{(1)}$ and $\hat{\mu}_i^{(2)}$ are, respectively, the $i$-th element of the $N$-dimensional output projection vectors $\hat{\bm{\mu}}^{(1)}$ and $\hat{\bm{\mu}}^{(2)}$ of the feature directions $\hat{\bm{\phi}}_1$ and $\hat{\bm{\phi}}_2$,
\begin{equation}
    \begin{aligned}
    \hat{\bm{\mu}}^{(1)} \, & = \, \cos(\theta/2) \frac{\bm{I}}{\bm{I}+\bm{W}} \, \vec{\bm{\phi}}_1
  + \sin(\theta/2) \frac{\bm{I}}{\bm{I}+\bm{W}} \, \vec{\bm{\phi}}_2 \; ,
\\
  \hat{\bm{\mu}}^{(2)} \, & = \, \cos(\theta/2) \frac{\bm{I}}{\bm{I}+\bm{W}} \, \vec{\bm{\phi}}_1
  - \sin(\theta/2) \frac{\bm{I}}{\bm{I}+\bm{W}} \, \vec{\bm{\phi}}_2 \; .
    \end{aligned}
    \label{eqd030525}
\end{equation}
If a single element $\hat{\mu}_i^{(1)}$ of the projection vector $\hat{\bm{\mu}}^{(1)}$ is much larger in magnitude in comparison with all the other elements, it means that the corresponding unit $i$ is mainly responding to the non-Gaussian feature $\hat{\bm{\phi}}_1$. To quantify the feature detection performance of the LPC system, similar to Eq.~(\ref{eq:Qoverlap}), we can define two order parameters $Q^{(1)}$ and $Q^{(2)}$ as
\begin{equation}
    \begin{aligned}
      Q^{(1)} \, & = \, \max\limits_{i} \biggl[ \frac{|\hat{\mu}_i^{(1)}|}{\sqrt{ \sum_{j=1}^N (\hat{\mu}_j^{(1)})^2}}\biggr] \; , \\
    Q^{(2)} \, & = \, \max\limits_{i} \biggl[ \frac{|\hat{\mu}_i^{(2)}|}{\sqrt{ \sum_{j=1}^N (\hat{\mu}_j^{(2)})^2}}\biggr] \; . 
    \end{aligned}
\end{equation}
\begin{figure*}
  \centering
  \subfigure[]{
    \includegraphics[angle=270,width=0.3\linewidth]{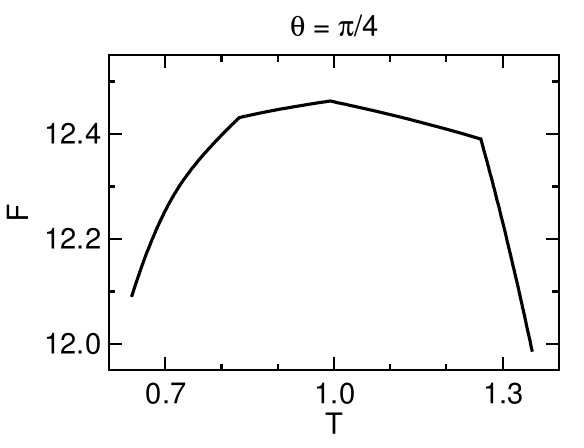}
    \label{fig:TF45N16TvsF}
  }
  \subfigure[]{
    \includegraphics[angle=270,width=0.3\linewidth]{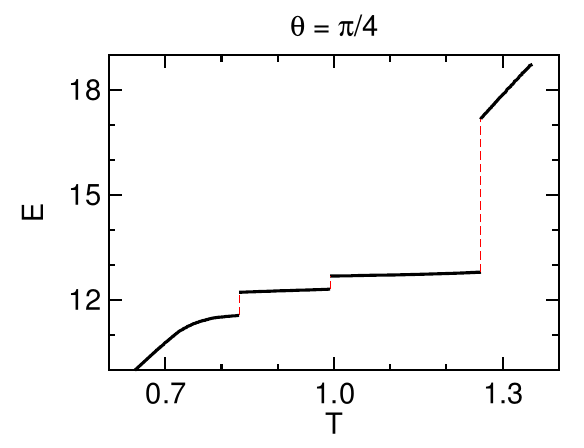}
    \label{fig:TF45N16TvsE}
  }
  \\
  \subfigure[]{
    \includegraphics[angle=270,width=0.3\linewidth]{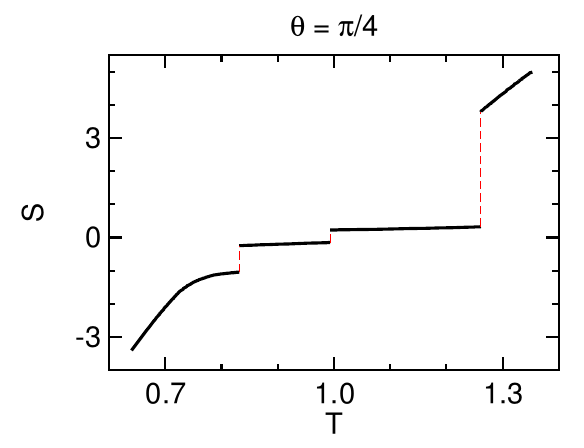}
    \label{fig:TF45N16TvsS}
  }
  \subfigure[]{
    \includegraphics[angle=270,width=0.3\linewidth]{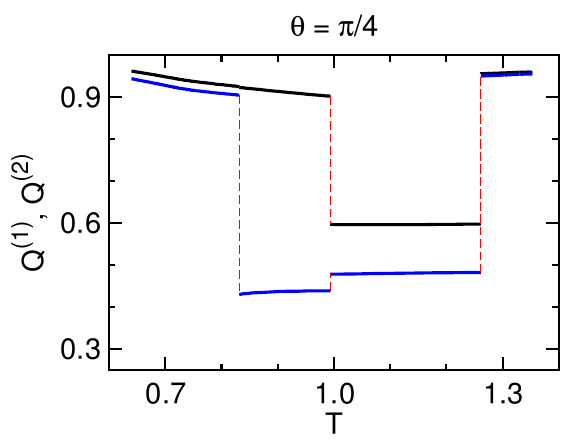}
    \label{fig:TF45N16TvsQ}
  }
  \caption{
 Thermodynamic quantities of optimal lateral predictive coding for input vectors (\ref{eq2finput}) containing two non-orthogonal random features (\ref{eq:hatphi12}) with angle $\theta = \pi/4$. The independent coefficients $a_1$ and $a_2$ follow the non-Gaussian distribution (\ref{eqPa1}) with parameter $p_0=0.6$. Network size $N=16$. We use temperature $T$ as the control parameter.  (a) Minimum free energy $F$. (b) Mean energy $E$. (c) Entropy $S$. (d) Order parameters $Q^{(1)}$ and $Q^{(2)}$. The vertical dashed lines at $T = 0.8316$, $0.9935$ and $1.2612$ mark the three discontinuous phase transitions.
  }
  \label{fig:TF45N16T}
\end{figure*}

In our computer simulations, we prepare two non-Gaussian feature directions $\hat{\bm{\phi}}_1$ and $\hat{\bm{\phi}}_2$ according to Eq.~(\ref{eq:hatphi12}), with the two orthogonal base vectors $\vec{\bm{\phi}}_1$ and $\vec{\bm{\phi}}_2$ being randomly picked from the $N$-dimensional unit sphere. We have tested several different angles $\theta \in \{\pi / 4, \pi / 3, \pi /2\}$, and the numerical results are qualitatively very similar. Here we show the representative results obtained on a system with $N=16$ units at $\theta = \pi /4$ (some additional results are described in the supplementary document~\cite{SI2025a}). The non-Gaussian coefficients $a_1$ and $a_2$ are distributed according to Eq.~(\ref{eqPa1}) with parameter $p_0 = 0.6$.

We perform energy $E$ minimization at many different fixed values $S$ of the entropy by stochastic search in the space of weight matrices $\bm{W}$ (see Sec.~\ref{subsec:algorithm} for technical details). Based on this large set of $(E, S)$ points, we then determine the optimal weight matrix at each fixed value of the temperature control parameter $T$ by locating the minimum value of the free energy $F = E - T S$.

Figure~\ref{fig:TF45N16TvsF} reveals that the continuous free energy function $F(T)$ is kinked at several critical temperature values $T = 0.8316$, $0.9935$, and  $1.2612$. These kinks are caused by sudden big changes in the optimal weight matrix $\bm{W}$. The mean energy $E$ and the entropy $S$ are discontinuous at these phase transition points  (Fig.~\ref{fig:TF45N16TvsE} and ~\ref{fig:TF45N16TvsS}). The discontinuous changes of the order parameters $Q^{(1)}$ and $Q^{(2)}$ (Fig.~\ref{fig:TF45N16TvsQ}) indicate that the optimal LPC system can successfully detect the two  feature directions $\hat{\bm{\phi}}_1$ and $\hat{\bm{\phi}}_2$ if the temperature is sufficiently low ($T < 0.8316$) or sufficiently high ($T > 1.2612$). The optimal weight matrix in the temperature range  $T \in (0.8316, 0.9935)$ successfully detects one feature direction but fails with the other one, and if $T \in (0.9935, 1.2612)$ the optimal weight matrix is unable to detect both feature directions.

\begin{figure*}[t]
\centering
   \subfigure[$\ S\, = \, -2$]{
    \includegraphics[angle=0,width=0.22\linewidth]{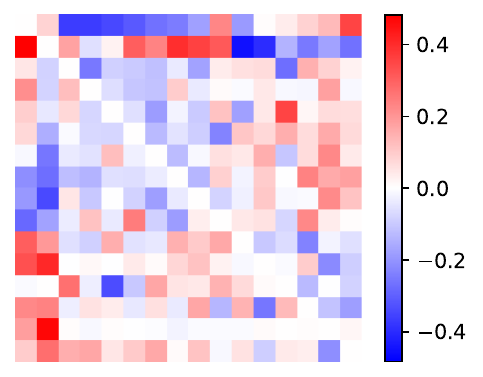}
    \label{fig:WTF45N16lD2}
  } 
  \subfigure[$\ S\, =\, -0.2$]{
    \includegraphics[angle=0,width=0.22\linewidth]{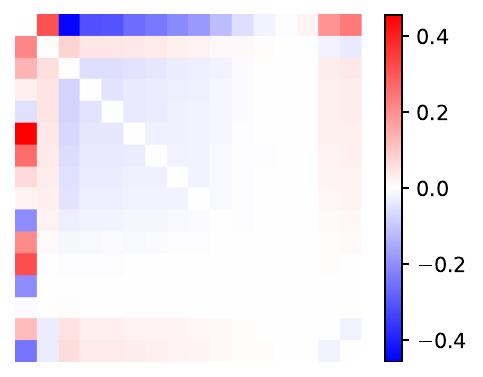}
    \label{fig:WTF45N16lD0p2}
  }
  \subfigure[$\ S\, =\, 0.3$]{
    \includegraphics[angle=0,width=0.22\linewidth]{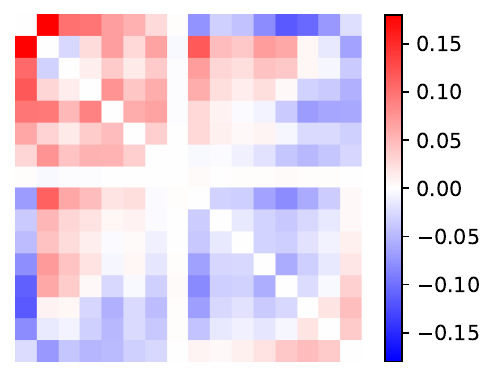}
    \label{fig:WTF45N16lDm0p3}
  } 
  \subfigure[$\ S\, =\, 4$]{
    \includegraphics[angle=0,width=0.22\linewidth]{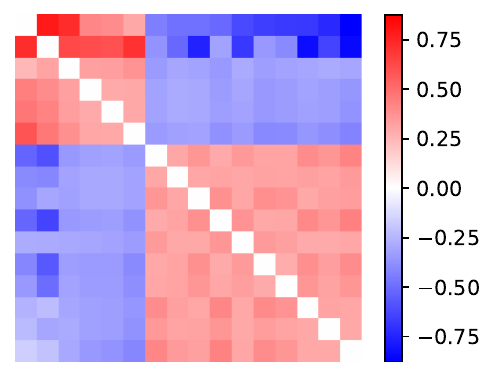}
    \label{fig:WTF45N16lDm4}
  }
  \caption{
Four representative optimal LPC matrices of Fig.~\ref{fig:TF45N16T}, with $S = -2.0$ at $T = 0.7053$ (a), $S = -0.2$ at $T = 0.9020$ (b), $S = 0.3$ at $T = 1.2110$ (c), and $S = 4.0$ at $T = 1.2753$ (d). These examples correspond to the four different phases of Fig.~\ref{fig:TF45N16T}. 
  }
  \label{fig:WTF45N16}
\end{figure*}

\begin{figure*}
  \centering
  \subfigure[$\ S\, = \, -2$]{
    \includegraphics[angle=270,width=0.23\linewidth]{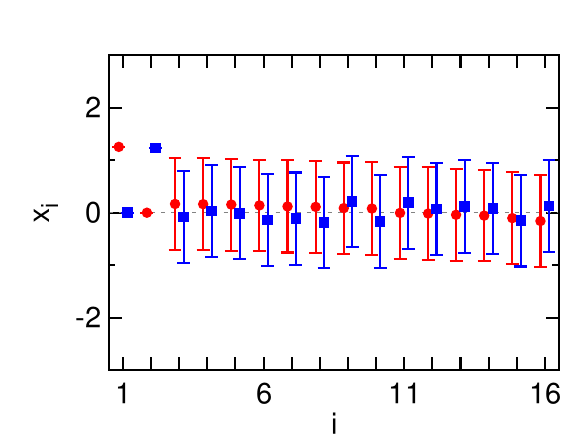}
    \label{fig:TF45N16lD2}
  }
  \subfigure[$\ S\, = \, -0.2$]{
    \includegraphics[angle=270,width=0.23\linewidth]{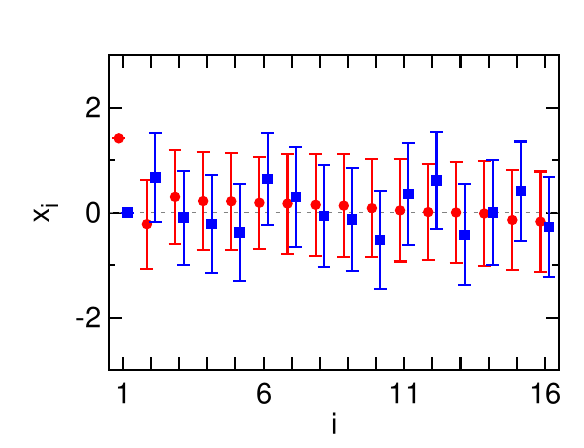}
    \label{fig:TF45N16lD0p2}
  }
  \subfigure[$\ S\, = \, 0.3$]{
    \includegraphics[angle=270,width=0.23\linewidth]{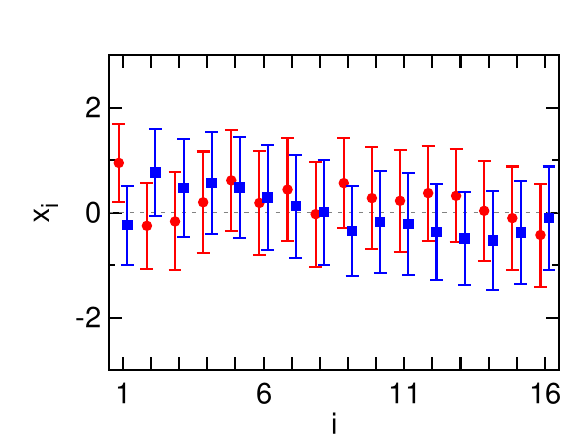}
    \label{fig:TF45N16lDm0p3}
  }
  \subfigure[$\ S\, = \, 4$]{
    \includegraphics[angle=270,width=0.23\linewidth]{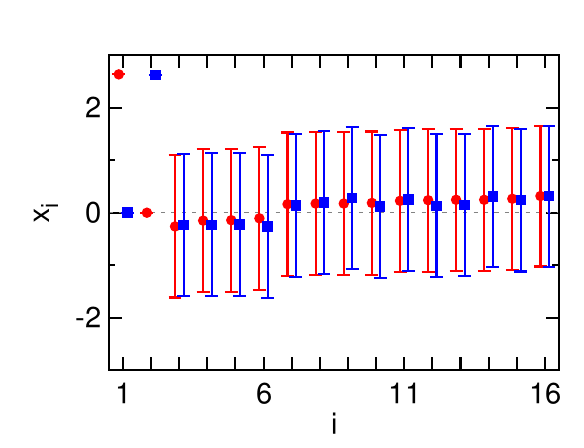}
    \label{fig:TF45N16lDm4}
  }
  \caption{
  The output signals $x_i$ of individual units $i$ produced by the four optimal LPC networks of Fig.~\ref{fig:WTF45N16}, with $S = -2.0$ (a), $S= - 0.2$ (b), $S = 0.3$ (c), and  $S=4.0$ (d). The output signal $x_i$ follows a Gaussian distribution for fixed values $a_1$ and $a_2$ of the input vectors (\ref{eq2finput}).  The means and standard deviations of the $N=16$ outputs $x_i$ are shown for $a_1 = 1/\sqrt{1-p_0} = 1.5811$ and $a_2=0$ (filled red circles) and for $a_1=0$ and $a_2 = 1.5811$ (filled blue squares). The two data points for each unit index $i$ are slightly displaced along the horizontal axis for better illustration. The dashed horizontal lines denote $x=0$.
    }
\label{fig:TF45N16S}
\end{figure*}

We choose four optimal weight matrices $\bm{W}$, one for each of the four phases revealed by Fig.~\ref{fig:TF45N16T}, for more detailed examination. These four different optimal weight matrices are shown in Fig.~\ref{fig:WTF45N16}, and we plot in Fig.~\ref{fig:TF45N16S} the statistical properties of individual output signals $x_i$. 

For the matrix with entropy $S= -2$ and mean energy $E = 10.8539$ at temperature $T = 0.7053$ (Figs.~\ref{fig:WTF45N16lD2} and ~\ref{fig:TF45N16lD2}), we find that there is one single unit (its index is assigned to be $i=1$) which is responding strongly if and only if the feature $\hat{\bm{\phi}}_1$ is present ($a_1 \neq 0$), and there is another different unit (with index $i=2$) which is selectively responding only to the feature $\hat{\bm{\phi}}_2$ ($a_2 \neq 0$) and is completely silent in all the other cases. The output states of all the remaining $(N-2)$ units only depend slightly on the values of $a_1$ and $a_2$ and they are fluctuating considerably, and therefore each of them does not contain much information about the presence or absence of the  non-Gaussian features $\hat{\bm{\phi}}_1$ and $\hat{\bm{\phi}}_2$. There are quite strong synaptic interactions between the subset of two selectively responsive units $1$ and $2$ and the subset of the remaining $(N-2)$ units (Fig.~\ref{fig:WTF45N16lD2}).


Similar output statistical properties are observed on the optimal network of entropy $S = 4$ and mean energy $E = 17.4366$ at $T = 1.2753$ (Fig.~\ref{fig:TF45N16lDm4}), but with the two selective units responding much more strongly in the presence of the corresponding features and the $(N-2)$ non-selective units having much larger output magnitudes and fluctuations. The optimal network of $S=4$ has very clear community structure and is relative symmetric (Fig.~\ref{fig:WTF45N16lDm4}): The two selectively responsive units $1$ and $2$ mutually inhibit each other and they are strongly excited by one group ($A$) of ten units and strongly inhibited by the other group ($B$) of four units; there are strong internal inhibitory interactions within group $A$ and group $B$ but these two groups mutually excite each other.

It is interesting to notice that, even if the two feature directions $\hat{\bm{\phi}}_1$ and $\hat{\bm{\phi}}_2$ are partially aligned (with $\theta \neq \pi /2$), each of the two selective units $1$ and $2$ is sensitive only to one of them and is non-responsive to the other. This essentially means that, for any input vector $\vec{\bm{s}}$, its projection in the subspace expanded by $\vec{\bm{\phi}}_1$ and $\vec{\bm{\phi}}_2$ will be decomposed into two non-orthogonal components  $\hat{\bm{\phi}}_1$ and $\hat{\bm{\phi}}_2$. The results of Fig.~\ref{fig:TF45N16lD2} and ~\ref{fig:TF45N16lDm4} demonstrate that, this function of  non-orthogonal feature extraction and separation (referred to as independent component decomposition~~\cite{Jutten-Herault-1991,Hyvarinen-Oja-2000}) is possible both for optimal LPC systems with relatively low energetic cost $E$ and for those optimal systems with relatively high entropy $S$.

When the temperature $T \in (0.8316, 0.9935)$ the optimal LPC network can only detect one of the non-Gaussian features. For example, the network with entropy $S = - 0.2$ and mean energy $E=12.2663$ at $T=0.9020$ detects the presence of the feature direction $\hat{\bm{\phi}}_1$ by the strong response of a single unit with index $i=1$ (Fig.~\ref{fig:TF45N16lD0p2}). This unit is completely silent when $\hat{\bm{\phi}}_1$ is absent, even if the other non-orthogonal feature direction $\hat{\bm{\phi}}_2$ is present. The system does not achieve a strong and localized response to the presence of the second feature $\hat{\bm{\phi}}_2$. The relatively strong synaptic interactions between the selectively responsive unit $i$ and the other $(N-1)$ units are not reciprocal: unit $i$ prefers to inhibit the remaining $(N-1)$ units and it is mainly excited in return (Fig.~\ref{fig:WTF45N16lD0p2}).

When the temperature $T \in (0.9935, 1.2612)$ the optimal LPC network performs even worse and it fails to detect either of the non-Gaussian features, see Fig.~\ref{fig:TF45N16lDm0p3} for the results obtained on the optimal network at $T=1.2110$, whose entropy $S = 0.3$ and mean energy $E=12.7693$. Here we see that all the $N$ units have relatively large fluctuations in their output states no matter whether the features $\hat{\bm{\phi}}_1$ and $\hat{\bm{\phi}}_2$ are present or absent. The corresponding weight matrix is largely symmetric (Fig.~\ref{fig:WTF45N16lDm0p3}).

\section{Discussion}

Phase transitions were recently discovered in deep neural networks (see, e.g., Refs.~\cite{Yoshino-2020,Wu-Fischer-2020}) and in lateral predictive coding with quadratic energetic cost~\cite{Huang-etal-2023}. Adding to this literature, our theoretical results demonstrated that the tradeoff between energetic cost and information robustness can drive the discontinuous emergence of feature detection function in the single-layered lateral predictive coding system. This work helps us appreciate an important biological function of LPC more deeply, and it resonates with the opinions of Refs.~\cite{Bialek-2024,Safavi-etal-2024,Sokolowski-etal-2025,Tatsukawa-Teramae-2025} that the optimization principle is a key to understand biological complexity. In the future one may consider the issue of multiple (more than two) non-Gaussian input feature signals and explore the capacity of the linear LPC system to perform independent component decomposition~\cite{Jutten-Herault-1991,Hyvarinen-Oja-2000}. 

In our present problem setting, as the non-Gaussian features and the Gaussian background noises have the same second moment, $\vec{\bm{\phi}}_1$ can not be detected if energy is the mean $L_2$-norm~\cite{Huang-etal-2023}. The $L_1$-norm property of the energy (\ref{eqL1E}) seems essential for the spontaneous emergence of feature detection function. An important point indicated by the results of Fig.~\ref{fig:E_Psi_N36} and Fig.~\ref{fig:Qiprofile} is that, at a given fixed level $S$ of information robustness, there are qualitatively distinct types of LPC matrices $\bm{W}$, and the minimization of the $L_1$-norm energetic cost helps to break their degeneracy in the feature detection task. Our work confirms that, besides the conventional strategy of energy minimization, information robustness (entropy $S$) maximization can also drive the emergence of feature detection function. We can combine the energetic and entropic effects by the free energy $F= E - T S$. Free energy minimization at sufficiently low and sufficiently high temperature values $T$ will stabilize optimal LPC matrices that are highly energy efficient and that are highly robust in information transmission, respectively. The biological brain is highly adaptable, and real-world LPC neural networks may have a great degree of diversity at a given level of information robustness to enable adaptation to different types of tradeoff demands. One natural extension of our work is to investigate the effect of the tradeoff between response speed and free energy as briefly mentioned in Sec.~\ref{subsec:spectrum}.

Another direction is to add memory effect or nonlinearity to the recursive dynamics (\ref{eq:lpc})~\cite{SI2025a}. For example, we may replace the linear effect $x_j$ of unit $j$ to unit $i$ by a nonlinear function $f(x_j)$ such as $ \tanh(x_j)$ or the rectified linear function $\max( 0, x_j)$. For such nonlinear LPC systems, Eq.~(\ref{eq:xss}) is no longer valid and the theoretical analysis will be more challenging.

In the present work, the optimal LPC matrix was achieved by a numerical optimization algorithm rather than through learning from samples of input signals. It is a future task to study more thoroughly the evolution dynamics of $\bm{W}$ under localized Hebbian learning rules~\cite{Huang-etal-2022,Tang-etal-2023b}. We expect that, because of the existence of discontinuous phase transitions, the adaptation of the weight matrix $\bm{W}$ will be a slow and discontinuous process. It is stimulating to notice that empirical evidence in the literature has indicated that, learning to recognize complex patterns or rules is indeed often a long and slow process with sudden huge elevation in performance~\cite{Hosenfeld-etal-1997,Boshuizen-2004,Collins-etal-2019,Rosenberg-etal-2021}.

As the entropy measure $S$ deviates more negatively away from the region of $S \approx 0$, the minimum value $\lambda_0$ of the real parts of eigenvalues of $(\bm{I}+\bm{W})$ gradually decreases and then stays at the lower-bound value $\lambda_0 \approx 0^+$. A concrete example of this decreasing trend, obtained for system size $N=100$, is shown in Fig.~\ref{fig:N100LambdaProfile}.  Weight matrices with vanishing $\lambda_0$ are said to be located at the edge of chaos~\cite{Sompolinsky-etal-1988,Qiu-Huang-2024,Calvo-etal-2024,Safavi-etal-2024,Barzon-etal-2025}. It is very interesting to study the dynamical properties of such critical optimal LPC networks. It may be important to consider heterogeneity in the parameters of single neural units (such as the time constant $\tau_0$)  to better account for the dynamical property of LPC systems~\cite{Wu-etal-2025}.

\begin{acknowledgments}
  The following funding supports are acknowledged: National Natural Science Foundation of China Grants No.~12247104 and No.~12447101. Numerical simulations were carried out at the HPC cluster of ITP-CAS and also at the BSCC-A3 platform of the National Supercomputer Center in Beijing.
\end{acknowledgments}

\begin{appendix}
  
\section{Entropy of the output signal}
\label{app:entropy}

Given the probability distribution $p_{\textrm{in}}(\vec{\bm{s}})$ of the input signal $\vec{\bm{s}}$, The marginal probability distribution $p_{\textrm{out}}(\vec{\bm{x}})$ of the output signal $\vec{\bm{x}}$ is
\begin{equation}
  p_{\textrm{out}}(\vec{\bm{x}}) \, =  \, \int \textrm{d} \vec{\bm{s}} \,
  p_{\textrm{in}}(\vec{\bm{s}}) \,
  \delta\Bigl(\vec{\bm{x}} - (\bm{I}+\bm{W})^{-1} \vec{\bm{s}} \Bigr) \; ,
  \label{eq:Qxdef}
\end{equation}
where $\delta(\bm{x})$ denotes the Dirac delta function, which is $\delta(\vec{\bm{x}}) \equiv \prod_{i=1}^{N} \delta( x_i)$ for a real vector $\vec{\bm{x}} = (x_1, \ldots, x_N)^\top$. From the definition (\ref{eq:Qxdef}) we obtain that
\begin{equation}
  p_{\textrm{out}}(\vec{\bm{x}}) \, = \, \textrm{det} ( \bm{I} + \bm{W}) \, p_{\textrm{in}}\Bigl( (\bm{I} + \bm{W} )\, \vec{\bm{x}} \Bigr) \; .    
\end{equation}

The entropy of the output signals $\vec{\bm{x}}$ is then
\begin{equation}
  \begin{aligned}
    &  H\bigl[ p_{\textrm{out}}(\vec{\bm{x}}) \bigr] \, \equiv \,
    - \int \textrm{d} \vec{\bm{x}} \,  p_{\textrm{out}}(\vec{\bm{x}})
    \ln p_{\textrm{out}}(\vec{\bm{x}} ) \\
    & \quad \, = \,
    -   \ln \Bigl[  \textrm{det}( \bm{I} + \bm{W}) \Bigr]
    -  \int \textrm{d} \vec{\bm{s}}\, p_{\textrm{in}}( \vec{\bm{s}} )
    \ln p_{\textrm{in}}( \vec{\bm{s}} ) \\
    & \quad \, = \,  - \ln \Bigl[ \textrm{det}( \bm{I} + \bm{W}) \Bigr]
    + H\bigl[ p_{\textrm{in}}(\vec{\bm{s}}) \bigr] \; ,
  \end{aligned}
  \label{eqHx}
\end{equation}
where $H\bigl[ p_{\textrm{in}}(\vec{\bm{s}}) \bigr]$ is the entropy of the input signals $\vec{\bm{s}}$. Since $H\bigl[ p_{\textrm{in}}(\vec{\bm{s}}) \bigr]$ is a constant independent of the weight matrix $\bm{W}$, the entropy difference $H\bigl[ p_{\textrm{out}}(\vec{\bm{x}}) \bigr] - H\bigl[ p_{\textrm{in}}(\vec{\bm{s}} ) \bigr]$  is referred to simply as the entropy of the output distribution $p_{\textrm{out}}(\vec{\bm{x}})$ and is denoted as $S$. From Eq.~(\ref{eqHx}) we obtain the explicit expression for $S$, which is Eq.~(\ref{entropy}).

\section{Information robustness}
\label{app:IR}

We now argue that the entropy $S$ as defined by Eq.~(\ref{entropy}) can serve as a robustness measure of information transmission.

Consider an additive noise vector $\vec{\bm{\epsilon}} = (\epsilon_1, \ldots, \epsilon_{N})^\top$ in the output $\vec{\bm{x}}$ for the input $\vec{\bm{s}}$, so
\begin{equation}
  \vec{\bm{x}} \, = \, (\bm{I} + \bm{W})^{-1}\, \vec{\bm{s}}
  + \vec{\bm{\epsilon}} \; .
  \label{eq250421a}
\end{equation}
All the elements $\epsilon_i$ are independent Gaussian random variables with zero mean and variance $\sigma_0^2$. Given an input signal $\vec{\bm{s}}$, the conditional distribution of the output signal $\vec{\bm{x}}$ is then
\begin{equation}
    p_{\textrm{out}}( \vec{\bm{x}} | \vec{\bm{s}} ) \, = \, 
    \frac{1}{(2 \pi \sigma_0^2)^{N/2}} \exp\Bigl[ - \frac{\bigl(\vec{\bm{x}}-
        (\bm{I}+\bm{W})^{-1} \vec{\bm{s}}\bigr)^2}{2 \sigma_0^2} \Bigr] \; .
  \label{eqpoutc}
\end{equation}
The mutual information between output $\vec{\bm{x}}$ and input $\vec{\bm{s}}$ is given by
\begin{equation}
    I\bigl[\vec{\bm{x}} ;  \vec{\bm{s}} \bigr]  \, \equiv  
    \, H\bigl[ p_{\textrm{out}}(\vec{\bm{x}}) \bigr]
    - H\bigl[ \vec{\bm{x}}  |  \vec{\bm{s}} \bigr] \; ,
\end{equation}
where $H\bigl[\vec{\bm{x}} | \vec{\bm{s}} \bigr]$ is the conditional entropy of the output $\vec{\bm{x}}$ given the input $\vec{\bm{s}}$:
\begin{equation}
\begin{aligned}
  H\bigl[\vec{\bm{x}} |  \vec{\bm{s}} \bigr] \, & \equiv  \,   
  - \int \textrm{d} \vec{\bm{s}} \, p_{\textrm{in}}(\vec{\bm{s}})
  \int \textrm{d} \vec{\bm{x}} \, p_{\textrm{out}}(\vec{\bm{x}} |  \vec{\bm{s}})
 \ln p_{\textrm{out}}(\vec{\bm{x}} | \vec{\bm{s}}) \\
 & = \,  N \ln \Bigl( \sqrt{ 2 \pi e \sigma_0^2} \Bigr) 
 \; .
 \end{aligned}
 \label{eq250421d}
 \end{equation}
Since this conditional entropy is independent of the weight matrix $\bm{W}$, we see that the mutual information $I\bigl[\vec{\bm{x}} ;  \vec{\bm{s}} \bigr]$ is equal to $H\bigl[ p_{\textrm{out}}(\vec{\bm{x}}) \bigr]$ up to a constant.

The entropy $H\bigl[ p_{\textrm{out}}(\vec{\bm{x}}) \bigr]$ is dependent on the noise variance $\sigma_0^2$. When $\sigma_0^2$ is small, we may assume $H\bigl[ p_{\textrm{out}}(\vec{\bm{x}}) \bigr]$ to be a smooth function of $\sigma_0^2$. As a zeroth-order approximation, we approximate the value of $H\bigl[ p_{\textrm{out}}(\vec{\bm{x}}) \bigr]$ by its limiting value at $\sigma_0^2 = 0$, which is Eq.~(\ref{eqHx}). The $\bm{W}$-dependent part of the mutual information $I\bigl[\vec{\bm{x}} ; \vec{\bm{s}} \bigr]$ is therefore approximated by
\begin{equation}
  I \bigl[ \vec{\bm{x}} ; \vec{\bm{s}} \bigr] \,  \approx \, 
  -  \ln \Bigl[ \textrm{det}( \bm{I} + \bm{W}) \Bigr] \, = \, S \; .
  \label{eq250421b}
 \end{equation}

When the output noise $\vec{\bm{\epsilon}}$ of Eq.~(\ref{eq250421a}) is not Gaussian, Eq.~(\ref{eq250421d}) will no longer hold exactly, and the mutual information $I\bigl[ \vec{\bm{x}}; \vec{\bm{s}} \bigr]$ may then have a more complicated dependence on $\bm{W}$. For such more realistic non-Gaussian scenarios, Eq.~(\ref{eq250421b}) may still serve as a simple approximate measure of information robustness to guide our search for close-to-optimal LPC matrices $\bm{W}$.

\section{Correlated Gaussian input}
\label{app:CGsystem}

We present some results for Gaussian input signals. Consider the input signal vector $\vec{\bm{s}}$ being Gaussian with correlations,
\begin{equation}
    \vec{\bm{s}} \, = \, \sqrt{c N }\ a_1 
    \begin{pmatrix}
    \frac{1}{\sqrt{N}} \\
    \vdots\\
    \frac{1}{\sqrt{N}}
    \end{pmatrix} 
    + \sqrt{1-c}\   \sum\limits_{k=1}^{N} b_k \vec{\bm{\phi}}_k \; ,
    \label{eq:s_gs}
\end{equation}
where $c \in [0, 1)$ is a constant, $a_1$ and $b_k$ are all Gaussian random variables of zero mean and unit variance, and $\vec{\bm{\phi}}_k$ are $N$ mutually orthogonal unit vectors. A simple recipe to generate this type of input signal vectors is to apply an external current of magnitude $\sqrt{c} a_1$ to all the $N$ units of the network~\cite{Wu-etal-2025}.

The correlation matrix of the input signals is denoted as $\bm{C} \equiv \langle \vec{\bm{s}} \vec{\bm{s}}^\top \rangle$.  For Gaussian inputs (\ref{eq:s_gs}), this matrix $\bm{C}$ is very simple: all its diagonal elements are equal to unity and all its non-diagonal elements are equal to $c$~\cite{Huang-etal-2023}. The output vector $\vec{\bm{x}}$ is a Gaussian random vector with zero mean, and its covariance matrix is
\begin{equation}
\bigl\langle  \vec{\bm{x}} \vec{\bm{x}}^\top \bigr\rangle \, = \, 
\frac{\bm{I} }{(\bm{I} + \bm{W})} \bm{C} \frac{\bm{I} }{(\bm{I} + \bm{W})^\top} \; .
\label{eqa280425}
\end{equation}
From Eq.~(\ref{eqa280425}) we obtain that the variance $\sigma_i^2$ of a single output $x_i$ is
\begin{equation}
\begin{aligned}
    \sigma_i^2 \, & =  \, 
    c\  \biggl[ \sum_{k=1}^N \Bigl(\frac{\bm{I}}{\bm{I}+\bm{W}}\Bigr)_{i k} \biggr]^2  \\
    & \quad \quad +  (1-c)\ \biggl[ 
    \frac{\bm{I}}{(\bm{I}+\bm{W})} \frac{\bm{I}}{(\bm{I}+\bm{W})^\top} \biggr]_{i i} \; .
    \end{aligned}
    \label{eqc280425}
\end{equation}
The $L_1$-norm mean energy of the system is
\begin{equation}
    E \, = \,  \sum\limits_{i=1}^{N} \bigl\langle | x_i | \bigr\rangle \, =  \, \sum\limits_{i=1}^{N} \sqrt{ \frac{ 2 \sigma_i^2}{\pi} } \; .
    \label{eqb280425}
\end{equation}

The mutual information measure (entropy $S$) is 
\begin{equation}
\begin{aligned}
    S \, & \equiv \, - \ln \Bigl[ \textrm{det}\bigl( \bm{I}+\bm{W} \bigr) \Bigr]  \\
    \, & = \, \frac{1}{2}\ln\Bigl[ \textrm{det}\bigl( \frac{\bm{I}}{(\bm{I}+\bm{W})}
    \bm{C} \frac{\bm{I}}{(\bm{I}+\bm{W})^\top} \bigr) \Bigr] 
    \\
    & \quad \quad \quad - \frac{1}{2}\ln\Bigl[ \textrm{det}\bigl( \bm{C} \bigr) \Bigr]  \; .
    \end{aligned}
\end{equation}
We can define an auxiliary real symmetric matrix as
\begin{equation}
\begin{aligned}
    \bm{Y} \, & = \,
    \textrm{Diag}\Bigl[ \frac{1}{\sqrt{\sigma_1}}, 
    \frac{1}{\sqrt{\sigma_2}}, \ldots,
    \frac{1}{\sqrt{\sigma_N}} \Bigr] 
    \frac{\bm{I}}{(\bm{I}+\bm{W})}
    \bm{C}  \\
    & \quad \quad 
    \frac{\bm{I}}{(\bm{I}+\bm{W})^\top}
     \textrm{Diag}\Bigl[ \frac{1}{\sqrt{\sigma_1}}, 
    \frac{1}{\sqrt{\sigma_2}}, \ldots,
    \frac{1}{\sqrt{\sigma_N}} \Bigr] \; ,
    \end{aligned}
\end{equation}
where $\textrm{Diag}[ \ldots ]$ means a diagonal matrix. A nice property of the matrix $\bm{Y}$ is that its $N$ diagonal elements are simply $\sigma_1, \ldots, \sigma_N$. Let us denote the $N$ positive eigenvalues of this matrix $\bm{Y}$ as $\tilde{\lambda}_1, \ldots, \tilde{\lambda}_N$, then we have
\begin{equation}
    \sum\limits_{i=1}^N \tilde{\lambda}_i \, = \, \sum\limits_{i=1}^N \sigma_i \; . 
\end{equation}
With the help of this auxiliary matrix $\bm{Y}$, we obtain the following upper-bound for the entropy $S$:
\begin{equation}
\begin{aligned}
 & S + \frac{1}{2}\ln\Bigl[ \textrm{det}\bigl( \bm{C} \bigr) \Bigr] \, = \, 
\frac{N}{2} \sum\limits_{i=1}^N \Bigl[ \frac{1}{N} \ln \sigma_i + \frac{1}{N} \ln \tilde{\lambda}_i \Bigr]  \\
& \quad \quad  \leq \, \frac{N}{2} \ln \Bigl( \frac{1}{N} \sum_{i} \sigma_i \Bigr) 
 + \frac{N}{2} \ln\Bigl( \frac{1}{N} \sum_{i} \tilde{\lambda}_i \Bigr)
  \\
 & \quad \quad  = \, N \ln\Bigl( \sqrt{ \frac{\pi }{2} } \frac{E}{N} \Bigr) \; .
  \end{aligned}
 \label{eqa010525}
\end{equation}
In deriving the second line of Eq.~(\ref{eqa010525}), we have used the Jensen inequality
\begin{equation}
\sum\limits_{i=1}^N \frac{1}{N} \ln h_i \, \leq \, \ln\Bigl( \frac{1}{N} \sum\limits_{i=1}^{N} h_i \Bigr) \; .
\label{eqb010525}
\end{equation}
\begin{figure*}
    \centering
    \subfigure[]{
\includegraphics[angle=270,width=0.3\linewidth]{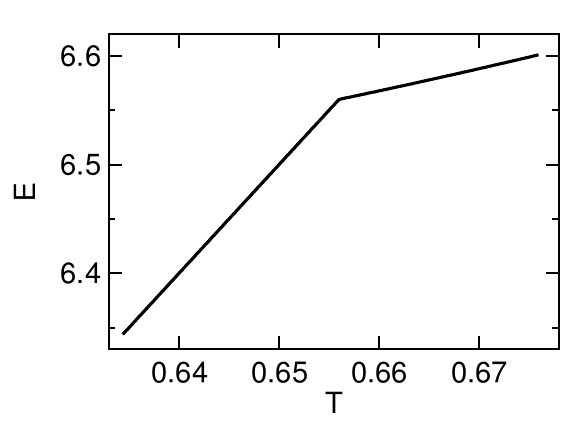}
\label{fig090525a}
}
\subfigure[]{
\includegraphics[angle=270,width=0.3\linewidth]{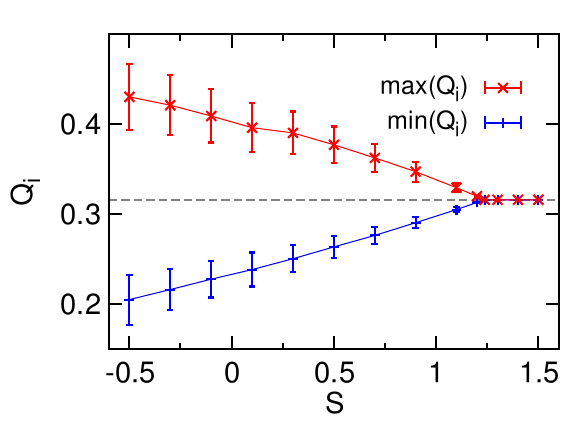}
\label{fig090525b}
}
    \caption{
        Continuous phase transition phenomenon for correlated Gaussian input signals (\ref{eq:s_gs}) with $c=0.6$ at $N=10$. (a) $L_1$-norm mean energy $E$ versus temperature $T$. The critical temperature $T^* = 0.6560$ and the critical entropy $S^* = 1.2373$. (b) The maximum and the minimum values of $Q_i$. The errorbars mark the standard deviations over $660$ independently sampled optimal LPC matrices.
    }
    \label{fig250422a}
\end{figure*}

The equality of Eq.~(\ref{eqb010525}) holds only if all the positive $h_i$ values are equal to each other. This means at, at a given value of mean $L_1$-norm energy $E$, the entropy $S$ achieves its maximum value if and only if all the $N$ variances $\sigma_i^2$ are equal and also all the $N$ eigenvalues $\tilde{\lambda}_i$ are equal. If these two conditions are satisfied simultaneously, then we have
\begin{equation}
   \frac{E}{N} \, = \,  \sqrt{\frac{2}{\pi}} \, \bigl(  \textrm{det} (\bm{C}) \bigr)^{\frac{1}{2N}} \,   \exp\Bigl(\frac{S}{N} \Bigr) \; , 
\end{equation}
and therefore the relationship between $E$ and temperature $T$ is 
\begin{equation}
    E \, = \, N\, T \; .
\end{equation}
The corresponding optimal weight matrix $\bm{W}$ satisfies
\begin{equation}
(\bm{I} + \bm{W}) (\bm{I}+\bm{W})^\top \,  = \, (2/\pi) T^{-2} \, \bm{C} \; .
\label{eqc010525}
\end{equation}
From this result and Eq.~(\ref{eqa280425}) we obtain that, $\langle x_i^2 \rangle = (\pi/2)^{1/2} T$ and $\langle x_i x_j \rangle = 0$ for  $i \neq j$. In other words, the output variables $x_i$ are governed by the same Gaussian distribution and they are mutually independent.

Because there is no self-interaction ($w_{i i}=0$), optimal weight matrices $\bm{W}$ with the property of Eq.~(\ref{eqc010525}) can only be constructed for systems containing $N\geq 3$ units, and only at temperatures $T$ lower than certain critical value $T^*$. Following the same derivation of Ref.~\cite{Huang-etal-2023}, we find that the analytical expression of $T^*$ is
\begin{equation}
T^* \, = \,  \sqrt{\frac{2}{\pi} }\, \frac{ \sqrt{1+(N-1) c} + (N-1) \sqrt{1-c}}{N}\; .
\end{equation}

We have confirmed these analytical results by numerical simulations. As a concrete example, we show in Fig.~\ref{fig250422a} the numerical results obtained on the system with $N=10$ units and at input correlation $c=0.6$. There is a continuous phase transition at $T^*=0.6560$, with the critical value of entropy being $S=S^*=1.2373$. At this phase transition point, the permutation symmetry of the optimal weight matrix $\bm{W}$ breaks down, leading to a kink in the mean $L_1$-norm energy $E$ (Fig.~\ref{fig090525a}). Similar to Eq.~(\ref{eq:Qi}), we can measure the projections $Q_i$ of the feature direction $\vec{\bm{\phi}}_1 = (1/\sqrt{N}, \ldots, 1/\sqrt{N})^\top$ of Eq.~(\ref{eq:s_gs}) to all the $N$ units $i$. We find that, after this phase transition, different units $x_i$ are responding differently to $\vec{\bm{\phi}}_1$ such that the maximum and minimum values of $Q_i$ deviate from each other as $S$ decreases from $S^*$ (Fig.~\ref{fig090525a}). Very interestingly, at each fixed value of $S < S^*$ there are many degenerate optimal matrices $\bm{W}$, all of them have the same minimum energy but the maximum and minimum $Q_i$ values are different. This high degree of degeneracy is the reason behind the errorbars of Fig.~\ref{fig090525b}. Notice that no unit $i$ is selectively responding only to the presence of $\vec{\bm{\phi}}_1$ in this Gaussian case.

\section{Single-unit conditional probability}
\label{app:pouti}

We derive the explicit expression (\ref{eq241217a}) for the conditional probability distribution of an output signal $x_i$. The output signal vector $\bm{x}$ is
\begin{equation}
    \vec{\bm{x}} \,  = \, a_1 \vec{\bm{\mu}} + \sum\limits_{j\geq 2} b_j \vec{\bm{\psi}}_j \; ,
    \label{eq0219a}
\end{equation}
where the output vectors $\vec{\bm{\mu}}$ and $\vec{\bm{\psi}}_j$ ($j\geq 2$) are, respectively, the transform of $\vec{\bm{\phi}}_1$ and $\vec{\bm{\phi}}_j$:
\begin{equation}
  \begin{aligned}
    \vec{\bm{\mu}} \, & = \,  \frac{\bm{I}}{\bm{I}+\bm{W}} \vec{\bm{\phi}}_1 \; , \\
    \vec{\bm{\psi}}_j \, & = \,
    \frac{\bm{I}}{\bm{I}+\bm{W}} \vec{\bm{\phi}}_j \, \quad \quad (j=2, \ldots, N)  \; .
  \end{aligned}
\end{equation}
The conditional mean vector of $\vec{\bm{x}}$ at fixed value of the non-Gaussian coefficient $a_1$ is simply $\langle \vec{\bm{x}} \rangle  \, = \, a_1 \vec{\bm{\mu}}$, and therefore $\langle x_i \rangle = a_1 \mu_i$ at fixed $a_1$ (Eq.~(\ref{eq:muiexp})).

For $N$ mutually orthogonal unit vectors $\vec{\bm{\phi}}_j$, we have $\sum_{j=1}^N \vec{\bm{\phi}}_j {\vec{\bm{\phi}}_j}^\top = \bm{I}$, and consequently, the correlation matrix of $\vec{\bm{x}}$ at fixed $a_1$ is
\begin{equation}
  \bigl\langle \vec{\bm{x}} {\vec{\bm{x}}}^\top \bigr\rangle 
  \, = \, (a_1^2-1) \, \vec{\bm{\mu}} {\vec{\bm{\mu}}}^\top  + 
  \frac{\bm{I}}{(\bm{I}+\bm{W})} \frac{\bm{I}}{(\bm{I}+\bm{W})^\top} \; .
  \label{eqxxt}
\end{equation}
At fixed $a_1$ the variance of $x_i$ is $\sigma_i^2 \equiv \langle x_i^2 \rangle - (a_1 \mu_i)^2$. Applying Eq.~(\ref{eqxxt}) we easily verify Eq.~(\ref{eq:sigmai2}).

Since the coefficients $b_j$ ($j \geq 2$) of Eq.~(\ref{eq0219a}) are Gaussian random variables, at fixed value of the non-Gaussian coefficient $a_1$, the conditional distribution $p_{\textrm{out}}(x_i | a_1)$ of $x_i$ must also be a Gaussian distribution, which is Eq.~(\ref{eq241217a}). The signal-to-noise ratio $\eta_i$ of this conditional distribution can be defined as the ratio between the mean and the standard deviation, namely
\begin{equation}
  \eta_i \, \equiv \, \frac{ |a_1 \mu_i|}{\sqrt{\sigma_i^2}}
  \, = \, \sqrt{\frac{a_1^2 \mu_i^2}{\sigma_i^2}} \; .
  \label{eq241217b}
\end{equation}

\section{Energy computation}
\label{app:mE}

The mean $L_1$-norm energy is
\begin{equation}
  E \,  = \, \sum\limits_{i=1}^N \int\int \textrm{d} a_1  \textrm{d} x_1
  q(a_1)\  p_{\textrm{out}}(x_1 | a_1) | x_i | \; .
\end{equation}
Performing the Gaussian integration, we obtain that
\begin{equation}
  \begin{aligned}
    E  \, = \, \sum\limits_{i=1}^{N}  \int \textrm{d} a_1 \, q( a_1)
    & \ \biggl[ \sqrt{\frac{2 \sigma_i^2}{\pi}} \,
      \exp\Bigl( - \frac{a_1^2 \mu_i^2}{2 \sigma_i^2} \Bigr) 
      \\
      & \quad  +  |a_1 \mu_i |\, \textrm{erf}\Bigl(
      \frac{|a_1 \mu_i|}{\sqrt{2 \sigma_i^2}} \Bigr) \biggr] \; ,
  \end{aligned}
  \label{eqL1Eexp}
\end{equation}
where $\textrm{erf}(x)$ is the error function:
\begin{equation}
  \textrm{erf}(x) \, = \,
  \frac{2}{\sqrt{\pi}} \int_0^x e^{-t^2}\, \textrm{d} t \; .
\end{equation}

For the discrete prior distribution (\ref{eqPa1}) with a parameter $p_0$, we can easily derive from Eq.~(\ref{eqL1Eexp}) the explicit expression (\ref{eqL1Edisc}) for the mean $L_1$-norm energy. The $\zeta_i$ quantity in Eq.~(\ref{eqL1Edisc}) is simply the rescaled signal-to-noise ratio $\eta_i$ at $a_1 = 1/\sqrt{1-p_0}$, namely $\zeta_i = \eta_i/\sqrt{2}$.

If the non-Gaussian coefficient $a_1$ follows the continuous Laplace distribution,
\begin{equation}
  q( a_1 ) \, = \, \frac{1}{\sqrt{2}} e^{-\sqrt{2} | a_1 |} \; ,
  \label{eq:250120a}
\end{equation}
the corresponding mean $L_1$-norm energy is
\begin{equation}
    E \,  = \, \sum\limits_{i=1}^N \biggl[ 
      \sqrt{ \frac{ 2 \sigma_i^2}{\pi} } 
      +\sqrt{\frac{\mu_i^2}{2}} 
      \exp\Bigl( \frac{\sigma_i^2}{\mu_i^2} \Bigr) \,
      \textrm{erfc}\Bigl( \sqrt{\frac{\sigma_i^2}{\mu_i^2}} \Bigr)
      \biggr] \; ,
  \label{eqPa1Lp}
\end{equation}
where $\textrm{erfc}(x)$ is the complementary error function:
\begin{equation}
  \textrm{erfc}(x) \, = \, \frac{2}{\sqrt{\pi}} \int_x^\infty e^{- t^2} \textrm{d} t \; .
\end{equation}
This energy expression (\ref{eqPa1Lp}) for the Laplace distribution is similar to Eq.~(\ref{eqL1Edisc}) for the discrete distribution (\ref{eqPa1}).

If the random coefficient $a_1$ follows a long-tailed power-law distribution, an explicit expression for the mean $L_1$-norm energy can also be derived, see Ref.~\cite{SI2025a} for details.

\section{The semicircle pattern of eigenvalues}
\label{app:semicircle}

We can express the matrix $(\bm{I}+\bm{W})^{-1}$ by singular-value decomposition as
\begin{equation}
  \frac{\bm{I}}{\bm{I}+\bm{W}} \, = \, \bm{U}
  \textrm{Diag}\bigl[v_1, v_2, \ldots, v_N \bigr] \bm{V}^\top \; .
  \label{eq250224a}
\end{equation}
Here $\bm{U}$ and $\bm{V}$ are two orthonormal real matrices with $\bm{U} \bm{U}^\top = \bm{U}^\top \bm{U} = \bm{I}$ and similarly for $\bm{V}$, and $v_i$ is the $i$-th singular value of $(\bm{I}+\bm{W})^{-1}$. We notice that 
\begin{equation}
  \frac{\bm{I}}{(\bm{I}+\bm{W})^\top (\bm{I}+\bm{W})} \, = \,
  \bm{U} \textrm{Diag}\bigl[v_1^2, v_2^2, \ldots, v_N^2\bigr] \bm{U}^\top \; ,
  \label{eq250224b}
\end{equation}
and therefore the entropy $S$ is
\begin{equation}
  S \, = \, \frac{1}{2} \sum\limits_{j=1}^{N} \ln v_j^2 \; .
  \label{eq250224c}
\end{equation}
We see that, fixing the entropy $S$ means fixing the product value $\prod_{j} v_j$. Energy minimization at fixed $S$ means minimizing the value of $\sum_j \bigl\langle |x_j| \bigr\rangle$ at fixed value of $\prod_j v_j$.

When a single unit (say with index $i_0=1$) is selectively responding to the feature $\vec{\bm{\phi}}_1$ very strongly and all the other units are indifferent to this feature direction, we find that the first column $\vec{\bm{v}}_1$ of $\bm{V}$ is almost identical to $\vec{\bm{\phi}}_1$, and the first column $\vec{\bm{u}}_1$ of $\bm{U}$ is almost identical to the column vector $(Q_1, Q_2, \ldots, Q_N)^\top$ with $Q_i$ being defined by Eq.~(\ref{eq:Qi}) and hence approximately $\vec{\bm{u}}_1 \approx (1, 0, \ldots, 0)$. The remaining $(N-1)$ column vectors of $\bm{V}$ therefore span the subspace orthogonal to $\vec{\bm{\phi}}_1$ and the remaining $(N-1)$ column vectors of $\bm{U}$ (approximately) span the subspace formed by the outputs $x_2, x_3, \ldots, x_N$. In other words, for indices $j \geq 2$, the mean $\mu_j$ as defined by Eq.~(\ref{eq:muiexp}) is approximately zero, and the variance $\sigma_j^2$ as defined by Eqs.~(\ref{eq:sigmai2}) only depends on the singular values $v_2, \ldots, v_N$ and but almost completely independent of $v_1$. We have $\sum_{j \geq 2} \sigma_j^2 \approx \sum_{j\geq 2} v_j^2$ according to Eq.~(\ref{eq250224b}). Under this constraint, the summed total energy of the units with $j\geq 2$, $\sum_{j\geq 2} \bigl\langle |x_j|\bigr\rangle$ for Gaussian random variables $x_j$ with mean $\mu_j \approx 0$ and variance $\sigma_j^2$, will achieves its global minimum when all the $\sigma_j^2$ values with indices $j\geq 2$ are equal. In other words, it is optimal to have all the singular values $v_j$ with $j\geq 2$ being equal to each other.

Under the constraint of fixed $S$, the singular value $v_1$ will be optimized to achieve the best balance between $\bigl\langle | x_1 |\bigr\rangle$ and $\sum_{j\geq 2} \bigl\langle | x_j | \bigr\rangle$.

From Eq.~(\ref{eq250224a}) we know that
\begin{equation}
  \bm{I}+\bm{W} \, = \, \bm{V}
  \textrm{Diag}\Bigl[\frac{1}{v_1}, \frac{1}{v_2}, \ldots, \frac{1}{v_N} \Bigr]
  \bm{U}^\top \; .
\end{equation}
When all the singular values $v_j$ with indices $j\geq 2$ are almost equal, the complex eigenvalues $\lambda_j$ with $j\geq 2$ will also be approximately equal in magnitude. This phenomenon has been demonstrated in Fig.~\ref{fig:LambdaRI} and \ref{fig:LambdaRadius}.

\end{appendix}


\begin{thebibliography}{44}%
\makeatletter
\providecommand \@ifxundefined [1]{%
 \@ifx{#1\undefined}
}%
\providecommand \@ifnum [1]{%
 \ifnum #1\expandafter \@firstoftwo
 \else \expandafter \@secondoftwo
 \fi
}%
\providecommand \@ifx [1]{%
 \ifx #1\expandafter \@firstoftwo
 \else \expandafter \@secondoftwo
 \fi
}%
\providecommand \natexlab [1]{#1}%
\providecommand \enquote  [1]{``#1''}%
\providecommand \bibnamefont  [1]{#1}%
\providecommand \bibfnamefont [1]{#1}%
\providecommand \citenamefont [1]{#1}%
\providecommand \href@noop [0]{\@secondoftwo}%
\providecommand \href [0]{\begingroup \@sanitize@url \@href}%
\providecommand \@href[1]{\@@startlink{#1}\@@href}%
\providecommand \@@href[1]{\endgroup#1\@@endlink}%
\providecommand \@sanitize@url [0]{\catcode `\\12\catcode `\$12\catcode
  `\&12\catcode `\#12\catcode `\^12\catcode `\_12\catcode `\%12\relax}%
\providecommand \@@startlink[1]{}%
\providecommand \@@endlink[0]{}%
\providecommand \url  [0]{\begingroup\@sanitize@url \@url }%
\providecommand \@url [1]{\endgroup\@href {#1}{\urlprefix }}%
\providecommand \urlprefix  [0]{URL }%
\providecommand \Eprint [0]{\href }%
\providecommand \doibase [0]{http://dx.doi.org/}%
\providecommand \selectlanguage [0]{\@gobble}%
\providecommand \bibinfo  [0]{\@secondoftwo}%
\providecommand \bibfield  [0]{\@secondoftwo}%
\providecommand \translation [1]{[#1]}%
\providecommand \BibitemOpen [0]{}%
\providecommand \bibitemStop [0]{}%
\providecommand \bibitemNoStop [0]{.\EOS\space}%
\providecommand \EOS [0]{\spacefactor3000\relax}%
\providecommand \BibitemShut  [1]{\csname bibitem#1\endcsname}%
\let\auto@bib@innerbib\@empty
\bibitem [{\citenamefont {Srinivasan}\ \emph {et~al.}(1982)\citenamefont
  {Srinivasan}, \citenamefont {Laughlin},\ and\ \citenamefont
  {Dubs}}]{Srinivasan-etal-1982}%
  \BibitemOpen
  \bibfield  {author} {\bibinfo {author} {\bibfnamefont {M.~V.}\ \bibnamefont
  {Srinivasan}}, \bibinfo {author} {\bibfnamefont {S.~B.}\ \bibnamefont
  {Laughlin}}, \ and\ \bibinfo {author} {\bibfnamefont {A.}~\bibnamefont
  {Dubs}},\ }\bibfield  {title} {\enquote {\bibinfo {title} {Predictive coding:
  a fresh view of inhibition in the retina},}\ }\href@noop {} {\bibfield
  {journal} {\bibinfo  {journal} {Proc. R. Soc. Lond. B}\ }\textbf {\bibinfo
  {volume} {216}},\ \bibinfo {pages} {427--459} (\bibinfo {year}
  {1982})}\BibitemShut {NoStop}%
\bibitem [{\citenamefont {Rao}\ and\ \citenamefont
  {Ballard}(1999)}]{Rao-Ballard-1999}%
  \BibitemOpen
  \bibfield  {author} {\bibinfo {author} {\bibfnamefont {R.~P.~N.}\
  \bibnamefont {Rao}}\ and\ \bibinfo {author} {\bibfnamefont {D.~H.}\
  \bibnamefont {Ballard}},\ }\bibfield  {title} {\enquote {\bibinfo {title}
  {Predictive coding in the visual cortex: a functional interpretation of some
  extra-classical receptive-field effects},}\ }\href@noop {} {\bibfield
  {journal} {\bibinfo  {journal} {Nature Neurosci.}\ }\textbf {\bibinfo
  {volume} {2}},\ \bibinfo {pages} {79--87} (\bibinfo {year}
  {1999})}\BibitemShut {NoStop}%
\bibitem [{\citenamefont {Huang}\ and\ \citenamefont
  {Rao}(2011)}]{Huang-Rao-2011}%
  \BibitemOpen
  \bibfield  {author} {\bibinfo {author} {\bibfnamefont {Y.}~\bibnamefont
  {Huang}}\ and\ \bibinfo {author} {\bibfnamefont {R.~P.~N.}\ \bibnamefont
  {Rao}},\ }\bibfield  {title} {\enquote {\bibinfo {title} {Predictive
  coding},}\ }\href@noop {} {\bibfield  {journal} {\bibinfo  {journal} {WIREs
  Cogn. Sci.}\ }\textbf {\bibinfo {volume} {2}},\ \bibinfo {pages} {580--593}
    (\bibinfo {year} {2011})}\BibitemShut {NoStop}%
\bibitem [{\citenamefont {Ali}\ \emph {et~al.}(2022)\citenamefont {Ali},
  \citenamefont {Ahmad}, \citenamefont {{de Groot}}, \citenamefont {{van
  Gerven}},\ and\ \citenamefont {Kietzmann}}]{Ali-etal-2022}%
  \BibitemOpen
  \bibfield  {author} {\bibinfo {author} {\bibfnamefont {A.}\
  \bibnamefont {Ali}}, \bibinfo {author} {\bibfnamefont {N.}\ \bibnamefont
  {Ahmad}}, \bibinfo {author} {\bibfnamefont {E.}\ \bibnamefont {{de
  Groot}}}, \bibinfo {author} {\bibfnamefont {M.~A.~J.}\
  \bibnamefont {{van Gerven}}}, \ and\ \bibinfo {author} {\bibfnamefont
  {T.~C.}\ \bibnamefont {Kietzmann}},\ }\bibfield  {title} {\enquote
  {\bibinfo {title} {Predictive coding is a consequence of energy efficiency in
  recurrent neural networks},}\ }\href@noop {} {\bibfield  {journal} {\bibinfo
  {journal} {Patterns}\ }\textbf {\bibinfo {volume} {3}},\ \bibinfo {pages}
    {100639} (\bibinfo {year} {2022})}\BibitemShut {NoStop}%
\bibitem [{\citenamefont {{van Zwol}}\ \emph {et~al.}(2024)\citenamefont {{van
  Zwol}}, \citenamefont {Jefferson},\ and\ \citenamefont {{van den
  Broek}}}]{VanZwol-etal-2024}%
  \BibitemOpen
  \bibfield  {author} {\bibinfo {author} {\bibfnamefont {B.}~\bibnamefont {{van
  Zwol}}}, \bibinfo {author} {\bibfnamefont {R.}~\bibnamefont {Jefferson}}, \
  and\ \bibinfo {author} {\bibfnamefont {E.~L.}\ \bibnamefont {{van den
  Broek}}},\ }\href@noop {} {\enquote {\bibinfo {title} {Predictive coding
  networks and inference learning: Tutorial and survey},}\ }\bibinfo
  {howpublished} {eprint arXiv:2407.04117 [cs.LG]} (\bibinfo {year}
  {2024})\BibitemShut {NoStop}%
\bibitem [{\citenamefont {Millidge}\ \emph {et~al.}(2022)\citenamefont
  {Millidge}, \citenamefont {Salvatori}, \citenamefont {Song}, 
  \citenamefont {Bogacz}, \ and\ \citenamefont
  {Lukasiewicz}}]{Millidge-etal-2022}%
  \BibitemOpen
  \bibfield  {author} {\bibinfo {author} {\bibfnamefont {B.}\ \bibnamefont
  {Millidge}}, \bibinfo {author} {\bibfnamefont {T.}\ \bibnamefont
  {Salvatori}}, \bibinfo {author} {\bibfnamefont {Y.}\ \bibnamefont
  {Song}}, \bibinfo {author} {\bibfnamefont {R.}\ \bibnamefont {Bogacz}}, \
  and\ \bibinfo {author} {\bibfnamefont {T.}\ \bibnamefont {Lukasiewicz}},\
  }\bibfield  {title} {\enquote {\bibinfo {title}  {Predictive coding: Towards a
        future of deep learning beyond backpropagation?}}\ 
}in\ \href@noop {} {\emph {\bibinfo {booktitle}
  {Proceedings of the 31st International Joint Conference on Artificial Intelligence (IJCAI), Vienna, Austria}}}\ (\bibinfo {year} {2022})\ pp.\ \bibinfo
             {pages} {5538--5545}\BibitemShut {NoStop}%
\bibitem [{\citenamefont {Huang}\ \emph {et~al.}(2022)\citenamefont {Huang},
  \citenamefont {Fan}, \citenamefont {Zhou},\ and\ \citenamefont
  {Zhou}}]{Huang-etal-2022}%
  \BibitemOpen
  \bibfield  {author} {\bibinfo {author} {\bibfnamefont {Z.-Y.}\ \bibnamefont
  {Huang}}, \bibinfo {author} {\bibfnamefont {X.-Y.}\ \bibnamefont {Fan}},
  \bibinfo {author} {\bibfnamefont {J.}\ \bibnamefont {Zhou}}, \ and\
  \bibinfo {author} {\bibfnamefont {H.-J.}\ \bibnamefont {Zhou}},\ }\bibfield
   {title} {\enquote {\bibinfo {title} {Lateral predictive coding revisited:
  internal model, symmetry breaking, and response time},}\ }\href@noop {}
  {\bibfield  {journal} {\bibinfo  {journal} {Commun. Theor. Phys.}\ }\textbf
  {\bibinfo {volume} {74}},\ \bibinfo {pages} {095601} (\bibinfo {year}
  {2022})}\BibitemShut {NoStop}%
\bibitem [{\citenamefont {Rozell}\ \emph {et~al.}(2008)\citenamefont {Rozell},
  \citenamefont {Johnson}, \citenamefont {Baraniuk},\ and\ \citenamefont
  {Olshausen}}]{Rozell-etal-2008}%
  \BibitemOpen
  \bibfield  {author} {\bibinfo {author} {\bibfnamefont {C.~J.}\
  \bibnamefont {Rozell}}, \bibinfo {author} {\bibfnamefont {D.~H.}\
  \bibnamefont {Johnson}}, \bibinfo {author} {\bibfnamefont {R.~G.}\
  \bibnamefont {Baraniuk}}, \ and\ \bibinfo {author} {\bibfnamefont {B.~A.}\
  \bibnamefont {Olshausen}},\ }\bibfield  {title} {\enquote {\bibinfo {title}
  {Sparse coding via theresholding and local competition in neural circuits},}\
  }\href@noop {} {\bibfield  {journal} {\bibinfo  {journal} {Neural
  Comput.}\ }\textbf {\bibinfo {volume} {20}},\ \bibinfo {pages}
    {2526--2563} (\bibinfo {year} {2008})}\BibitemShut {NoStop}%
\bibitem [{\citenamefont {Yu}\ \emph {et~al.}(2018)\citenamefont {Yu},
  \citenamefont {Shen}, \citenamefont {Wang},\ and\ \citenamefont
  {Yu}}]{Yu-etal-2018}%
  \BibitemOpen
  \bibfield  {author} {\bibinfo {author} {\bibfnamefont {L.}\
  \bibnamefont {Yu}}, \bibinfo {author} {\bibfnamefont {Z.}\ \bibnamefont
  {Shen}}, \bibinfo {author} {\bibfnamefont {C.}\ \bibnamefont {Wang}}, \
  and\ \bibinfo {author} {\bibfnamefont {Y.}\ \bibnamefont {Yu}},\
  }\bibfield  {title} {\enquote {\bibinfo {title} {Efficient coding and energy
  efficiency are promoted by balanced excitatory and inhibitory synaptic
  currents in neuronal network},}\ }\href@noop {} {\bibfield  {journal}
  {\bibinfo  {journal} {Front. Cell. Neurosci.}\ }\textbf {\bibinfo {volume}
  {12}},\ \bibinfo {pages} {123} (\bibinfo {year} {2018})}\BibitemShut
             {NoStop}%
\bibitem [{\citenamefont {Yang}\ \emph {et~al.}(2017)\citenamefont {Yang},
  \citenamefont {Zhou},\ and\ \citenamefont {Zhou}}]{Yang-Zhou-etal-2017}%
  \BibitemOpen
  \bibfield  {author} {\bibinfo {author} {\bibfnamefont {D.-P.}\ \bibnamefont
  {Yang}}, \bibinfo {author} {\bibfnamefont {H.-J.}\ \bibnamefont {Zhou}}, \
  and\ \bibinfo {author} {\bibfnamefont {C.}~\bibnamefont {Zhou}},\ }\bibfield
  {title} {\enquote {\bibinfo {title} {Co-emergence of multi-scale cortical
  activities of irregular firing, oscillations and avalanches achieves
  cost-efficient information capacity},}\ }\href@noop {} {\bibfield  {journal}
  {\bibinfo  {journal} {PLoS Comput. Biol.}\ }\textbf {\bibinfo {volume}
  {13}},\ \bibinfo {pages} {e1005384} (\bibinfo {year} {2017})}\BibitemShut
  {NoStop}%
\bibitem [{\citenamefont {Tang}\ \emph {et~al.}(2023)\citenamefont {Tang},
  \citenamefont {Salvatori}, \citenamefont {Millidge}, \citenamefont {Song},
  \citenamefont {Lukasiewicz},\ and\ \citenamefont {Bogacz}}]{Tang-etal-2023b}%
  \BibitemOpen
  \bibfield  {author} {\bibinfo {author} {\bibfnamefont {M.}\ \bibnamefont
  {Tang}}, \bibinfo {author} {\bibfnamefont {T.}\ \bibnamefont
  {Salvatori}}, \bibinfo {author} {\bibfnamefont {B.}\ \bibnamefont
  {Millidge}}, \bibinfo {author} {\bibfnamefont {Y.}\ \bibnamefont {Song}},
  \bibinfo {author} {\bibfnamefont {T.}\ \bibnamefont {Lukasiewicz}}, \
  and\ \bibinfo {author} {\bibfnamefont {R.}\ \bibnamefont {Bogacz}},\
  }\bibfield  {title} {\enquote {\bibinfo {title} {Recurrent predictive coding
  models for associative memory employing covariance learning},}\ }\href@noop
  {} {\bibfield  {journal} {\bibinfo  {journal} {PLOS Comput. Biol.}\ }\textbf
  {\bibinfo {volume} {19 (4)}},\ \bibinfo {pages} {e1010719} (\bibinfo {year}
  {2023})}\BibitemShut {NoStop}%
\bibitem [{\citenamefont {Hyv\"arinen}\ \emph {et~al.}(2009)\citenamefont
  {Hyv\"arinen}, \citenamefont {Hurri},\ and\ \citenamefont
  {Hoyer}}]{Hyvarinen-etal-2009}%
  \BibitemOpen
  \bibfield  {author} {\bibinfo {author} {\bibfnamefont {A.}\ \bibnamefont
  {Hyv\"arinen}}, \bibinfo {author} {\bibfnamefont {J.}\ \bibnamefont
  {Hurri}}, \ and\ \bibinfo {author} {\bibfnamefont {P.~O.}\ \bibnamefont
  {Hoyer}},\ }\href@noop {} {\emph {\bibinfo {title} {Natural Image Statistics:
  A Probabilistic Approach to Early Computational Vision}}}\ (\bibinfo
  {publisher} {Springer},\ \bibinfo {address} {London, UK},\ \bibinfo {year}
  {2009})\BibitemShut {NoStop}%
\bibitem [{\citenamefont {Huang}\ \emph {et~al.}(2024)\citenamefont {Huang},
  \citenamefont {Zhou}, \citenamefont {Huang},\ and\ \citenamefont
  {Zhou}}]{Huang-etal-2023}%
  \BibitemOpen
  \bibfield  {author} {\bibinfo {author} {\bibfnamefont {Z.-Y.}\ \bibnamefont
  {Huang}}, \bibinfo {author} {\bibfnamefont {R.}\ \bibnamefont {Zhou}},
  \bibinfo {author} {\bibfnamefont {M.}\ \bibnamefont {Huang}}, \ and\
  \bibinfo {author} {\bibfnamefont {H.-J.}\ \bibnamefont {Zhou}},\ }\bibfield
   {title} {\enquote {\bibinfo {title} {Energy--information trade-off induces
  continuous and discontinuous phase transitions in lateral predictive
  coding},}\ }\href@noop {} {\bibfield  {journal} {\bibinfo  {journal} {Science
  China: Phys. Mech. Astron.}\ }\textbf {\bibinfo {volume} {67}},\ \bibinfo
     {pages} {260511} (\bibinfo {year} {2024})}\BibitemShut {NoStop}%
\bibitem [{\citenamefont {Barlow}(1972)}]{Barlow-1972}%
  \BibitemOpen
  \bibfield  {author} {\bibinfo {author} {\bibfnamefont {H.~B.}\ \bibnamefont
  {Barlow}},\ }\bibfield  {title} {\enquote {\bibinfo {title} {Single units and
  sensation: A neuron doctrine for perceptual psychology?}}\ }\href@noop {}
  {\bibfield  {journal} {\bibinfo  {journal} {Perception}\ }\textbf {\bibinfo
  {volume} {1}},\ \bibinfo {pages} {371--394} (\bibinfo {year}
    {1972})}\BibitemShut {NoStop}%
\bibitem [{\citenamefont {Bialek}(2024)}]{Bialek-2024}%
  \BibitemOpen
  \bibfield  {author} {\bibinfo {author} {\bibfnamefont {W.}\ \bibnamefont
  {Bialek}},\ }\bibfield  {title} {\enquote {\bibinfo {title} {Ambitions for
  theory in the physics of life},}\ }\href@noop {} {\bibfield  {journal}
  {\bibinfo  {journal} {SciPost Phys. Lect. Notes}\ ,\ \bibinfo {pages} {84}}
  (\bibinfo {year} {2024})}\BibitemShut {NoStop}%
\bibitem [{\citenamefont {Jutten}\ and\ \citenamefont
  {Herault}(1991)}]{Jutten-Herault-1991}%
  \BibitemOpen
  \bibfield  {author} {\bibinfo {author} {\bibfnamefont {C.}\
  \bibnamefont {Jutten}}\ and\ \bibinfo {author} {\bibfnamefont {J.}\
  \bibnamefont {Herault}},\ }\bibfield  {title} {\enquote {\bibinfo {title}
  {Blind separation of sources, part I: An adaptive algorithm based on
  neuromimetic architecture},}\ }\href@noop {} {\bibfield  {journal} {\bibinfo
  {journal} {Signal Processing}\ }\textbf {\bibinfo {volume} {24}},\ \bibinfo
    {pages} {1--10} (\bibinfo {year} {1991})}\BibitemShut {NoStop}%
\bibitem [{\citenamefont {Hosenfeld}\ \emph {et~al.}(1997)\citenamefont
  {Hosenfeld}, \citenamefont {{van den Maas}},\ and\ \citenamefont {{van den
  Boom}}}]{Hosenfeld-etal-1997}%
  \BibitemOpen
  \bibfield  {author} {\bibinfo {author} {\bibfnamefont {B.}\ \bibnamefont
  {Hosenfeld}}, \bibinfo {author} {\bibfnamefont {H.~L.~J.}\ \bibnamefont
  {{van den Maas}}}, \ and\ \bibinfo {author} {\bibfnamefont {D.~C.}\
  \bibnamefont {{van den Boom}}},\ }\bibfield  {title} {\enquote {\bibinfo
  {title} {Indicators of discontinuous change in the development of analogical
  reasoning},}\ }\href@noop {} {\bibfield  {journal} {\bibinfo  {journal} {J.
  Exper. Child Psychol.}\ }\textbf {\bibinfo {volume} {64}},\ \bibinfo {pages}
    {367--395} (\bibinfo {year} {1997})}\BibitemShut {NoStop}%
\bibitem [{\citenamefont {Boshuizen}(2004)}]{Boshuizen-2004}%
  \BibitemOpen
  \bibfield  {author} {\bibinfo {author} {\bibfnamefont {H.~P.~A.}\
  \bibnamefont {Boshuizen}},\ }\enquote {\bibinfo {title} {Does practice make
  perfect? a slow and discontinuous process},}\ in\ \href@noop {} {\emph
  {\bibinfo {booktitle} {Professional Learning: Gaps and Transitions on the Way
  from Novice to Expert}}},\ \bibinfo {editor} {edited by\ \bibinfo {editor}
  {\bibfnamefont {H.~P.~A.}\ \bibnamefont {Boshuizen}}, \bibinfo {editor}
  {\bibfnamefont {R.}\ \bibnamefont {Bromme}}, \ and\ \bibinfo {editor}
  {\bibfnamefont {H.}\ \bibnamefont {Gruber}}}\ (\bibinfo  {publisher}
  {Kluwer Academic Publishers},\ \bibinfo {address} {New York},\ \bibinfo
  {year} {2004})\ Chap.~\bibinfo {chapter} {5}, pp.\ \bibinfo {pages}
  {73--96}\BibitemShut {NoStop}%
\bibitem [{\citenamefont {Collins}\ \emph {et~al.}(2019)\citenamefont
  {Collins}, \citenamefont {Regenbrecht}, \citenamefont {Langlotz},
  \citenamefont {Can}, \citenamefont {Ersoy},\ and\ \citenamefont
  {Butson}}]{Collins-etal-2019}%
  \BibitemOpen
  \bibfield  {author} {\bibinfo {author} {\bibfnamefont {J.}\ \bibnamefont
  {Collins}}, \bibinfo {author} {\bibfnamefont {H.}\ \bibnamefont
  {Regenbrecht}}, \bibinfo {author} {\bibfnamefont {T.}\ \bibnamefont
  {Langlotz}}, \bibinfo {author} {\bibfnamefont {Y.~S.}\ \bibnamefont
  {Can}}, \bibinfo {author} {\bibfnamefont {C.}\ \bibnamefont {Ersoy}}, \ and\
  \bibinfo {author} {\bibfnamefont {R.}\ \bibnamefont {Butson}},\
  }\bibfield  {title} {\enquote {\bibinfo {title} {Measuring cognitive load and
  insight: A methodology exemplified in a virtual reality learning context},}\
  }in\ \href {\doibase 10.1109/ISMAR.2019.00033} {\emph {\bibinfo {booktitle}
  {Proceedings of 2019 IEEE International Symposium on Mixed and Augmented
  Reality (ISMAR), Beijing, China}}}\ (\bibinfo {year} {2019})\ pp.\ \bibinfo
             {pages} {351--362}\BibitemShut {NoStop}%
\bibitem [{\citenamefont {Rosenberg}\ \emph {et~al.}(2021)\citenamefont
  {Rosenberg}, \citenamefont {Zhang}, \citenamefont {Perona},\ and\
  \citenamefont {Meister}}]{Rosenberg-etal-2021}%
  \BibitemOpen
  \bibfield  {author} {\bibinfo {author} {\bibfnamefont {M.}\ \bibnamefont
  {Rosenberg}}, \bibinfo {author} {\bibfnamefont {T.}\ \bibnamefont {Zhang}},
  \bibinfo {author} {\bibfnamefont {P.}\ \bibnamefont {Perona}}, \ and\
  \bibinfo {author} {\bibfnamefont {M.}\ \bibnamefont {Meister}},\
  }\bibfield  {title} {\enquote {\bibinfo {title} {Mice in a labyrinth show
  rapid learning, sudden insight, and efficient exploration},}\ }\href
  {\doibase 10.7554/eLife.66175} {\bibfield  {journal} {\bibinfo  {journal}
  {eLife}\ }\textbf {\bibinfo {volume} {10}},\ \bibinfo {pages} {e66175}
    (\bibinfo {year} {2021})}\BibitemShut {NoStop}%
 \bibitem [{\citenamefont {Bretas}\ \emph {et~al.}(2020)\citenamefont {Bretas},
  \citenamefont {Yamazaki},\ and\ \citenamefont {Iriki}}]{Bretas-etal-2020}%
  \BibitemOpen
  \bibfield  {author} {\bibinfo {author} {\bibfnamefont {R.~V.}\
  \bibnamefont {Bretas}}, \bibinfo {author} {\bibfnamefont {Y.}\
  \bibnamefont {Yamazaki}}, \ and\ \bibinfo {author} {\bibfnamefont {A.}\
  \bibnamefont {Iriki}},\ }\bibfield  {title} {\enquote {\bibinfo {title}
  {Phase transitions of brain evolution that produced human language and
  beyond},}\ }\href@noop {} {\bibfield  {journal} {\bibinfo  {journal}
  {Neurosci. Res.}\ }\textbf {\bibinfo {volume} {161}},\ \bibinfo
    {pages} {1--7} (\bibinfo {year} {2020})}\BibitemShut {NoStop}%
\bibitem [{\citenamefont {Ginsburg}\ and\ \citenamefont
  {Jablonka}(2021)}]{Ginsburg-Jablonka-2021}%
  \BibitemOpen
  \bibfield  {author} {\bibinfo {author} {\bibfnamefont {S.}\ \bibnamefont
  {Ginsburg}}\ and\ \bibinfo {author} {\bibfnamefont {E.}\ \bibnamefont
  {Jablonka}},\ }\bibfield  {title} {\enquote {\bibinfo {title} {Evolutionary
  transitions in learning and cognition},}\ }\href {\doibase
  10.1098/rstb.2019.0766} {\bibfield  {journal} {\bibinfo  {journal}
  {Phil. Trans. R. Soc. B}\ }\textbf {\bibinfo
  {volume} {376}},\ \bibinfo {pages} {20190766} (\bibinfo {year}
    {2021})}\BibitemShut {NoStop}%
\bibitem [{\citenamefont {Niven}(2016)}]{Niven-2016}%
  \BibitemOpen
  \bibfield  {author} {\bibinfo {author} {\bibfnamefont {J.~E.}\
  \bibnamefont {Niven}},\ }\bibfield  {title} {\enquote {\bibinfo {title}
  {Neuronal energy consumption: biophysics, efficiency and evolution},}\
  }\href@noop {} {\bibfield  {journal} {\bibinfo  {journal} {Curr. Opin.
  Neurobiol.}\ }\textbf {\bibinfo {volume} {41}},\ \bibinfo {pages} {129--135}
    (\bibinfo {year} {2016})}\BibitemShut {NoStop}%
\bibitem [{\citenamefont {Howarth}\ \emph {et~al.}(2012)\citenamefont
  {Howarth}, \citenamefont {Gleeson},\ and\ \citenamefont
  {Attwell}}]{Howarth-etal-2012}%
  \BibitemOpen
  \bibfield  {author} {\bibinfo {author} {\bibfnamefont {C.}\ \bibnamefont
  {Howarth}}, \bibinfo {author} {\bibfnamefont {P.}\ \bibnamefont
  {Gleeson}}, \ and\ \bibinfo {author} {\bibfnamefont {D.}\ \bibnamefont
  {Attwell}},\ }\bibfield  {title} {\enquote {\bibinfo {title} {Updated energy
  budgets for neural computation in the neocortex and cerebellum},}\
  }\href@noop {} {\bibfield  {journal} {\bibinfo  {journal} {J. Cereb. Blood
  Flow Metabol.}\ }\textbf {\bibinfo {volume} {32}},\ \bibinfo {pages}
    {1222--1232} (\bibinfo {year} {2012})}\BibitemShut {NoStop}%
\bibitem [{\citenamefont {Bell}\ and\ \citenamefont
  {Sejnowski}(1995)}]{Bell-Sejnowski-1995}%
  \BibitemOpen
  \bibfield  {author} {\bibinfo {author} {\bibfnamefont {A.~J.}\
  \bibnamefont {Bell}}\ and\ \bibinfo {author} {\bibfnamefont {T.~J.}\
  \bibnamefont {Sejnowski}},\ }\bibfield  {title} {\enquote {\bibinfo {title}
  {An information-maximization approach to blind separation and blind
  deconvolution},}\ }\href@noop {} {\bibfield  {journal} {\bibinfo  {journal}
  {Neural Comput.}\ }\textbf {\bibinfo {volume} {7}},\ \bibinfo {pages}
    {1129--1159} (\bibinfo {year} {1995})}\BibitemShut {NoStop}%
\bibitem [{\citenamefont {Seoane}\ and\ \citenamefont
  {Sol\'e}(2015)}]{Seoane-Sole-2015}%
  \BibitemOpen
  \bibfield  {author} {\bibinfo {author} {\bibfnamefont {L.~F.}\
  \bibnamefont {Seoane}}\ and\ \bibinfo {author} {\bibfnamefont {R.}\
  \bibnamefont {Sol\'e}},\ }\bibfield  {title} {\enquote {\bibinfo {title}
  {Phase transitions in pareto optimal complex networks},}\ }\href {\doibase
  10.1103/PhysRevE.92.032807} {\bibfield  {journal} {\bibinfo  {journal}
  {Phys. Rev. E}\ }\textbf {\bibinfo {volume} {92}},\ \bibinfo {pages}
    {032807} (\bibinfo {year} {2015})}\BibitemShut {NoStop}%
\bibitem [{\citenamefont {Qian}(2024)}]{Qian-2022}%
  \BibitemOpen
  \bibfield  {author} {\bibinfo {author} {\bibfnamefont {H.}\ \bibnamefont
  {Qian}},\ }\bibfield  {title} {\enquote {\bibinfo {title} {Internal energy,
  fundamental thermodynamic relation, and gibbs' ensemble theory as emergent
  laws of statistical counting},}\ }\href@noop {} {\bibfield  {journal}
  {\bibinfo  {journal} {Entropy}\ }\textbf {\bibinfo {volume} {26}},\ \bibinfo
  {pages} {1091} (\bibinfo {year} {2024})}\BibitemShut {NoStop}%
\bibitem [{\citenamefont {Koçillari}\ \emph {et~al.}(2018)\citenamefont
  {Koçillari}, \citenamefont {Fariselli}, \citenamefont {Trovato},
  \citenamefont {Seno},\ and\ \citenamefont {Maritan}}]{Kocillari-etal-2018}%
  \BibitemOpen
  \bibfield  {author} {\bibinfo {author} {\bibfnamefont {L.}\ \bibnamefont
  {Koçillari}}, \bibinfo {author} {\bibfnamefont {P.}\ \bibnamefont
  {Fariselli}}, \bibinfo {author} {\bibfnamefont {A.}\ \bibnamefont
  {Trovato}}, \bibinfo {author} {\bibfnamefont {F.}\ \bibnamefont {Seno}},
  \ and\ \bibinfo {author} {\bibfnamefont {A.}\ \bibnamefont {Maritan}},\
  }\bibfield  {title} {\enquote {\bibinfo {title} {Signature of pareto
  optimization in the escherichia coli proteome},}\ }\href@noop {} {\bibfield
  {journal} {\bibinfo  {journal} {Sci. Rep.}\ }\textbf {\bibinfo
  {volume} {8}},\ \bibinfo {pages} {9141} (\bibinfo {year} {2018})}\BibitemShut
             {NoStop}%
\bibitem [{\citenamefont {Wang}\ and\ \citenamefont
  {Lu}(2019)}]{Wang-Lu-2019}%
  \BibitemOpen
  \bibfield  {author} {\bibinfo {author} {\bibfnamefont {C.}\ \bibnamefont
  {Wang}}\ and\ \bibinfo {author} {\bibfnamefont {Y.~M.}\ \bibnamefont
  {Lu}},\ }\bibfield  {title} {\enquote {\bibinfo {title} {The scaling limit of
  high-dimensional online independent component analysis},}\ }\href {\doibase
  10.1088/1742-5468/ab39d6} {\bibfield  {journal} {\bibinfo  {journal} {J.
  Stat. Mech. Theor. Exp.}\ }\textbf {\bibinfo {volume}
  {2019}},\ \bibinfo {pages} {124011} (\bibinfo {year} {2019})}\BibitemShut
             {NoStop}%
\bibitem [{SI2()}]{SI2025a}%
  \BibitemOpen
  \href@noop {} {}\bibinfo {note} {Online supplementary information
  accompanying this work. It contains addtional technical details of the
  theoretical derivation, and additional numerical data to support the main
  conclusions of this work.}\BibitemShut {Stop}%
\bibitem [{\citenamefont {Lan}\ \emph {et~al.}(2012)\citenamefont {Lan},
  \citenamefont {Sartori}, \citenamefont {Neumann}, \citenamefont {Sourjik},\
  and\ \citenamefont {Tu}}]{Lan-etal-2012}%
  \BibitemOpen
  \bibfield  {author} {\bibinfo {author} {\bibfnamefont {G.}\ \bibnamefont
  {Lan}}, \bibinfo {author} {\bibfnamefont {P.}\ \bibnamefont {Sartori}},
  \bibinfo {author} {\bibfnamefont {S.}\ \bibnamefont {Neumann}}, \bibinfo
  {author} {\bibfnamefont {V.}\ \bibnamefont {Sourjik}}, \ and\ \bibinfo
  {author} {\bibfnamefont {Y.}\ \bibnamefont {Tu}},\ }\bibfield  {title}
  {\enquote {\bibinfo {title} {The energy-speed-accuracy trade-off in sensory
  adaptation},}\ } {\bibfield  {journal}
  {\bibinfo  {journal} {Nature Phys.}\ }\textbf {\bibinfo {volume} {8}},\
  \bibinfo {pages} {422--428} (\bibinfo {year} {2012})}\BibitemShut {NoStop}%
\bibitem [{\citenamefont {Nicoletti}\ and\ \citenamefont
  {Busiello}(2024)}]{Nicoletti-Busiello-2024}%
  \BibitemOpen
  \bibfield  {author} {\bibinfo {author} {\bibfnamefont {G.}\ \bibnamefont
  {Nicoletti}}\ and\ \bibinfo {author} {\bibfnamefont {D.~M.}\
  \bibnamefont {Busiello}},\ }\bibfield  {title} {\enquote {\bibinfo {title}
  {Tuning transduction from hidden observables to optimize information
  harvesting},}\ }\href {\doibase 10.1103/PhysRevLett.133.158401} {\bibfield
  {journal} {\bibinfo  {journal} {Phys. Rev. Lett.}\ }\textbf {\bibinfo
  {volume} {133}},\ \bibinfo {pages} {158401} (\bibinfo {year}
  {2024})}\BibitemShut {NoStop}%
\bibitem [{\citenamefont {Olsen}\ \emph {et~al.}(2024)\citenamefont {Olsen},
  \citenamefont {Gupta}, \citenamefont {Mori},\ and\ \citenamefont
  {Krishnamurthy}}]{Olsen-etal-2024}%
  \BibitemOpen
  \bibfield  {author} {\bibinfo {author} {\bibfnamefont {K.~S.}\
  \bibnamefont {Olsen}}, \bibinfo {author} {\bibfnamefont {D.}\
  \bibnamefont {Gupta}}, \bibinfo {author} {\bibfnamefont {F.}\
  \bibnamefont {Mori}}, \ and\ \bibinfo {author} {\bibfnamefont {S.}\
  \bibnamefont {Krishnamurthy}},\ }\bibfield  {title} {\enquote {\bibinfo
  {title} {Thermodynamic cost of finite-time stochastic resetting},}\ }\href
  {\doibase 10.1103/PhysRevResearch.6.033343} {\bibfield  {journal} {\bibinfo
  {journal} {Phys. Rev. Res.}\ }\textbf {\bibinfo {volume} {6}},\ \bibinfo
    {pages} {033343} (\bibinfo {year} {2024})}\BibitemShut {NoStop}%
\bibitem [{\citenamefont {Sompolinsky}\ \emph {et~al.}(1988)\citenamefont
  {Sompolinsky}, \citenamefont {Crisanti},\ and\ \citenamefont
  {Sommers}}]{Sompolinsky-etal-1988}%
  \BibitemOpen
  \bibfield  {author} {\bibinfo {author} {\bibfnamefont {H.}~\bibnamefont
  {Sompolinsky}}, \bibinfo {author} {\bibfnamefont {A.}~\bibnamefont
  {Crisanti}}, \ and\ \bibinfo {author} {\bibfnamefont {H.~J.}\ \bibnamefont
  {Sommers}},\ }\bibfield  {title} {\enquote {\bibinfo {title} {Chaos in random
  neural networks},}\ }\href@noop {} {\bibfield  {journal} {\bibinfo  {journal}
  {Phys. Rev. Lett.}\ }\textbf {\bibinfo {volume} {61}},\ \bibinfo {pages}
    {259--262} (\bibinfo {year} {1988})}\BibitemShut {NoStop}%
\bibitem [{\citenamefont {Qiu}\ and\ \citenamefont
  {Huang}(2024)}]{Qiu-Huang-2024}%
  \BibitemOpen
  \bibfield  {author} {\bibinfo {author} {\bibfnamefont {J.}\ \bibnamefont
  {Qiu}}\ and\ \bibinfo {author} {\bibfnamefont {H.}\ \bibnamefont
  {Huang}},\ }\bibfield  {title} {\enquote {\bibinfo {title} {An
  optimization-based equilibrium measure describing fixed points of
  non-equilibrium dynamics: application to the edge of chaos},}\ }\href
  {\doibase 10.1088/1572-9494/ad8126} {\bibfield  {journal} {\bibinfo
  {journal} {Commun. Theor. Phys.}\ }\textbf {\bibinfo {volume} {77}},\
    \bibinfo {pages} {035601} (\bibinfo {year} {2024})}\BibitemShut {NoStop}%
\bibitem [{\citenamefont {Calvo}\ \emph {et~al.}(2024)\citenamefont {Calvo},
  \citenamefont {Martorell}, \citenamefont {Morales}, \citenamefont {{Di
  Santo}},\ and\ \citenamefont {Mu{\~n}oz}}]{Calvo-etal-2024}%
  \BibitemOpen
  \bibfield  {author} {\bibinfo {author} {\bibfnamefont {R.}\ \bibnamefont
  {Calvo}}, \bibinfo {author} {\bibfnamefont {C.}\ \bibnamefont
  {Martorell}}, \bibinfo {author} {\bibfnamefont {G.~B.}\ \bibnamefont
  {Morales}}, \bibinfo {author} {\bibfnamefont {S.}\ \bibnamefont {{Di
  Santo}}}, \ and\ \bibinfo {author} {\bibfnamefont {M.~A.}\ \bibnamefont
  {Mu{\~n}oz}},\ }\bibfield  {title} {\enquote {\bibinfo {title}
  {Frequency-dependent covariance reveals critical spatiotemporal patterns of
  synchronized activity in the human brain},}\ }\href@noop {} {\bibfield
  {journal} {\bibinfo  {journal} {Phys. Rev. Lett.}\ }\textbf {\bibinfo
  {volume} {133}},\ \bibinfo {pages} {208401} (\bibinfo {year}
  {2024})}\BibitemShut {NoStop}%
\bibitem [{\citenamefont {Safavi}\ \emph {et~al.}(2024)\citenamefont {Safavi},
  \citenamefont {Chalk}, \citenamefont {Logothetis},\ and\ \citenamefont
  {Levina}}]{Safavi-etal-2024}%
  \BibitemOpen
  \bibfield  {author} {\bibinfo {author} {\bibfnamefont {S.}\ \bibnamefont
  {Safavi}}, \bibinfo {author} {\bibfnamefont {M.}\ \bibnamefont {Chalk}},
  \bibinfo {author} {\bibfnamefont {N.~K.}\ \bibnamefont {Logothetis}}, \
  and\ \bibinfo {author} {\bibfnamefont {A.}\ \bibnamefont {Levina}},\
  }\bibfield  {title} {\enquote {\bibinfo {title} {Signatures of criticality in
  efficient coding networks},}\ }\href {\doibase 10.1073/pnas.2302730121}
  {\bibfield  {journal} {\bibinfo  {journal} {Proc. Natl. Acad. Sci. USA}\
  }\textbf {\bibinfo {volume} {121}},\ \bibinfo {pages} {e2302730121} (\bibinfo
    {year} {2024})}\BibitemShut {NoStop}%
\bibitem [{\citenamefont {Barzon}\ \emph {et~al.}(2025)\citenamefont {Barzon},
  \citenamefont {Busiello},\ and\ \citenamefont
  {Nicoletti}}]{Barzon-etal-2025}%
  \BibitemOpen
  \bibfield  {author} {\bibinfo {author} {\bibfnamefont {G.}\ \bibnamefont
  {Barzon}}, \bibinfo {author} {\bibfnamefont {D.~M.}\ \bibnamefont
  {Busiello}}, \ and\ \bibinfo {author} {\bibfnamefont {G.}\ \bibnamefont
  {Nicoletti}},\ }\bibfield  {title} {\enquote {\bibinfo {title}
  {Excitation-inhibition balance controls information encoding in neural
  populations},}\ }\href {\doibase 10.1103/PhysRevLett.134.068403} {\bibfield
  {journal} {\bibinfo  {journal} {Phys. Rev. Lett.}\ }\textbf {\bibinfo
  {volume} {134}},\ \bibinfo {pages} {068403} (\bibinfo {year}
  {2025})}\BibitemShut {NoStop}%
\bibitem [{\citenamefont {Hyv\"arinen}\ and\ \citenamefont
  {Oja}(2000)}]{Hyvarinen-Oja-2000}%
  \BibitemOpen
  \bibfield  {author} {\bibinfo {author} {\bibfnamefont {A.}~\bibnamefont
  {Hyv\"arinen}}\ and\ \bibinfo {author} {\bibfnamefont {E.}~\bibnamefont
  {Oja}},\ }\bibfield  {title} {\enquote {\bibinfo {title} {Independent
  component analysis: algorithms and applications},}\ }\href@noop {} {\bibfield
   {journal} {\bibinfo  {journal} {Neural Networks}\ }\textbf {\bibinfo
  {volume} {13}},\ \bibinfo {pages} {411--430} (\bibinfo {year}
   {2000})}\BibitemShut {NoStop}%
\bibitem [{\citenamefont {Yoshino}(2020)}]{Yoshino-2020}%
  \BibitemOpen
  \bibfield  {author} {\bibinfo {author} {\bibfnamefont {H.}~\bibnamefont
  {Yoshino}},\ }\bibfield  {title} {\enquote {\bibinfo {title} {From complex to
  simple: hierarchical free-energy landscape renormalized in deep neural
  networks},}\ }\href@noop {} {\bibfield  {journal} {\bibinfo  {journal}
  {SciPost Phys. Core}\ }\textbf {\bibinfo {volume} {2}},\ \bibinfo {pages}
    {005} (\bibinfo {year} {2020})}\BibitemShut {NoStop}%
\bibitem [{\citenamefont {Wu}\ and\ \citenamefont
  {Fischer}(2020)}]{Wu-Fischer-2020}%
  \BibitemOpen
  \bibfield  {author} {\bibinfo {author} {\bibfnamefont {T.}\ \bibnamefont
  {Wu}}\ and\ \bibinfo {author} {\bibfnamefont {I.}\ \bibnamefont {Fischer}},\
  }\bibfield  {title} {\enquote {\bibinfo {title} {Phase transitions for the
  information bottleneck in representation learning},}\ }in\ \href
  {https://openreview.net/forum?id=HJloElBYvB} {\emph {\bibinfo {booktitle}
  {International Conference on Learning Representations}}}\ (\bibinfo {year}
  {2020})\BibitemShut {NoStop}%
\bibitem [{\citenamefont {Sokolowski}\ \emph {et~al.}(2025)\citenamefont
  {Sokolowski}, \citenamefont {Gregor}, \citenamefont {Bialek},\ and\
  \citenamefont {Tkačik}}]{Sokolowski-etal-2025}%
  \BibitemOpen
  \bibfield  {author} {\bibinfo {author} {\bibfnamefont {T.~R.}\
  \bibnamefont {Sokolowski}}, \bibinfo {author} {\bibfnamefont {T.}\
  \bibnamefont {Gregor}}, \bibinfo {author} {\bibfnamefont {W.}\
  \bibnamefont {Bialek}}, \ and\ \bibinfo {author} {\bibfnamefont {G.}\
  \bibnamefont {Tkačik}},\ }\bibfield  {title} {\enquote {\bibinfo {title}
  {Deriving a genetic regulatory network from an optimization principle},}\
  }\href {\doibase 10.1073/pnas.2402925121} {\bibfield  {journal} {\bibinfo
  {journal} {Proc. Natl. Acad. Sci. USA}\ }\textbf {\bibinfo {volume} {122}},\
  \bibinfo {pages} {e2402925121} (\bibinfo {year} {2025})}\BibitemShut
             {NoStop}%
\bibitem [{\citenamefont {Tatsukawa}\ and\ \citenamefont {nosuke
  Teramae}(2024)}]{Tatsukawa-Teramae-2025}%
  \BibitemOpen
  \bibfield  {author} {\bibinfo {author} {\bibfnamefont {T.}\
  \bibnamefont {Tatsukawa}}\ and\ \bibinfo {author} {\bibfnamefont {J.-n.}\
  \bibnamefont {Teramae}},\ }\href@noop {} {\enquote {\bibinfo {title}
  {Energy-information trade-off makes the cortical critical power law the
  optimal coding},}\ }\bibinfo {howpublished} {eprint arXiv:2407.16215
    [q-bio.NC]} (\bibinfo {year} {2024})\BibitemShut {NoStop}%
\bibitem [{\citenamefont {Wu}\ \emph {et~al.}(2025)\citenamefont {Wu},
  \citenamefont {Huang}, \citenamefont {Wang}, \citenamefont {Chen},
  \citenamefont {Zhou},\ and\ \citenamefont {Yang}}]{Wu-etal-2025}%
  \BibitemOpen
  \bibfield  {author} {\bibinfo {author} {\bibfnamefont {S.}\
  \bibnamefont {Wu}}, \bibinfo {author} {\bibfnamefont {H.}\ \bibnamefont
  {Huang}}, \bibinfo {author} {\bibfnamefont {S.}\ \bibnamefont {Wang}},
  \bibinfo {author} {\bibfnamefont {G.}\ \bibnamefont {Chen}}, \bibinfo
  {author} {\bibfnamefont {C.}\ \bibnamefont {Zhou}}, \ and\ \bibinfo
  {author} {\bibfnamefont {D.}\ \bibnamefont {Yang}},\ }\bibfield
  {title} {\enquote {\bibinfo {title} {Neural heterogeneity enhances reliable
  neural information processing: Local sensitivity and globally input-slaved
  transient dynamics},}\ }\href {\doibase 10.1126/sciadv.adr3903} {\bibfield
  {journal} {\bibinfo  {journal} {Science Adv.}\ }\textbf {\bibinfo
  {volume} {11}},\ \bibinfo {pages} {eadr3903} (\bibinfo {year}
  {2025})}\BibitemShut {NoStop}%
\end{thebibliography}

%

\clearpage

\onecolumngrid

\setcounter{equation}{0}
\setcounter{section}{0}
\setcounter{figure}{0}
\setcounter{table}{0}
\setcounter{page}{1}

\renewcommand{\theequation}{S\arabic{equation}}
\renewcommand{\thesection}{S\arabic{section}}  
\renewcommand{\thesubsection}{S\arabic{section}.\arabic{subsection}}

\renewcommand{\thefigure}{S\arabic{figure}}

\renewcommand{\bibnumfmt}[1]{[S#1]}
\renewcommand{\citenumfont}[1]{#1}

\begin{center}
    \textbf{\large Discontinuous phase transitions of feature detection in lateral predictive coding}
\end{center}
\vskip 0.3cm
\begin{center}
  Zhen-Ye Huang, Weikang Wang, and Hai-Jun Zhou
\end{center}

\vskip 0.5cm

\begin{center}
{\large Supplementary Information}
\end{center}

\vskip 1.0cm

To simplify the notation, we will use lower-case bold form to denote a real-valued column vector. Some examples are the input signal $\bm{s} = (s_1, s_2, \ldots, s_N)^\top$ and the output signal (internal state vector)  $\bm{x} = (x_1, x_2, \ldots, x_N)^\top$. Notice that such vectors are denoted as $\vec{\bm{s}}$ and $\vec{\bm{x}}$ in the main text.

 \clearpage
 
\section{Entropy of the output signal}

This supplementary section is an expanded version of Appendix A of the main text.

Let us denote by $p_{\textrm{in}}(\bm{s})$ the probability distribution of the input signal $\bm{s}$. The marginal probability distribution $p_{\textrm{out}}(\bm{x})$ of the output signal $\bm{x}$ is then
\begin{equation}
  p_{\textrm{out}}(\bm{x}) \, =  \, \int \textrm{d} \bm{s} p_{\textrm{in}}(\bm{s}) \,
  \delta\Bigl(\bm{x} - (\bm{I}+\bm{W})^{-1} \bm{s} \Bigr) \; ,
  \label{eq:Qxdef}
\end{equation}
where $\delta(\bm{x})$ denotes the Dirac delta function,  which is $\delta(\bm{x}) \equiv \prod_{i=1}^{N} \delta( x_i)$ for a real vector $\bm{x} = (x_1, \ldots, x_N)^\top$.  A convenient alternative form for this delta function is
\begin{equation}
  \delta( \bm{x} ) \, = \, \lim\limits_{\sigma_0\rightarrow 0}
  \frac{1}{(2 \pi \sigma_0^2)^{N/2}} \exp\Bigl[ - \frac{\bm{x}^2}{2 \sigma_0^2} \Bigr] \; ,
  \label{eqdelta0}
\end{equation}
where $\sigma_0$ is the standard deviation of a random Gaussian noise. Then we can rewrite Eq.~(\ref{eq:Qxdef}) as
\begin{equation}
  \begin{aligned}
    p_{\textrm{out}}(\bm{x}) \, & = \, \lim\limits_{\sigma_0 \rightarrow 0}
    \frac{1}{(2 \pi \sigma_0^2)^{N/2}}
    \int \textrm{d} \bm{s} \,  p_{\textrm{in}}(\bm{s})
    \exp\Bigl[ - \frac{\bigl( \bm{x} -
        (\bm{I}+\bm{W})^{-1} \bm{s}\bigr)^2}{2 \sigma_0^2} \Bigr]  \\
    & = \,  \lim\limits_{\sigma_0 \rightarrow 0}
    \frac{1}{(2\pi \sigma_0^2)^{N/2}} \int \textrm{d} \bm{s} \,  p_{\textrm{in}}(\bm{s}) 
    \exp\Bigl[ - \frac{\bm{x}^2}{2 \sigma_0^2}
      -\frac{1}{2 \sigma_0^2} \bm{s}^\top \frac{\bm{I}}{(\bm{I}+\bm{W})^\top}
      \frac{\bm{I}}{(\bm{I}+\bm{W})}\, \bm{s}
      + \frac{2}{2 \sigma_0^2} \bm{s}^\top
      \frac{\bm{I}}{(\bm{I}+\bm{W})^\top} \, \bm{x} \Bigr] \; .
  \end{aligned}
  \label{eqQxexp1}
\end{equation}

To simplify this expression, let us perform the following eigen-decomposition:
\begin{equation}
  \frac{\bm{I}}{(\bm{I}+\bm{W})^\top} \frac{\bm{I}}{(\bm{I}+\bm{W})}
 \,  = \,  \bm{U}\,  \textrm{Diag}\Bigl[ \frac{1}{\lambda_1}, 
  \frac{1}{\lambda_2}, \ldots, \frac{1}{\lambda_N}\Bigr] \, \bm{U}^\top \; ,
\end{equation}
where $\lambda_1, \ldots, \lambda_N$ are the $N$ eigenvalues of the symmetric real matrix $(\bm{I}+\bm{W}) (\bm{I}+\bm{W})^\top$ and the matrix $\bm{U}$ are formed by the $N$ corresponding eigenvectors. Notice that $\bm{U}$ is an orthogonal matrix, so we have $\bm{U} \bm{U}^\top = \bm{U}^\top \bm{U} = \bm{I}$, and $\bigl| \textrm{det}( \bm{U} ) \bigr| = 1$.  Let us introduce an auxiliary vector $\bm{z}$ as
\begin{equation}
  \bm{z} \, = \,  \bm{U}^\top\frac{\bm{I}}{(\bm{I}+\bm{W})^\top}\, \bm{x} \; .
  \label{eqzvec}
\end{equation}
We notice that
\begin{equation}
  \begin{aligned}
    \sum_j \lambda_j z_j^2
    \, & = \, \textrm{Tr}\Bigl[ \bm{x}^\top
      \frac{\bm{I}}{(\bm{I}+\bm{W})}\, \bm{U}\,
      \textrm{Diag}\bigl(\lambda_1, \ldots, \lambda_N\bigr)\,
      \bm{U}^\top \frac{\bm{I}}{(\bm{I}+\bm{W})^\top} \, \bm{x}
      \Bigr] \\
   & = \, \textrm{Tr} \Bigl[ \bm{x}^\top
      \frac{\bm{I}}{(\bm{I}+\bm{W})} (\bm{I} + \bm{W}) (\bm{I}+\bm{W})^\top
      \frac{\bm{I}}{(\bm{I}+\bm{W})^\top} \, \bm{x}
      \Bigr] \\
    & =  \, \textrm{Tr} \bigl[ \bm{x}^\top \bm{x} \bigr] \, = \, \sum_j x_j^2 \; ,
  \end{aligned}
\end{equation}
It is also easy to prove that
\begin{equation}
  \bm{U} \,  \textrm{Diag}\bigl[ \lambda_1, \lambda_2, \ldots, \lambda_N\bigr]
  \bm{z} \, = \, (\bm{I}+\bm{W}) \bm{x} \; ,
    \label{eqUlz}
\end{equation}
simply by replacing $\bm{z}$ by the expression of Eq.~(\ref{eqzvec}). Let us make the transform
\begin{equation}
  \bm{y}  \, = \, \bm{U}^\top \bm{s} \; ,  \quad \quad 
  \bm{s}  \, = \, \bm{U} \bm{y} \; .
\end{equation}
Then Eq.~(\ref{eqQxexp1}) is rewritten as
\begin{equation}
\begin{aligned}
  p_{\textrm{out}}(\bm{x}) \, & = \, \lim\limits_{\sigma_0\rightarrow 0}
  \frac{1}{(2 \pi \sigma_0^2)^{N/2}} \int \textrm{d} \bm{y}\, p_{\textrm{in}}( \bm{U} \bm{y} )
  \exp\Bigl[
    - \frac{\bm{x}^2}{2 \sigma_0^2} - \sum\limits_{j}
    \frac{(y_j - \lambda_j z_j)^2}{2 \lambda_j \sigma_0^2} +
    \sum_j \frac{\lambda_j z_j^2}{2 \sigma_0^2}
    \Bigr] \\
  & =  \, \lim\limits_{\sigma_0\rightarrow 0}
  \sqrt{\lambda_1 \lambda_2 \ldots \lambda_N} \int \textrm{d} \bm{y} \,
  p_{\textrm{in}} ( \bm{U} \bm{y} ) 
  \prod\limits_j \frac{\exp\bigl[-(y_j - \lambda_j z_j)^2/
      (2 \lambda_j \sigma_0^2) \bigr]}{\sqrt{ 2 \pi \sigma_0^2 \lambda_j}} \\
  & =  \,  \sqrt{\lambda_1 \lambda_2 \ldots \lambda_N} \int \textrm{d} \bm{y} \,
  p_{\textrm{in}}( \bm{U} \bm{y} ) \prod\limits_{j} \delta\bigl( y_j - \lambda_j z_j \bigr)
  \\
  & =  \, \sqrt{\lambda_1 \lambda_2 \ldots \lambda_N} \, 
  p_{\textrm{in}}\Bigl( \bm{U}\, \textrm{Diag}\bigl[ \lambda_1, \ldots, \lambda_N \bigr] \, \bm{z} \Bigr)
  \\ 
  & = \,   \sqrt{\lambda_1 \lambda_2 \ldots \lambda_N} \, 
  p_{\textrm{in}}\Bigl( (\bm{I}+\bm{W}) \, \bm{x} \Bigr) \; .
\end{aligned}
\label{eqQexp2}
\end{equation}

From the last line of Eq.~(\ref{eqQexp2}) we obtain the desired result that
\begin{equation}
  p_{\textrm{out}}(\bm{x}) \, = \, \bigl| \textrm{det} ( \bm{I} + \bm{W}) \bigr| \, p_{\textrm{in}}\bigl( \bm{s} \bigr)\;  \quad\quad \textrm{with} \quad \bm{s} \, = \, (\bm{I} + \bm{W} )\, \bm{x} \; .    
\end{equation}
The entropy of the output signal $\bm{x}$ is then
\begin{equation}
  \begin{aligned}
    H\bigl[ p_{\textrm{out}}(\bm{x}) \bigr] &
    \equiv - \int \textrm{d} \bm{x} \,  p_{\textrm{out}}(\bm{x}) \ln p_{\textrm{out}}(\bm{x} ) \\
    & = - \int \textrm{d} \bm{x}\, p_{\textrm{out}}(\bm{x} )
    \ln \Bigl( \bigl| \textrm{det}( \bm{I} + \bm{W}) \bigr| \Bigr)
    -  \int \textrm{d} \bm{x}\, \bigl| \textrm{det} (
    \bm{I} + \bm{W} ) \bigr| \, p_{\textrm{in}}\bigl( (\bm{I} + \bm{W}) \bm{x} \bigr)
    \ln p_{\textrm{in}} \bigl( (\bm{I} + \bm{W} ) \bm{x} \bigr) \\
    & = -   \ln \Bigl( \bigl| \textrm{det}( \bm{I} + \bm{W}) \bigr| \Bigr)
    -  \int \textrm{d} \bm{s}\, p_{\textrm{in}}( \bm{s} ) \ln p_{\textrm{in}}( \bm{s} ) \\
    & =  - \ln \Bigl( \bigl| \textrm{det}( \bm{I} + \bm{W}) \bigr| \Bigr)
    + H\bigl[ p_{\textrm{in}}(\bm{s}) \bigr] \; ,
    \end{aligned}
    \label{eqHx}
\end{equation}
where $H\bigl[ p_{\textrm{in}}(\bm{s}) \bigr]$ is the entropy of the input signal $\bm{s}$. Since $H\bigl[ p_{\textrm{in}}(\bm{s}) \bigr]$ is a constant independent of the weight matrix $\bm{W}$, the entropy difference $H\bigl[ p_{\textrm{out}}(\bm{x}) \bigr] - H\bigl[ p_{\textrm{in}}(\bm{s} ) \bigr]$  is referred to simply as the entropy of the output distribution $p_{\textrm{out}}(\bm{x})$ and is denoted as $S$:
\begin{equation}
  S \, \equiv \, -  \ln \Bigl[ \bigl| \textrm{det}( \bm{I} + \bm{W}) \bigr|
    \Bigr]  \; .
\end{equation}

\clearpage

\section{Explicit analytical expression for the mean energy cost}

This supplementary section is an expanded version of Appendix D and Appendix E of the main text.

First, we list some basic results concerning Gaussian random variables. The Gaussian (normal) distribution for a real variable $x$ is
\begin{equation}
  p(x)\, = \, \frac{1}{\sqrt{2 \pi \sigma^2}}
  \exp\Bigl( - \frac{x^2}{2 \sigma^2} \Bigr) \; .
\end{equation}
The mean value of such a Gaussian variable is zero and its variance is $\sigma^2$.  The mean of the absolute value $|x|$ is
\begin{equation}
    \bigl\langle \, | x | \, \bigr\rangle \,
    \equiv \, \int_{-\infty}^{\infty} p(x) |x|\, \textrm{d} x 
    \, = \,  2 \int_{0}^\infty
    \frac{x}{\sqrt{2 \pi \sigma^2}} \exp\Bigl( - \frac{x^2}{2 \sigma^2} \Bigr)
    \, \textrm{d} x  \,    = \, \sqrt{ \frac{2 \sigma^2}{\pi} } \; .
\end{equation}

The Gaussian distribution of a random real variable $x$ with positive mean $x_0$ ($> 0$) and variance $\sigma^2$ is
\begin{equation}
  p(x) \, = \, \frac{1}{\sqrt{2 \pi \sigma^2}}
  \exp\Bigl( - \frac{(x -x_0) ^2}{2 \sigma^2} \Bigr) \; .
\end{equation}
The mean value of $|x|$ is
\begin{equation}
  \begin{aligned}
    \bigl\langle \, | x | \, \bigr\rangle & = \, \int_{-x_0}^\infty
    \frac{x_0 + \Delta}{\sqrt{2 \pi \sigma^2}}
    \exp\Bigl( - \frac{\Delta^2}{2 \sigma^2} \Bigr)\, \textrm{d} \Delta 
    \, + \, \int_{x_0}^\infty \frac{- x_0 + \Delta}{\sqrt{2 \pi \sigma^2}}
    \exp\Bigl( - \frac{\Delta^2}{2 \sigma^2} \Bigr)\, \textrm{d} \Delta \\
    & =  \sqrt{ \frac{2 \sigma^2}{\pi} } e^{- x_0^2/(2 \sigma^2)}  +
    \frac{ 2 x_0}{\sqrt{\pi}} \int_{0}^{\frac{x_0}{\sqrt{2 \sigma^2}}} e^{-y^2}  
    \textrm{d} y \\
    & = \,  \sqrt{ \frac{2 \sigma^2}{\pi} }
    \exp\Bigl( - \frac{x_0^2}{2 \sigma^2} \Bigr) \, + \,
    x_0 \, \textrm{erf}\Bigl(  \frac{x_0}{\sqrt{2 \sigma^2}} \Bigr) \; ,
  \end{aligned}
\end{equation}
where $\textrm{erf}(x)$ is the error function defined by
\begin{equation}
  \textrm{erf}(x) \, = \,
  \frac{2}{\sqrt{\pi}} \int_0^x e^{-t^2}\, \textrm{d} t \; .
\end{equation}

Second, we derive the explicit expression for the conditional probability distribution of an output signal. The output signal vector $\bm{x}$ is expressed as
\begin{equation}
  \begin{aligned}
    \bm{x} & = a_1  \frac{\bm{I}}{\bm{I}+\bm{W}} \bm{\phi}_1
    + \sum\limits_{j=2}^{N} b_j \frac{\bm{I}}{\bm{I}+\bm{W}} \bm{\phi}_j \\
    & = a_1 \bm{\mu} + \sum\limits_{j\geq 2} b_j \bm{\psi}_j \; ,
  \end{aligned}
\end{equation}
where the output vector $\bm{\mu} \equiv (\mu_1, \ldots, \mu_N)^\top$ and $\bm{\psi}_j$ ($j\geq 2$) are, respectively,  the transform of $\bm{\phi}_1$ and $\bm{\phi}_j$:
\begin{equation}
\bm{\mu} \, = \,  \frac{\bm{I}}{\bm{I}+\bm{W}} \bm{\phi}_1 \; , \quad \quad
  \bm{\psi}_j = \frac{\bm{I}}{\bm{I}+\bm{W}} \bm{\phi}_j \quad \quad (j=2, \ldots, N)  \; .
\end{equation}
Since all the coefficients $b_j$ with indices $j = 2, \ldots, N$ are independent Gaussian random variables with zero mean and unit variance, the conditional mean vector of $\bm{x}$ at fixed value of the non-Gaussian coefficient $a_1$ is simply
\begin{equation}
  \langle \bm{x} \rangle  \, = \, a_1 \bm{\mu} \; .
\end{equation}
The second-moment matrix of $\bm{x}$ at fixed $a_1$ is
\begin{equation}
  \begin{aligned}
    \bigl\langle \bm{x} \bm{x}^\top \bigr\rangle 
    & = \, a_1^2  \frac{\bm{I}}{\bm{I}+\bm{W}}  \bm{\phi}_1 \bm{\phi}_1^\top
    \frac{\bm{I}}{(\bm{I}+\bm{W})^\top} 
    + \sum\limits_{j=2}^{N} \frac{\bm{I}}{\bm{I}+\bm{W}}
    \bm{\phi}_j \bm{\phi}_j^\top \frac{\bm{I}}{(\bm{I}+\bm{W})^\top} \\
    & = (a_1^2 - 1)  \frac{\bm{I}}{\bm{I}+\bm{W}}  \bm{\phi}_1 \bm{\phi}_1^\top
    \frac{\bm{I}}{(\bm{I}+\bm{W})^\top} \, + \, 
    \sum\limits_{j=1}^{N} \frac{\bm{I}}{\bm{I}+\bm{W}}
    \bm{\phi}_j \bm{\phi}_j^\top \frac{\bm{I}}{(\bm{I}+\bm{W})^\top} \\
    & = \, (a_1^2-1) \, \bm{\mu} \bm{\mu}^\top  + 
    \frac{\bm{I}}{(\bm{I}+\bm{W})} \frac{\bm{I}}{(\bm{I}+\bm{W})^\top} \; .
  \end{aligned}
  \label{eqxxt}
\end{equation}
In deriving the last line of the above equation, we have used the property that, for $N$ mutually orthogonal vectors $\bm{\phi}_j$, the following identity holds:
\begin{equation}
  \sum\limits_{j=1}^N \bm{\phi}_j \bm{\phi}_j^\top \, = \, \bm{I} \; .
\end{equation}

At fixed value of the non-Gaussian coefficient $a_1$, the conditional distribution of the $i$-th element $x_i$ of the output vector $\bm{x}$ is a Gaussian distribution with mean $a_1 \mu_i$ and variance $\sigma_i^2$:
\begin{equation}
  p_{\textrm{out}}\bigl(x_i | a_1 \bigr) \, = \, \frac{1}{\sqrt{2 \pi \sigma_i^2}}
  \exp\Bigl( - \frac{(x_i - a_1 \mu_i)^2}{2 \sigma_i^2} \Bigr) \; ,
  \label{eq241217a}
\end{equation}
and $\mu_i$ and $\sigma_i^2$ are computed through
\begin{equation}
  \mu_i \,  =  \,
  \Bigl[ \frac{\bm{I}}{\bm{I}+\bm{W}} \bm{\phi}_1 \Bigr]_{i} \; ,
  \quad \quad 
  \sigma_i^2 \,  =  \,
  \Bigl[ \frac{\bm{I}}{(\bm{I}+\bm{W})} \frac{\bm{I}}{(\bm{I}+\bm{W})^\top}
    \Bigr]_{i i} - \mu_i^2 \; .
\end{equation}
The signal-to-noise ratio $\eta_i$ of the conditional distribution (\ref{eq241217a}) can be defined by the ratio between the mean and the standard deviation, namely
\begin{equation}
  \eta_i \, \equiv \, \frac{ |a_1 \mu_i|}{\sqrt{\sigma_i^2}}
  \, = \, \sqrt{\frac{a_1^2 \mu_i^2}{\sigma_i^2}} \; .
  \label{eq241217b}
\end{equation}

Finally, with these preparations, we can derive the analytical expression for the mean $L_1$-norm energy as
\begin{equation}
  \begin{aligned}
    E & \, = \, \sum\limits_{i=1}^{N} \bigl\langle \, | x_i | \, \bigr\rangle 
    \, = \,    \int \textrm{d} a_1 \, q( a_1) \,
    \sum\limits_{i=1}^{N} \int_{-\infty}^\infty \frac{| x_i |}{\sqrt{2 \pi \sigma_i^2}}
    \exp\Bigl( - \frac{(x_i - a_1 \mu_i)^2}{2 \sigma_i^2} \Bigr)
    \, \textrm{d} x_i \\
    & \, = \,   \sum\limits_{i=1}^{N}  \int \textrm{d} a_1 \, q( a_1) \,
    \biggl[ \sqrt{\frac{2 \sigma_i^2}{\pi}} \,
      \exp\Bigl( - \frac{a_1^2 \mu_i^2}{2 \sigma_i^2} \Bigr)
      \, + \, |a_1 \mu_i |\, \textrm{erf}\Bigl(
      \frac{|a_1 \mu_i|}{\sqrt{2 \sigma_i^2}} \Bigr) \biggr] \; .
  \end{aligned}
  \label{eqL1Eexp}
\end{equation}

As one concrete example, we consider the following discrete distribution for the non-Gaussian coefficient $a_1$:
\begin{equation}
  q( a_1 ) \, = \,  \left\{
  \begin{array}{cl}
    \frac{1-p_0}{2} & \quad \quad a_1 = \frac{1}{\sqrt{1-p_0}} \; , \\
    p_0 & \quad \quad a_1 = 0 \; , \\
    \frac{1-p_0}{2} & \quad \quad a_1 = -\frac{1}{\sqrt{1-p_0}} \; .
  \end{array}
  \right.
  \label{eq:250118a}
\end{equation}
This prior distribution has a parameter $p_0$. We can easily check that the mean value of $a_1$ is zero and its variance is unity. For such a distribution, the mean $L_1$-norm energy is then
\begin{equation}
  \begin{aligned}
    E \, & = \, \sum\limits_{i=1}^{N} \biggl[ \sqrt{\frac{2 \sigma_i^2}{\pi}} \Bigl( 
      p_0 + (1-p_0) \exp\bigl( - \frac{\mu_i^2}{2 (1 - p_0) \sigma_i^2}
      \bigr)  \Bigr) \, + \, \sqrt{(1 - p_0) \mu_i^2} \,
      \textrm{erf}\Bigl( \frac{ | \mu_i|}{\sqrt{2 (1 - p_0) \sigma_i^2}}
      \Bigr) \biggr] \\
    & = \,  \sum\limits_{i=1}^{N} \Bigl[ \sqrt{\frac{2 \sigma_i^2}{\pi}} \bigl( 
      p_0 + (1-p_0) e^{- \zeta_i^2} \bigr)  \, + \,  \sqrt{(1-p_0) \mu_i^2} \, 
      \textrm{erf}( \zeta_i ) \Bigr] \; ,
  \end{aligned}
  \label{eqE1withp0}
\end{equation}
where $\zeta_i$ is computed through
\begin{equation}
  \zeta_i = \sqrt{\frac{ \mu_i^2}{2 (1 - p_0) \sigma_i^2}} \; .
\end{equation}
Notice that $\zeta_i$ is simply the (rescaled) signal-to-noise ratio $\eta_i$ (with $\zeta_i = \eta_i/\sqrt{2}$) as defined by Eq.~(\ref{eq241217b}) for the special case of $a_1 = 1/\sqrt{1-p_0}$.

As another concrete example, we assume the non-Gaussian coefficient $a_1$ is a continuous random variable sampled from the Laplace distribution, 
\begin{equation}
    q( a_1 ) \, = \, \frac{1}{\sqrt{2}} \exp\biggl( - \sqrt{2 a_1^2} \biggr) \; .
    \label{eq:250120a}
\end{equation}
It is again easy to check that the mean of $a_1$ is zero and the variance of $a_1$ is unity. The $L_1$-norm mean energy of this system, following Eq.~(\ref{eqL1Eexp}), can be computed through
\begin{equation}
\begin{aligned}
    E \, & = \, \sum\limits_{i=1}^N \biggl[ 
    \sqrt{ \frac{ 2 \sigma_i^2}{\pi} }
    +\sqrt{\frac{2 \mu_i^2}{\pi}} \exp\Bigl( \frac{\sigma_i^2}{\mu_i^2} \Bigr)  \int_{\sqrt{\sigma_i^2/\mu_i^2}}^\infty \textrm{d} t \, e^{-t^2} 
    \biggr] \\
    & = \, \sum\limits_{i=1}^N \biggl[ 
    \sqrt{ \frac{ 2 \sigma_i^2}{\pi} } 
    +\sqrt{\frac{\mu_i^2}{2}} 
    \exp\Bigl( \frac{\sigma_i^2}{\mu_i^2} \Bigr) \, \textrm{erfc}\Bigl( \sqrt{\frac{\sigma_i^2}{\mu_i^2}} \Bigr)
  \biggr] \; ,
    \end{aligned}
    \label{eqPa1Lp}
\end{equation}
where $\textrm{erfc}(z)$ is the complementary error function defined by
\begin{equation}
    \textrm{erfc}(z) \, \equiv \, \frac{2}{\sqrt{\pi}} \int_z^\infty e^{- t^2} \textrm{d} t \; .
\end{equation}
The energy expression (\ref{eqPa1Lp}) for the Laplace distribution is similar to Eq.~(\ref{eqE1withp0}) for the discrete distribution (\ref{eq:250118a}). The correctness of Eq.~(\ref{eqPa1Lp}) can be verified by noticing that
\begin{eqnarray}
    & & \sqrt{\frac{\sigma_i^2}{\pi}} \int_{-\infty}^{\infty} \textrm{d} a_1 \, e^{- \sqrt{2 a_1^2}} \exp\Bigl( - \frac{\mu_i^2 a_1^2}{2 \sigma_i^2} \Bigr) \, = \, \sqrt{\frac{8 \sigma_i^4}{\pi \mu_i^2}} \, 
    \exp\Bigl( \frac{\sigma_i^2}{ \mu_i^2} \Bigr) \int_{\sqrt{\sigma_i^2 /\mu_i^2}}^\infty e^{- y^2} \textrm{d} y \; , 
    \\
& &    \sqrt{\frac{8 \mu_i^2}{\pi}} \int_0^\infty \textrm{d} a_1 \, a_1 e^{-\sqrt{2} a_1} \int_0^{\mu_i a_1 /\sqrt{2 \sigma_i^2}} \textrm{d} t \, e^{-t^2} \, = \,  \sqrt{\frac{8 \mu_i^2}{\pi}} \int_0^\infty \textrm{d} t \, e^{-t^2} \int_{\sqrt{ 2 \sigma_i^2/\mu_i^2} t}^\infty \textrm{d} a_1 \, a_1 e^{-\sqrt{2} a_1} 
\nonumber \\
&  & \quad \quad = \,  \sqrt{\frac{8 \mu_i^2}{\pi}} \int_0^\infty \textrm{d} t \, e^{-t^2} 
\biggl[ \sqrt{\frac{\sigma_i^2}{\mu_i^2}}\, t \exp\Bigl( - \frac{2 \sigma_i}{\mu_i} t \Bigr) +
\frac{1}{2} \exp\Bigl( - \frac{2 \sigma_i}{\mu_i} t \Bigr) \biggr] 
\nonumber \\
& & \quad \quad = \,  \sqrt{\frac{2 \sigma_i^2}{\pi}} 
- \sqrt{\frac{8 \sigma_i^4}{\pi \mu_i^2}} \exp\Bigl( \frac{\sigma_i^2}{\mu_i^2} \Bigr) 
\int_{\sigma_i/\mu_i}^\infty \textrm{d} t \, e^{-t^2} 
+\sqrt{\frac{2 \mu_i^2}{\pi}} \exp\Bigl( \frac{\sigma_i^2}{\mu_i^2} \Bigr)  \int_{\sigma_i/\mu_i}^\infty \textrm{d} t \, e^{-t^2} \; .
\end{eqnarray}

As a third concrete example, we consider the non-Gaussian coefficient $a_1$ has discrete values
\begin{equation}
    a_1 \, = \, \pm\, c_0 2^n \quad \quad (n = 0, 1, \ldots, 9) \; ,
    \label{eq:250120b}
\end{equation}
and the probability of $n$ is
\begin{equation}
    p(n) \, = \, \frac{1}{Z} 2^{- n \gamma} \quad \quad (n = 0, 1\, \ldots, 9) \; , \quad
    \quad \quad Z \, = \, 
    \sum\limits_{n=0}^{9} 2^{- n \gamma} \; . 
\end{equation}
The value of $c_0$ is fixed by the requirement that the variance of $a_1$ should be equal to unity. We can easily check the discrete coefficient $a_1$ following the power-law with decay exponent $\gamma$:
\begin{equation}
    q( a_1 ) \, \propto |a_1|^{-\gamma} \; .
    \label{eqPa1SF}
\end{equation}

For such a power-law distribution, the mean $L_1$-norm energy $E$ is written down following  Eq.~(\ref{eqL1Eexp}) as
\begin{equation}
  E = \frac{1}{\sum_{n=0}^{9} 2^{- n \gamma} } \sum\limits_{n=0}^{9}
  2^{- n \gamma} \,     \biggl[ \sqrt{\frac{2 \sigma_i^2}{\pi}} \,
      \exp\Bigl( - \frac{c_0^2 2^{2 n} \mu_i^2}{2 \sigma_i^2} \Bigr)
      \, + \, |c_0 2^{n}  \mu_i |\, \textrm{erf}\Bigl(
      \frac{|c_0 2^n \mu_i|}{\sqrt{2 \sigma_i^2}} \Bigr) \biggr] \; .
      \label{eq:Epl0}
\end{equation}

\clearpage

\section{An example phase diagram for a small system}

\begin{figure}[h]
    \centering
    \includegraphics[width=0.33\linewidth]{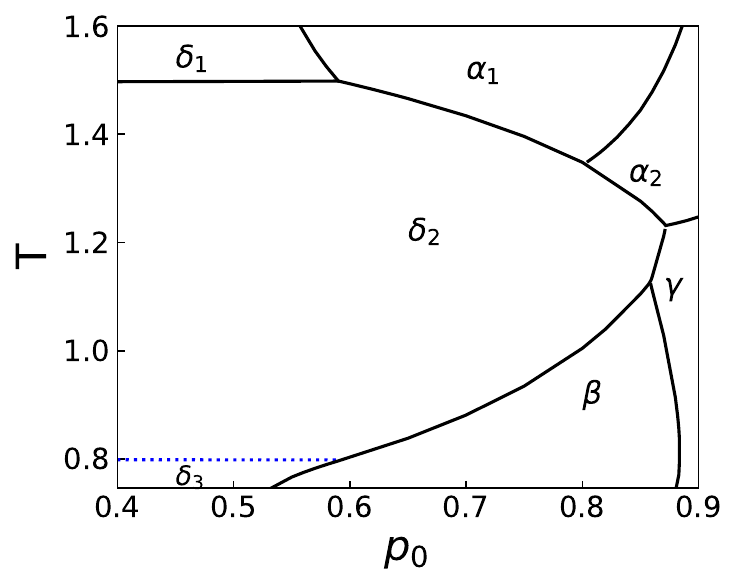}
    \caption{Phase diagram for the system of size $N=10$. The distribution $q(a_1)$ is described by Eq.~(\ref{eq:Pa1S}) with parameter $p_0$, and the feature vector $\bm{\phi}_1 = \frac{1}{\sqrt{N}} (1, \ldots, 1)^\top$. The dotted line indicates a continuous phase transition, and the solid lines denote discontinuous phases transitions. Phases $\delta_1$, $\delta_2$, and $\delta_3$ are unable to detect the hidden feature direction $\bm{\phi}_1$. In phases $\alpha_1$, $\alpha_2$, and $\beta$, one unit responds selectively to the feature direction $\bm{\phi}_1$. In the $\gamma$ phase, one unit responds very strongly to the feature direction $\bm{\phi}_1$ and another unit also partially detects the feature direction.}
    \label{fig:N10PhaseDiagram}
\end{figure}

\begin{figure}[h]
    \centering
    \subfigure[$\delta_1$]{
    \includegraphics[width=0.23\linewidth]{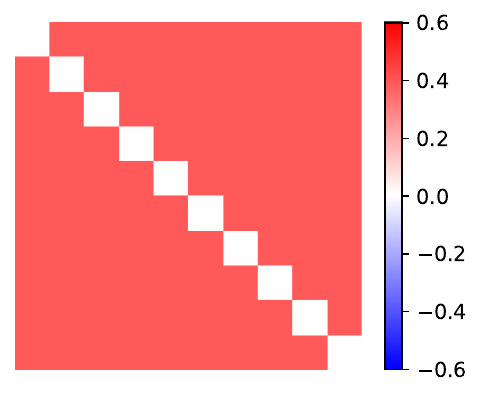} 
    \label{fig:EWn10a}
    }
    \subfigure[$\delta_2$]{
    \includegraphics[width=0.236\linewidth]{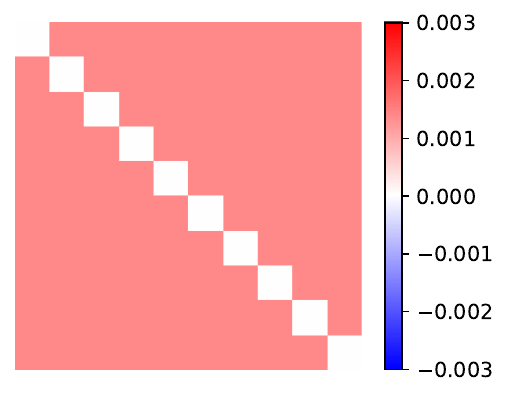}
    \label{fig:EWn10b}
    }
    \subfigure[$\delta_3$]{
    \includegraphics[width=0.234\linewidth]{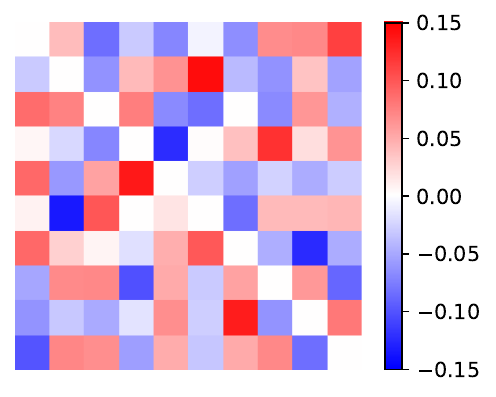}
    \label{fig:EWn10c}
    } \\
    
    \subfigure[$\alpha_1$]{
    \includegraphics[width=0.233\linewidth]{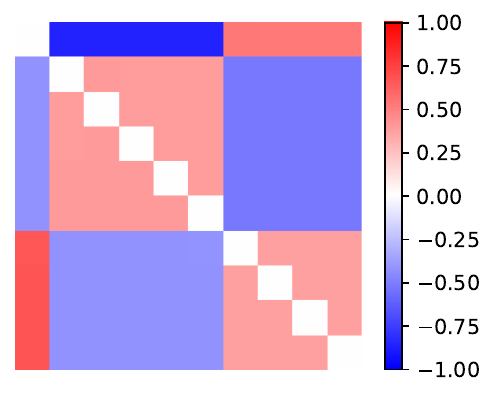}
    \label{fig:EWn10d}
    }
    \subfigure[$\alpha_2$]{
    \includegraphics[width=0.233\linewidth]{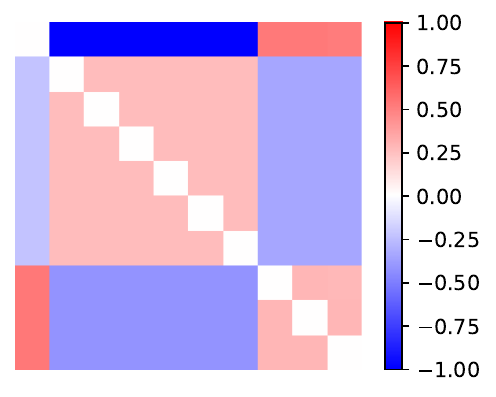} 
        \label{fig:EWn10e}
    }
    \subfigure[$\beta$]{
    \includegraphics[width=0.233\linewidth]{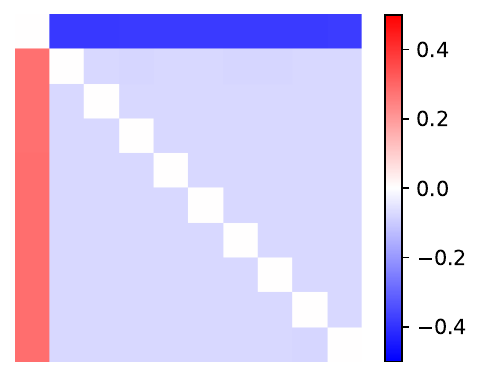}
    \label{fig:EWn10f}
}
    \subfigure[$\gamma$]{
    \includegraphics[width=0.233\linewidth]{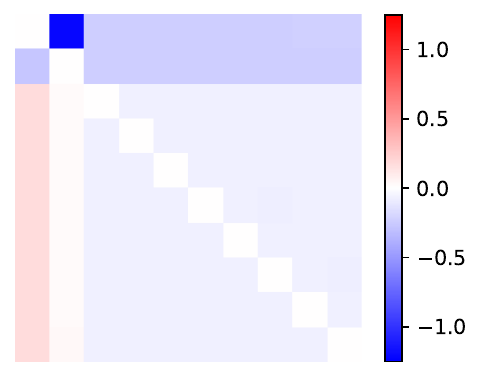} 
    \label{fig:EWn10g}
}
\caption{Example optimal weight matrices of size $N=10$ for different phases: (a) $\delta_1$ at $p_0=0.5$, $T=1.583$ with $Q = 0.316$; (b) $\delta_2$ at $p_0=0.5$, $T=1.401$ with $Q=0.316$; (c) $\delta_3$ at $p_0=0.5$, $T=0.782$ with $Q=0.316$; (d) $\alpha_1$ at $p_0=0.7$, $T=1.507$ with $Q= 0.933$; (e) $\alpha_2$ at $p_0=0.9$, $T=1.306$ with $Q=0.951$; (f) $\beta$ at $p_0=0.7$, $T=0.822$ with $Q= 0.872$;  (g) $\gamma$ at $p_0=0.9$, $T=0.871$ with $Q=0.861$.}
\label{fig:ExampleWeightsofEachPhase}
\end{figure}

Assuming the non-Gaussian coefficient $a_1$ is described by the discrete probability distribution

\begin{equation}
  q( a_1 ) \, = \,
  \left\{
  \begin{array}{ll}
    (1-p_0)/2 \; ,
    & \quad \quad a_1 = 1/ \sqrt{1-p_0} \; , \\
    p_0 \; , & \quad \quad a_1 = 0 \; , \\  
    (1-p_0) / 2 \; , & \quad\quad a_1 = -1 / \sqrt{1-p_0} \; ,
  \end{array}
  \right.
  \label{eq:Pa1S}
\end{equation}
and setting the feature direction as $\bm{\phi}_1 = \frac{1}{\sqrt{N}}(1, 1, \ldots, 1)^\top$, we obtain the phase diagram for a small system of size $N=10$ using $p_0$ and the tradeoff temperature $T$ as control parameters (Fig.~\ref{fig:N10PhaseDiagram}). We briefly describe this phase diagrams together with some example optimal weight matrices (Fig.~\ref{fig:ExampleWeightsofEachPhase}).


In phases denoted as $\delta_1$, $\delta_2$, and $\delta_3$, the system is unable to detect the hidden feature $\bm{\phi}_1$. It is observed that the temperature range within which the system fails to extract the feature decreases as $p_0$ increases. In the $\delta_1$ phase, the weights are permutation symmetric such that all the weights $w_{i j}$ are the same, rendering the system incapable of feature detection (Fig.~S\ref{fig:EWn10a}). For instance, at $T = 1.583$ and $p_0=0.5$, the overlap value of the optimal network is $Q = 0.316$, which is very close to the lower-bound $10^{-\frac{1}{2}}$. In the $\delta_2$ phase, the weights are also permutation symmetric, but the elements are very small (Fig.~S\ref{fig:EWn10b}). In the $\delta_3$ phase, the weights lack permutation symmetry (Fig.~S\ref{fig:EWn10c}). The system remains unable to detect the feature. For example, at $T = 0.782$ and $p_0=0.5$, the overlap value is also $Q = 0.316$.

In the $\alpha_1$ and $\alpha_2$ phases, one unit becomes selective to the feature, while the remaining units primarily represent noise and are divided into different groups. In the $\alpha_1$ phase, one single unit detects the feature (Fig.~S\ref{fig:EWn10d}). The interactions between it and a group $A$ of five units are all excitatory (negative $w_{i j}$),  while the interactions with the remaining group $B$ of four units are inhibitory (positive $w_{i j}$).  The units within the groups $A$ and $B$  inhibit each other, while units from different groups excite each other. The overlap is very high. For example, at $T=1.507$ and $p_0=0.7$, $Q = 0.933$. In the $\alpha_2$ phase, the network consists of one single unit detecting the feature and two other groups of units (see Fig.~S\ref{fig:EWn10e}), similar to the $\alpha_1$ phase. However, in the $\alpha_2$ phase, one group $A$ contains six units, and the other group $B$ contains three units. At the point $T=1.306$ and $p_0=0.9$, the overlap is $Q = 0.951$.

In the $\beta$ phase, a single unit (say unit $i=1$) extracts the feature and all the other units from a single group $A$ (Fig.~S\ref{fig:EWn10f}).  Unit $1$ inhibits all the units of group $A$ and it is excited by group $A$. The nine units of group $A$ weakly excite each other. At the point $T=0.822$ and $p_0=0.7$, the overlap $Q= 0.872$.

In the $\gamma$ phase, one unit (say unit $i=1$) is highly selective to the feature, and another unit (unit $j=2$) is partially selective (Fig.~S\ref{fig:EWn10g}). These two units inhibit the other eight units and are excited by them. The other eight neurons weakly excite each other. At the point $p_0=0.9$ and $T=0.871$, the overlap is $Q=0.861$.  Besides the order parameter $Q$, we may also consider the signal ratio, defined as $\hat{\mu}_i = \sqrt{\mu_i^2 / (\sigma_i^2 + \mu_i^2)}$, to characterize the proportion of feature signal in the output of unit $i$. The signal ratios $\hat{\mu}_i$ for the ten units are, in descending order,  $1$, $0.807$, $0.078$, $0.077$, $0.077$, $0.077$, $0.077$, $0.077$, $0.077$, $0.076$.

We note that Fig.~\ref{fig:N10PhaseDiagram} shows only part of the phase diagram. Here, we focus on the temperature range of $T \in (0.75, 1.6)$ to demonstrate the influence of $p_0$ on the feature detection capability. As the temperature increases beyond $T=1.6$ or decreases below $T=0.75$, more phase transitions may occur. For instance, we find that, as the temperature $T$ decreases, the symmetry of the nine non-selective units in the $\beta$ phase will break. With a further decrease in the temperature $T$, the minimum value $\lambda_0$ of the real parts of the eigenvalues of the matrix $\bm{I}+\bm{W}$ will reach and stay at the lower-bound value (set to be $10^{-5}$). 

\clearpage

\section{More numerical results on the median-sized system}

In addition to the results shown in the main text, here we present more numerical results for the median-sized ($N=36$) system.

\begin{figure}[h]
  \centering
  \subfigure[]{
    \includegraphics[angle=270,width=0.31\linewidth]{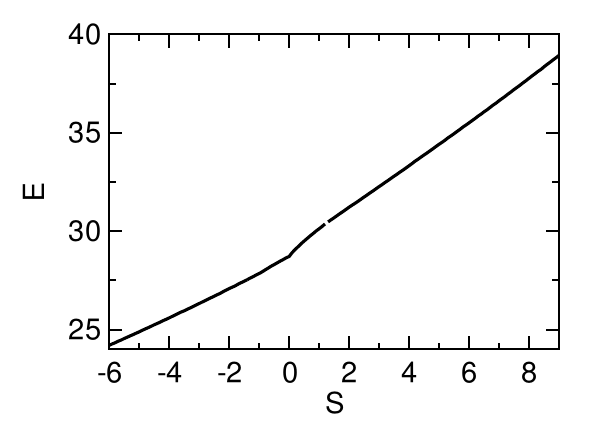}
    \label{fig:S_E_N36_R0}
  } 
  \subfigure[]{
    \includegraphics[angle=270,width=0.31\linewidth]{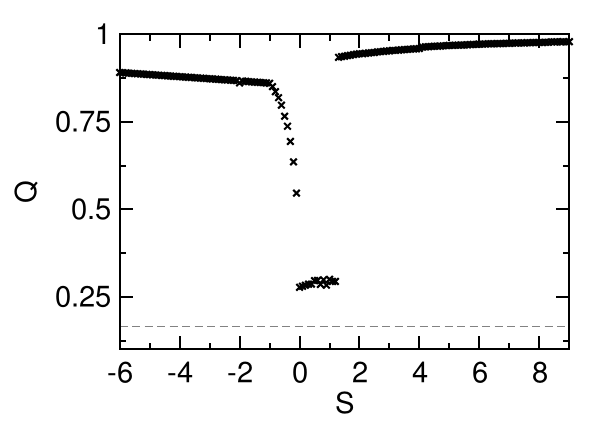}
    \label{fig:S_Q_N36_R0}
  } 
  \subfigure[]{
    \includegraphics[angle=270,width=0.31\linewidth]{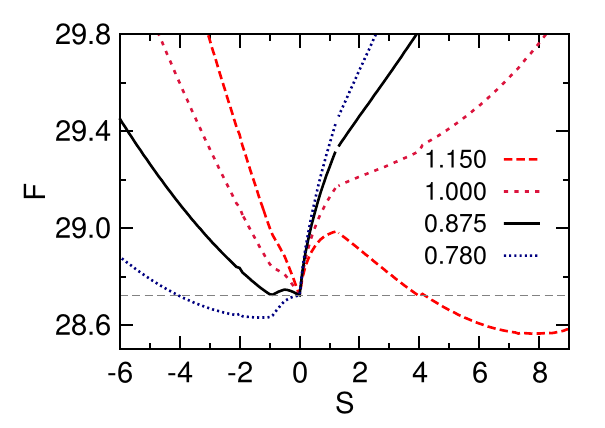}
    \label{fig:S_F_N36_R0}
  }
  \caption{
    Thermodynamic quantities for the case of $N=36$ and $p_0=0.7$ with a random feature direction $\bm{\phi}_1$. 
    (a) Minimum energy $E$ versus entropy $S$. (b) Overlap $Q$ versus $S$. (c) Free energy $F = E - T S$ versus $S$ at $T=0.78$, $0.875$, $1.0$, and $1.15$. 
  }
  \label{fig:ESN36R0}
\end{figure}

First, we investigate whether the feature direction $\bm{\phi}_1$ will have a qualitative influence of the property of the system. For this purpose, we generate many random feature directions $\bm{\phi}_1 = (\phi_{1, 1}, \phi_{2, 1}, \ldots, \phi_{N, 1})^\top$ by sampling $\phi_{j, 1}$ independently and uniformly randomly from the interval $(-1, 1)$. Each generated $\bm{\phi}_1$ is then rescaled to the unit length, that is, $\sum_j \phi_{j, 1}^2  = 1$. We then solve the optimal LPC weight matrix problem assuming the non-Gaussian coefficient $a_1$ is distributed according to Eq.~(\ref{eq:Pa1S}) with $p_0 = 0.7$. 

\begin{figure}[h]
  \centering
  \subfigure[]{
    \includegraphics[angle=270,width=0.31\linewidth]{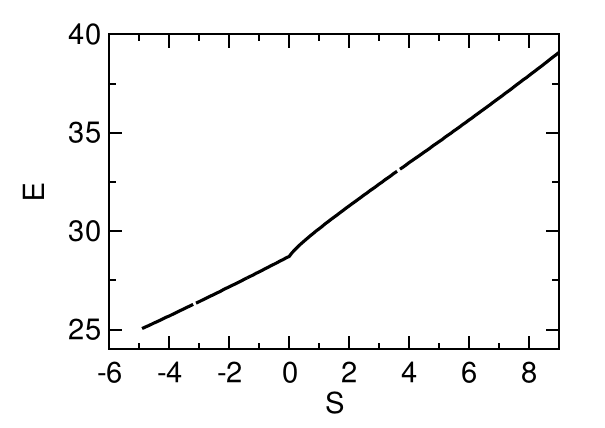}
    \label{fig:S_E_N36_p0p6}
  } 
  \subfigure[]{
    \includegraphics[angle=270,width=0.31\linewidth]{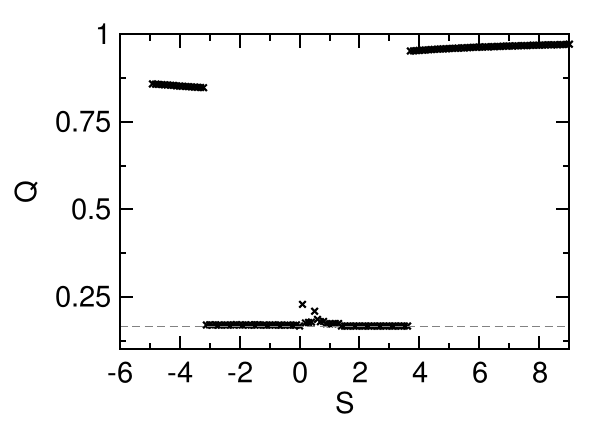}
    \label{fig:S_Q_N36_p0p6}
  } 
  \caption{
Thermodynamic quantities for the case of $N=36$ and $p_0=0.6$ with the feature direction being uniform, $\bm{\phi}_1 = (1/6, 1/6, \ldots, 1/6)^{\top}$.    (a) Minimum energy $E$ versus entropy $S$. (b) Overlap $Q$ versus  entropy $S$.
  }
  \label{fig:ESN36p0p6}
\end{figure}

The numerical results for all these sampled random feature directions $\bm{\phi}_1$ are qualitatively similar, indicating that the discontinuous emergence of feature detection function is a general property of the linear LPC network. As a concrete example, we show in Fig.~\ref{fig:ESN36R0} the results obtained for a single random feature direction $\bm{\phi}_1$. In comparison with Fig.~3 of the main text, the only major difference may be that the overlap $Q$ at $S\in (0, 1.3)$ is elevated to $Q \approx 0.3$. 

Second, we consider the effect of decreasing the value of $p_0$. As $p_0$ is decreased, the probability distribution $q(a_1)$ become less deviated from being Gaussian. In agreement with Fig.~\ref{fig:N10PhaseDiagram}, we find that as $p_0$ decreases, the onset of feature detection occurs at larger absolute values of $S$. An concrete example is shown in Fig.~\ref{fig:ESN36p0p6} for $p_0=0.6$. In comparison with Fig.~3 of the main text, we see that at  $p_0 = 0.6$, feature detection is possible only at much lower $S$ values ($S  < - 3.1$) or much higher values ($S > 3.6$). The range of failure to graph the hidden feature direction is enlarged ($-3.1 \leq S \leq 3.6$).

\clearpage

\section{Analysis of the Laplace-distributed feature}

\begin{figure}[h]
    \centering
    \subfigure[]{
    \includegraphics[width=0.23\linewidth]{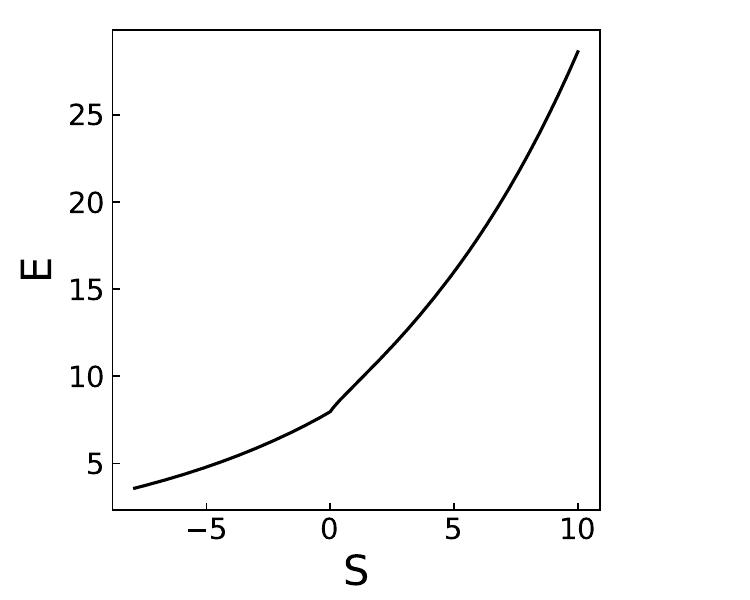}
    }
    \subfigure[]{
    \includegraphics[width=0.23\linewidth]{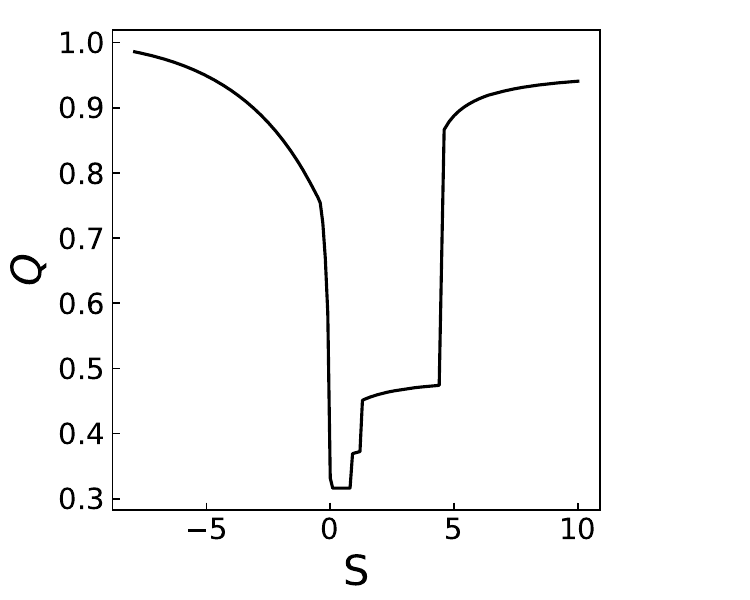}
    \label{fig:250120b}
    }
     \subfigure[]{
    \includegraphics[width=0.23\linewidth]{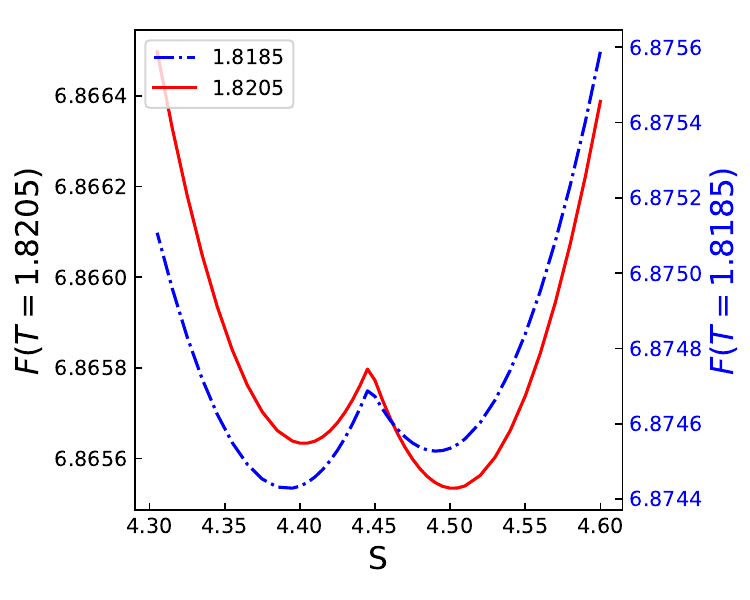}
    \label{fig:250120c}
    }
     \subfigure[]{
    \includegraphics[width=0.23\linewidth]{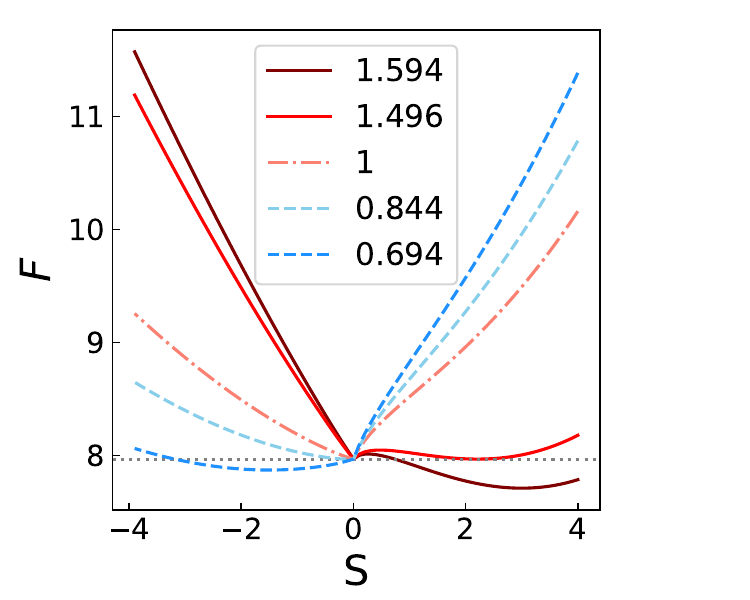}
    \label{fig:250120d}
    }
    \caption{Thermodynamic quantities for the case of $N=10$ with the Laplace distribution (\ref{eqPa1Lp}). (a)  Energy $E$ versus  entropy $S$. (b) overlap $Q$ versus $S$.  (c) Free  energy $F = E - T S$ versus $S$ at $T=1.8185$ (dashed line) and $T=1.8205$ (solid line). (d) Free energy $F$ at several other tradeoff temperatures $T = 0.694$, $0.844$, $1.0$, $1.496$, and $1.594$. The feature direction $\bm{\phi}_1$ is uniform with all its elements taking the same value.}
    \label{fig:Laplacefeature}
\end{figure}

When the non-Gaussian coefficient $a_1$ follows the continuous Laplace distribution Eq.~(\ref{eq:250120a}), the mean energy $E$ can be computed through Eq.~(D5) of the main text.  Figure~\ref{fig:Laplacefeature} reports the numerical results obtained for this problem ensemble with $N=10$ units. These results closely resemble those of the ensembles with discrete $a_1$ values.

\begin{figure}[h]
    \centering
    \subfigure[$\ S=-2$]{
    \includegraphics[width=0.3\linewidth]{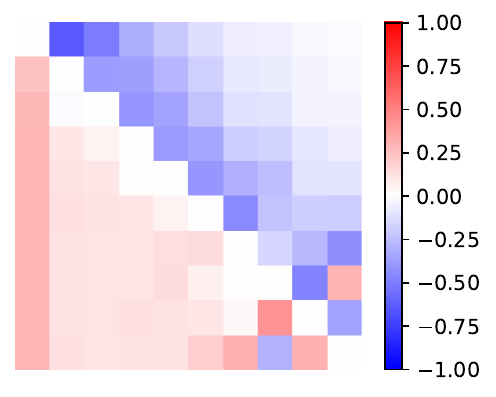} 
    \label{fig:250121a}
    }
\subfigure[$\ S=-1$]{
    \includegraphics[width=0.3\linewidth]{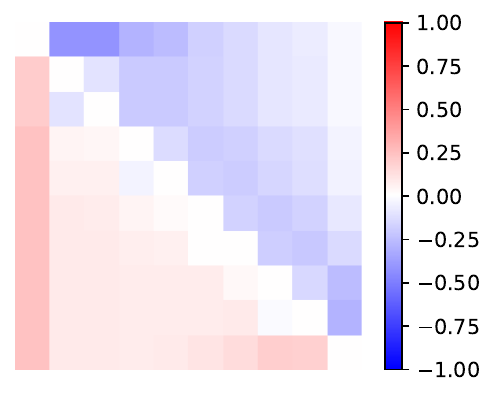}
    \label{fig:250121b}
    } 
    \subfigure[$\ S=8$]{
    \includegraphics[width=0.3\linewidth]{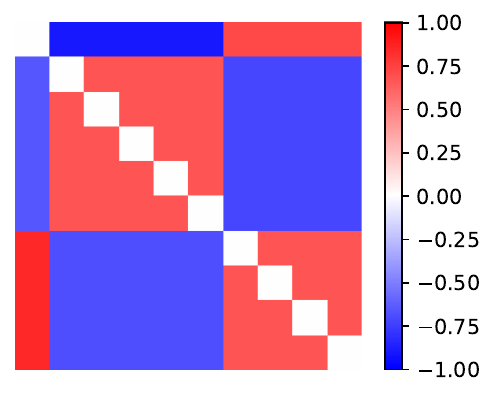}
    \label{fig:250121c}
    }
    \caption{Optimal weight matrices for the system with Laplace-distributed  coefficient $a_1$ and size $N=10$. The entropy value is $S=-2$ (a), $-1$ (b), and $8$ (c).}
        \label{fig:N=10-Laplace-W}
\end{figure}

Both at the low entropy ($S < -0.41$) and the high entropy ($S > 4.5$) regions, the optimal LPC matrix is capable of detect the non-Gaussian feature direction $\bm{\phi}_1$, while at the intermediate region of $S \in (-0.41, 4.5)$ the overlap order parameter $Q$ is relatively small (Fig.~S\ref{fig:250120b}). 

If the tradeoff temperature $T$ is used as the control parameter, we find that when $T > 1.8195$, there is only one global minimum of $F$ and the overlap $Q$ is very large. At $T = 1.8195$, two degenerate optimal solutions emerge: one at $S=4.495$ with $Q=0.858$, and the other at $S=4.395$ with $Q=0.474$. The optimal system switches from one solution branch to the other, characterizing a discontinuous phase transition (Fig.~S\ref{fig:250120c}). As the temperature further decreases to  $T=1.496$, the global minimum energy shifts from the branch at $S=2.21$, $Q=0.327$ to the other branch at $S=0$, $Q=0.325$ (Fig.~S\ref{fig:250120d}). Within the temperature range of $(0.844, 1.496)$, the system becomes stuck in the optimal solution at $S=0$ and small $Q=0.325$.   When the temperature drops to $T = 0.844$, the overlap suddenly jumps to a value $Q =0.548$ as the free energy minimum position changes to  $S=-0.07$. As the temperature further decreases, $Q$ rapidly increases, and then at $T=0.781$ (and $S=-0.41$) the optimal weight matrix experiences a continuous phase transition  with a kink of the overlap $Q$ (Fig.~S\ref{fig:250120b}).

Some example weight matrices are shown in Fig.~\ref{fig:N=10-Laplace-W}. At high entropy levels, the optimal weight matrices exhibit grouping and a high degree of symmetry. For example, at $S = 8$ (Fig.~S\ref{fig:250121c}), a single unit detects the feature direction $\bm{\phi}_1$, while the other five units form a group (say $A$) and the remaining four units form another group (say $B$). The selective unit and units of group $A$ mutually excite each other, while the selective unit and units of group $B$ inhibit each other. Units of group $A$ and units of group $B$ mutually excite each other. The interactions within group $A$ and group $B$ are all inhibitory. Overall, it shows a high degree of symmetry in this high entropy system. Conversely, when the entropy $S$ is weakly negative, the optimal weight matrices display a lower degree of symmetry, as depicted in Figs.~S\ref{fig:250121a} and S\ref{fig:250121b}. In the optimal network, the selective unit strongly inhibits the other units and is excited by them. The weights $w_{i j}$ between the remaining units are not symmetric. The lower the entropy, the lower the degree of symmetry.

\clearpage

\section{The case of power-law distribution for the non-Gaussian coefficient}

We consider the power-law distribution Eq.~(\ref{eqPa1SF}) for the non-Gaussian coefficient $a_1$. For computational simplicity the values of $a_1$ are restricted to only $20$ different values as specified by Eq.~(\ref{eq:250120b}). The mean energy of such a system is then computed through Eq.~(\ref{eq:Epl0}). For simplicity we assign the feature direction as $\bm{\phi}_1=(\frac{1}{\sqrt{10}}, \ldots, \frac{1}{\sqrt{10}})^\top$.

\begin{figure}[h]
    \centering
    \subfigure[$\ \gamma = 1$]{
    \includegraphics[width=0.23\linewidth]{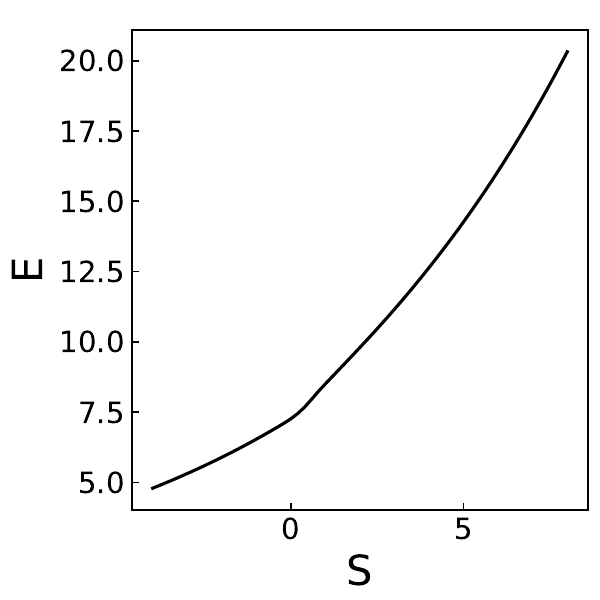} }
    \subfigure[$\ \gamma = 1$]{
    \includegraphics[width=0.235\linewidth]{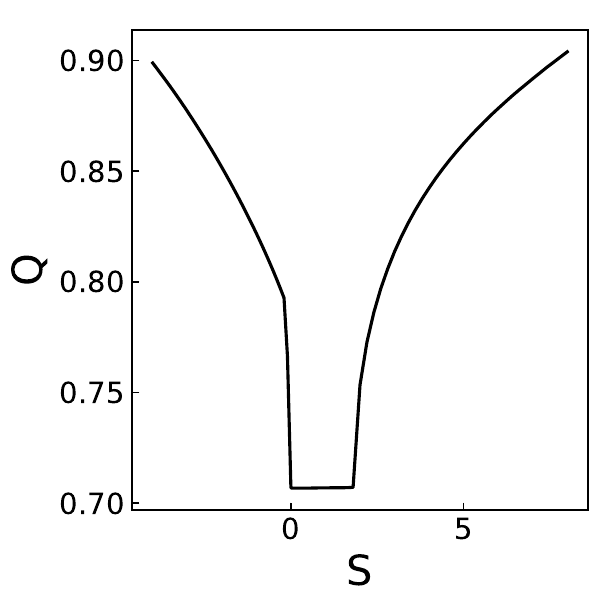} 
    \label{fig:250122a}
    }
    \subfigure[$\ \gamma = 1.5$]{
    \includegraphics[width=0.23\linewidth]{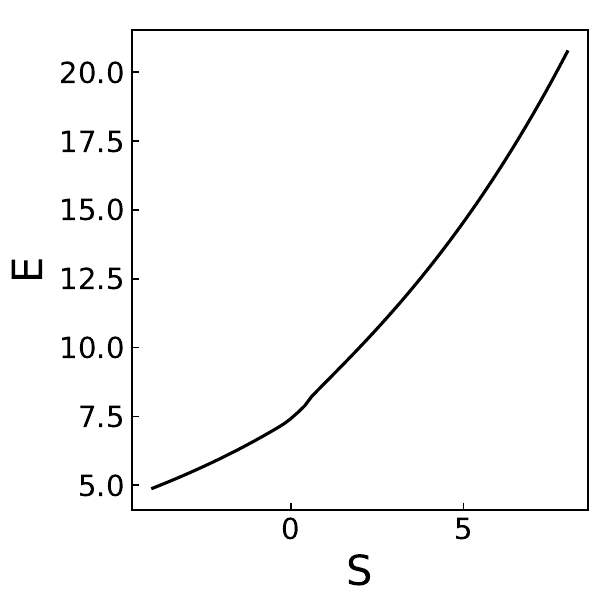} }
    \subfigure[$\ \gamma = 1.5$]{
      \includegraphics[width=0.23\linewidth]{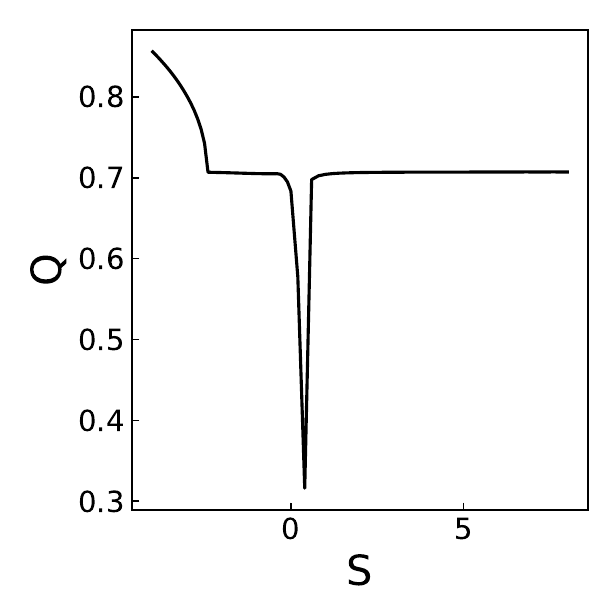}
    }
    \caption{Results for power law distributed features. The energy versus entropy for $\gamma = 1$ (a) and $\gamma = 1.5$ (c). The overlap parameter $Q$ for $\gamma = 1$ (b) and $\gamma = 1.5$ (d). The system size is $N=10$. The feature direction $\bm{\phi}_1$ is uniform with all its elements taking the same value.}
    \label{fig:N=10-powerlaw}
\end{figure}
\begin{figure}[h]
    \centering
    \subfigure[$\ S=-2$]{
      \includegraphics[width=0.3\linewidth]{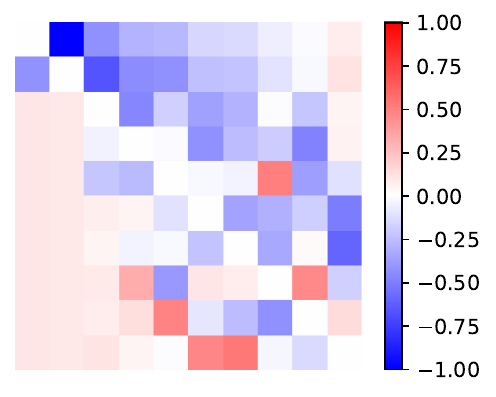}
    }
    \subfigure[$\ S=0$]{
      \includegraphics[width=0.3\linewidth]{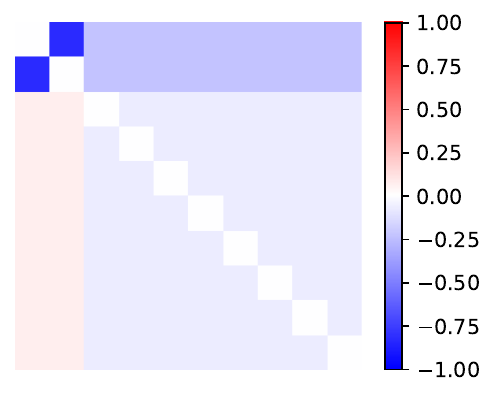}
    }
    \subfigure[$\ S=4$]{
      \includegraphics[width=0.3\linewidth]{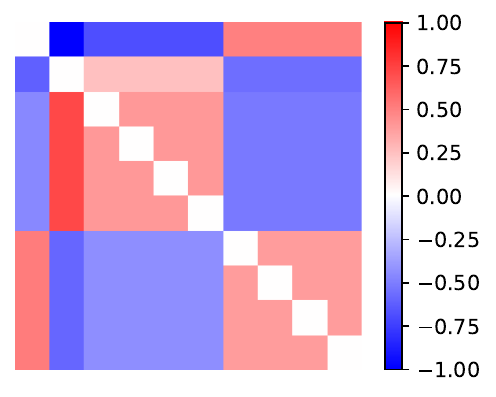}
    }
    \caption{Several example optimal weight matrices of size $N=10$, obtained for the power-law distribution of coefficient $a_1$ with exponent $\gamma=1$. The entropy values are $S=-2$ (a), $S=0$ (b), and $S=4$ (c), which are located respectively at the three different regions of Fig.~S\ref{fig:250122a}.
    }
    \label{fig:N=10-powerlaw-W}
\end{figure}

The numerical results for power-law distributed coefficient $a_1$  are similar to those discussed in the main text and in the preceding subsections. We present these results in Fig.~\ref{fig:N=10-powerlaw} for system size $N=10$ and power-law exponent $\gamma = 1$ and $\gamma = 1.5$.  In the case of $\gamma = 1$, a single unit in the system detects the feature at both low entropy (e.g., $S = -4$ with $Q=0.899$) and high entropy (e.g., $S = 8$ with $Q=0.904$). At a median entropy range $(0, 1.8)$, two units have the same $\mu_i$, while the other units have $\mu_i$ near zero, and the overlap order parameter is also relatively high ($Q \approx 0.71$), indicating that two units in the system jointly represent the non-Gaussian feature direction $\bm{\phi}_1$. For $\gamma = 1.5$, one unit detects the feature $\bm{\phi}_1$ at low entropy (e.g., $S = -4$ with $Q=0.856$). However, at high entropy $S$, two units again jointly represent the feature, similar to the cases of $S \in (0, 1.8)$  for $\gamma = 1$. In a small range of entropy around $S = 0.4$, the system cannot detect the feature (e.g., $S = 0.4$ with $Q=0.316$).

We present some example optimal weight matrices of size $N=10$  obtained for the case of $\gamma =1$ in Fig.~\ref{fig:N=10-powerlaw-W}. We see that at entropy $S$ close to zero, two units (say unit $1$ and $2$) have the same large value of $\mu_1=\mu_2$ and the other eight units have small $\mu_i$ values. For example, at $S=0$, $\mu_1 = \mu_2 = 2.234$ while $\mu_i = 0.029$ for all the other eight units. The overlap order parameter is $Q=0.7068$, close to $\frac{1}{\sqrt{2}} = 0.7071$.  As entropy $S$ increase or decrease from zero ($S > 1.8$ or $S < 0$), the symmetry of the two units $1$ and $2$ break and only one of them is responding strongly and selectively to the feature direction $\bm{\phi}_1$, and hence the system will have very higher level of $Q > \frac{1}{\sqrt{2}}$.  

When $S = 4$ the ten units of the network form three major groups: unit $1$ is selectively responding to the feature direction $\bm{\phi}_1$, units $2$-$6$ form group $A$, and units $7$-$10$ form group $B$. Group $A$ can be divided into two subgroups, namely unit $2$ on one side and units $3$-$6$ on the other side.

When $S = -2$ the optimal weight matrix does not have clear hierarchical structure, but we can still group unit $1$ and $2$ together and regard the other eight units as forming a single group. A major difference with the optimal matrix at $S=0$ is that the symmetry between units $1$ and $2$ is broken and the symmetry within the other eight units is also broken.  This symmetry-breaking enables unit $1$ to be most selectively responding to the feature direction $\bm{\phi}_1$.

If the power-law exponent $\gamma$ becomes large, e.g., $\gamma = 3$, we find that the optimal LPC network fails to detect the non-Gaussian feature direction $\bm{\phi}_1$ for the entropy $S$ range examined in our numerical simulations. The reason is that the coefficient $a_1$ becomes too concentrated at very small values. 

\clearpage
\section{Detection of two orthogonal non-Gaussian features}

The main text has demonstrated in Fig.~6 some results concerning the detection and separation of two non-orthogonal features (with angle $\theta = \pi/4$ between them). Here we show the qualitatively similar results obtained with two orthogonal features (with angle $\theta = \pi /2$). The two random orthogonal base vectors $\vec{\bm{\phi}}_1$ and $\vec{\bm{\phi}}_2$ are the same as used in Fig.~6, as well as the same system size $N=16$ and the same non-Gaussian parameter $p_0=0.6$.

\begin{figure}[h]
    \centering
    \subfigure[]{
      \includegraphics[angle=270,width=0.3\linewidth]{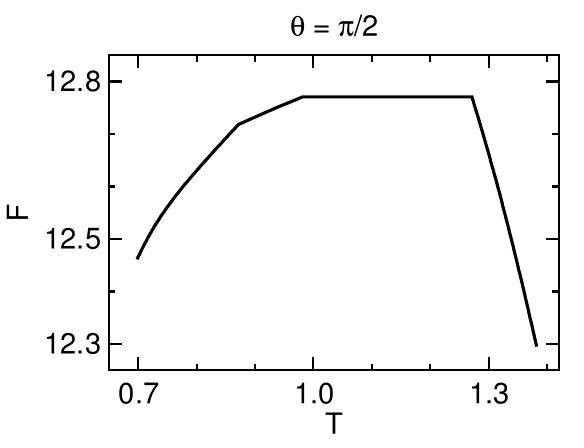}
      \label{fig:TF90TvsF}
      }
    \subfigure[]{
    \includegraphics[angle=270,width=0.3\linewidth]{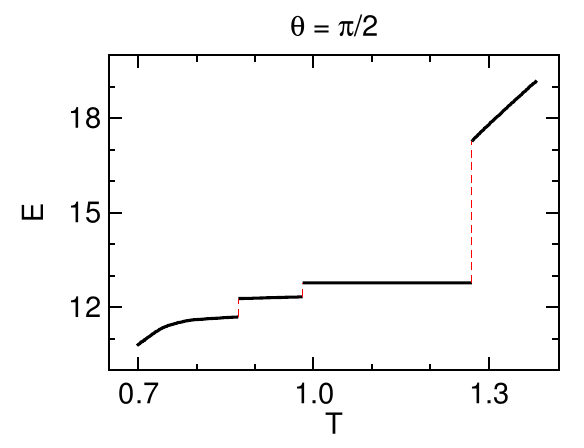}
    \label{fig:TF90TvsE}
    }
    \\
    \subfigure[]{
    \includegraphics[angle=270,width=0.3\linewidth]{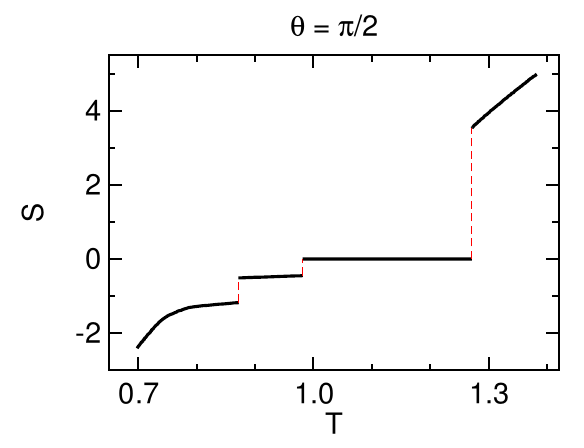} 
    \label{fig:TF90TvsS}
    }
    \subfigure[]{
    \includegraphics[angle=270,width=0.3\linewidth]{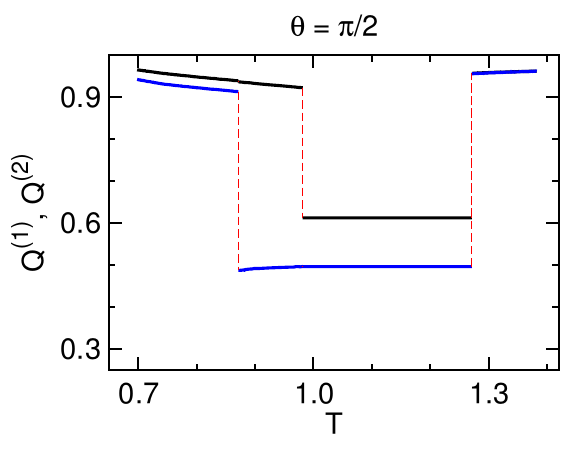}
    \label{fig:TF90TvsQ}
    }
    \caption{Thermodynamic quantities of optimal lateral predictive coding for input vectors containing two orthogonal random features (with angle $\theta = \pi/2$). The independent coefficients $a_1$ and $a_2$ follow the non-Gaussian distribution (\ref{eq:250118a}) with parameter $p_0=0.6$. Network size $N=16$. We use temperature $T$ as the control parameter.  (a) Minimum free energy $F$. (b) Mean energy $E$. (c) Entropy $S$. (d) Order parameters $Q^{(1)}$ and $Q^{(2)}$. The vertical dashed lines at $T = 0.8715$, $0.9816$ and $1.2711$ mark the three discontinuous phase transitions.
    }
    \label{fig:TF90N16}
\end{figure}

We see from Fig.~\ref{fig:TF90N16} that there are three discontinuous phase transitions at $T = 0.8715$, $0.9816$, $1.2711$. 

Both at low temperatures $T < 0.8715$ ($S < -1.174$, $E < 11.6946$) and at high temperatures $T > 1.2711$ ($S > 3.535$, $E > 17.2640$) the system is capable of detecting the two orthogonal non-Gaussian features and separating them by two different single units. 

In the temperature range $T \in (0.8715, 0.9816)$ and consequently $S \in (-0.51, -0.45)$ and $E \in (12.2733, 12.3288)$, the system can detect one of the two non-Gaussian features by the strong response of a single unit. The order parameter $Q^{(2)} \approx 0.5$ for the other feature direction is much weaker.

In the temperature range $T \in (0.9816, 1.2711)$ the optimal weight matrix is not changed and it has entropy $S = 0$ and energy $E=12.7705$, and this system fails to selectively respond to the two non-Gaussian features by single units (both order parameters $Q^{(1)}$ and $Q^{(2)}$ are much less than unity).

\clearpage

\section{Extension to include memory effect and nonlinearity}

Here we briefly mention two simple extensions of the present lateral predictive coding model.

One extension is to consider memory effect. We can we can modify the recursive dynamical process to the following form:
\begin{equation}
\tau_0 \frac{\textrm{d} x_i(t)}{\textrm{d}t} \, = \, s_i(t) - x_i(t) - \sum_{j \neq i} w_{i j} f[x_j(t)] \; ,
\label{eqRa260425}
\end{equation}
where $f[x_j(t)]$ is a functional of the internal state $x_j$ of unit $j$ up to time $t$. Convenient choices might be the exponentially decaying memory kernel
\begin{equation}
    f[x_j(t)] \, = \, \frac{1}{\tau_\textrm{m}} \int_0^\infty e^{-t^\prime /\tau_{\textrm{m}}} \ x_j(t-t^\prime)\, \textrm{d} t^\prime \; ,
\end{equation}
or the bell-shaped memory kernel
\begin{equation}
    f[x_j(t)] \, = \, \frac{1}{\tau_\textrm{m}^2} \int_0^\infty t^\prime e^{-t^\prime /\tau_{\textrm{m}}} \ x_j(t-t^\prime)\, \textrm{d} t^\prime \; .
\end{equation}
Notice that, if the memory time constant $\tau_m$ is much shorter than the time scale $\tau_0$ of Eq.~(\ref{eqRa260425}) while the time constant of the external input $\vec{\bm{s}}(t)$ is much longer than $\tau_0$, then the steady-state of Eq.~(\ref{eqRa260425}) can be well approximated by Eq.~(2) of the main text, and the results of our present work are also applicable.

If the memory effect is negligible but there is strong nonlinearity, we may assume $f[x_j(t)]$ to be a bounded function such as $f[ x_j(t)] = \tanh x_j(t)$, or be the rectified linear function $f[x_j(t)] = \max\bigl( 0, x_j(t) \bigr)$. Theoretical investigations on nonlinear LPC systems within the theoretical framework of energy--information tradeoff is still an open issue.

\end{document}